\documentstyle[preprint,eqsecnum,aps,epsf]{revtex}

\tightenlines
\renewcommand{\theequation}{\arabic{section}.\arabic{equation}}

\begin{document}
\draft

\preprint{Report number: UT-Komaba 96-16}

\title{T-Duality Transformation and Universal Structure\\
of Non-Critical String Field Theory}

\author{
Takashi Asatani
\footnote{\tt asatani@hep1.c.u-tokyo.ac.jp}
\, \,
Tsunehide Kuroki
\footnote{\tt kuroki@hep1.c.u-tokyo.ac.jp}
\, \,
Yuji Okawa
\footnote{\tt okawa@hep1.c.u-tokyo.ac.jp}
\footnote{JSPS Research Fellow} \\
Fumihiko Sugino
\footnote
{Present address: KEK Theory Group, Tsukuba, Ibaraki 305, Japan
and \tt sugino@theory.kek.jp
}
\, \, Tamiaki Yoneya
\footnote{\tt tam@hep1.c.u-tokyo.ac.jp}
}

\address{ Institute of Physics, University of Tokyo, Komaba,
Meguro-ku, Tokyo 153, Japan}

\maketitle

\begin{abstract}
We discuss a T-duality transformation for
the $c=1/2$ matrix model for the purpose of
studying duality transformations in a possible
 toy example of nonperturbative frameworks of string theory.
Our approach is to first investigate the scaling limit of the
 Schwinger-Dyson equations and
the stochastic Hamiltonian
in terms of the dual variables and then compare the
results
with those using the original spin variables.
It is shown that the $c=1/2$ model in the scaling
limit is T-duality symmetric in the sphere approximation.
The duality symmetry is however violated
when the higher-genus
effects are taken into account, owing to the
existence of global $\mbox{\boldmath $Z$}_2$ vector fields corresponding
to nontrivial homology cycles.
Some universal properties
of the stochastic Hamiltonians which play an
important role in discussing the scaling limit and have been
discussed in a previous work by the last two authors
are refined
in both the original and dual formulations.
We also report a number of new explicit
results for various amplitudes containing macroscopic
loop operators.

\end{abstract}


\narrowtext
\section{Introduction}
Recently, dual transformations in various different forms
 are playing increasingly important roles in string theory.
Historically, one of the first useful examples of the dual transformations
was the Kramers-Wannier duality \cite{KW} for the two-dimensional Ising
model on the square lattice, which allows us to locate exactly the
critical temperature for the order-disorder phase transition.
The familiar T-dual transformation in closed string theories compactified on a
torus is basically the same as the Kramers-Wannier
dual transformation on  discretized world sheets
for a compact $U(1)$-spin model (an XY model
with a Villain-type action, see, e.g. \cite{sw} and
Appendix A), where
the circle as the target space becomes the spin variable and
the role of the compactification radius is replaced by the inverse temperature.
A remarkable feature
of the string T-duality is that the duality symmetry
is exact for any world sheet of arbitrary fixed genus,
owing to the symmetrical roles  \cite{KY} played by the momentum
and winding modes of closed strings.
On the other
hand, the conjectured S-duality transformation of string
theory is a duality with respect to
the target space itself and is a generalization of the electric-magnetic
duality of gauge field theories, exchanging the coupling
constant and its inverse as required by the Dirac condition.
Complete understanding of all the duality transformations, especially
the S-dualities, in
string theory would obviously require some nonperturbative framework.
The recent developments \cite{Pollec}
 of the idea of Dirichlet branes in fact
suggest that the T-dual transformations should also play some important
roles in nonperturbative physics.

In view of this situation,  it seems worthwhile to formulate the
duality symmetries in a simplest setting for
nonperturbative string theory. Only known effective formulation
of string theory which is in principle nonperturbative is the
matrix model corresponding to $c\le 1$ conformal matter
coupled with the world-sheet metric. In the present paper, as a
simple exercise towards nonperturbative understanding
of dualities in string theory,
we study the T-dual transformation of the double-scaled
two-matrix model
which is the Ising model at the critical point on random surfaces.
Naively, we expect that the model should be self-dual, since
the Ising system sits precisely at the critical point.
We will in fact argue
that the self-duality is valid in the sphere approximation.
In higher genera, however, the duality symmetry is violated
owing to the
effect of nontrivial homology cycles. Here, there is no analogue of
the symmetry  between the momentum and winding modes
which is responsible for the exact duality symmetry
in the torus-compactified model.  Rather, the dual theory
must be described by vector fields residing
on links. Only in the special case of sphere,
the vector fields can be expressed in
terms of the dual spin fields.

Another motivation for studying this model is to further pursue
the remarkable universal structure of non-critical string field
theory \cite{IK} which is relevant in taking the scaling limit
and for studying background independence,
using both the original
and dual variables and to
complement the results of the previous work \cite{SY} by two
of the present authors.

The plan of the present paper is as follows.
In the next section, we first explain the
Kramers-Wannier duality transformation
of the two-matrix model.
Although the dual transformation of the two-matrix model
has been discussed in the literature,\footnote{
In particular, we mention the reference \cite{COT} in which
the disk amplitude in the dual basis was derived in the
sphere approximation.
 }
we here explain some elementary details of the construction
for the purpose of making our interpretation to be
sufficiently definite.  In particular, we emphasize that the
nontrivial homology cycles generally
violate the duality symmetry between
the spin and dual-spin variables,
since they
allow global vector fields.
In section III, following the method of the
previous paper \cite{SY}, we derive the stochastic
Hamiltonian \cite{RJ} in terms of the dual variables.
In section IV, we proceed to discuss the double
scaling limit of the dual matrix model by
examining the structure of the Schwinger-Dyson
equations and the stochastic Hamiltonian
using the dual-matrix variables.
We will explicitly evaluate several
lower-point correlators and
determine their scaling behaviors.
In section V, based on the results of the
previous section, we exhibit
commutativity property between the operator mixing
and the merging-splitting interaction
of the macroscopic loop fields, in terms
of both the original and dual matrix variables.
The results generalize those of the
previous paper \cite{SY}.
In section VI, we discuss the
duality symmetry of the stochastic Hamiltonians.
We show that the Hamiltonians are duality
symmetric in the sphere approximation,
although they are not symmetric for higher genus
owing to the existence of global vector
fields on the world sheets with
nontrivial homology cycles, in conformity with the
discussion in section II.
Section VII is devoted to concluding remarks.
In Appendix A, we present a very brief discussion
on how the  exact T-duality symmetry is understood
as the Kramers-Wannier duality on arbitrary
discretized world sheets.
Some of the computational details are discussed in
Appendices B and C.

\section{Kramers-Wannier dual transformation in
the two-matrix model}

The partition function we study in the present paper is
the $N\times N$ hermitian two-matrix model given by
\begin{equation}
Z= \int d^{N^2}A d^{N^2}B \exp [-S(A, B)]
\end{equation}
\begin{equation}
S(A,B)= N{\rm tr}({1\over 2}A^2 + {1\over 2}B^2  -cAB
-{g\over 3}A^3 -{g\over 3}B^3) .
\label{twomatrixmodel}
\end{equation}
As is well known, this represents an Ising system on
random surface, where the Boltzmann factors are given by
$
 {1 \over 1-c^2}
$
and
$
 {c\over 1-c^2}
$
for the nearest neighbor Ising links,
connecting the spin sites with
 the same or opposite spins, respectively.

The inverse temperature $\beta$ for the Ising
system is then
\begin{equation}
\beta = -{1\over 2} \ln c.
\end{equation}
Note that the vertices of
Feynman diagrams are nothing but the surface elements
in the standard random surface interpretation and the Ising
spins live on the center of the surface elements. Hence the
Ising links just correspond to the propagators of Feynman diagrams.
The geometrical Boltzmann factor for the elementary surface element is
$g \sim {\rm e}^{ -\lambda}$ with $\lambda$ being the bare
cosmological constant. In what follows,
we use the terminologies, `sites', `links', in the sense of lattice
spin systems, and `vertices', `edges', in the sense of triangulations
of the random surface.

Now let us consider the Kramers-Wannier
dual transformation for this system.
A standard way of performing the dual transformation is to first reinterpret
the Boltzmann factors ${\rm e}^{\pm\beta}$ for the links as those
for the dual links
${\rm e}^{\pm\tilde{\beta}}$ by making
the $\mbox{\boldmath $Z$}_2$ Fourier transformation as
\begin{eqnarray}
{\rm e}^{\beta}= K({\rm e}^{\tilde {\beta}} + {\rm e}^{-\tilde{\beta}})
\label{fouriertransf1}\\
{\rm e}^{-\beta}= K({\rm e}^{\tilde {\beta}} - {\rm e}^{-\tilde{\beta}})
\label{fouriertransf2}
\end{eqnarray}
where $K$ is an overall normalization constant
$K = 1/ ({\rm e}^{2\tilde{\beta}}-{\rm e}^{-2\tilde{\beta}})^{1/2}$.
This fixes the dual inverse temperature $\tilde{\beta}$ as
\begin{equation}
\tilde{\beta}= -{1\over 2} \ln \tilde{c}
\end{equation}
with
\begin{equation}
\tilde{c}= {1-c \over 1+ c}.
\end{equation}
For a fixed lattice, the Boltzmann factor
$\exp \tilde{\beta}v_{\tilde{\ell}}$ would
correspond to dual links $\tilde{\ell}$ with
parallel (``stick": $v_{\tilde{\ell}}=+1$ ) dual spins or anti-parallel
(``flip":$v_{\tilde{\ell}}=-1$) dual spins,
respectively. In the present case, however, we will not
introduce dual spin variables for reasons explained
below. Usually, the dual
spin variables are introduced as the solutions for the
constraint for the dual link variables after integrating over the original
spin configurations.  The constraints take the form
\begin{equation}
\prod_{{\tilde\ell} \subset \tilde{C}} v_{\tilde{\ell}} =1
\label{constraint0}
\end{equation}
for all elementary (namely smallest) closed loops $\tilde{C}$
of dual links encircling
the spin sites of the original lattice.  The solution for these
constraints is locally given as
$v_{\tilde{\ell}}=\tilde{s}_{\tilde{\ell_1}}\cdot \tilde{s}_{\tilde{\ell_2}}$
by introducing the dual spin variables
$\tilde{s}_{\tilde{\ell_1}},  \tilde{s}_{\tilde{\ell_2}}$ ($=\pm 1$)
at the center of
all dual surface elements. The fixed square lattice with Ising spins is
then duality symmetric under the interchange $\tilde{\beta} \leftrightarrow
\beta$.

For the present model,  there are at least
two reasons that the system cannot  in general be duality symmetric.
The first reason is due to the nontrivial homology
cycles of the general random surfaces.
The constraint (\ref{constraint0}) allows nontrivial
solutions which cannot be reduced to the product of the dual spin
variables, owing to the existence of global $\mbox{\boldmath $Z$}_2$
vector fields,
associated with the nontrivial homology cycles.
For this reason, we will treat the dual transformed
model by dealing only with the dual link variables without
introducing the dual spin variables explicitly.
Secondly, the random surfaces of the two-matrix model
(\ref{twomatrixmodel})
are discretized by triangles corresponding to the
cubic vertices. Hence the original spin sites have always
three links attached. After the above dual transformation,
the dual spin sites can now have an arbitrary number of dual links
attached. For this reason, the critical point cannot be
fixed to be the self-dual point in contrary to the case of the fixed
square lattice.  We can however expect that in the continuum limit
this difference might be washed out and the model should exhibit
partial duality symmetry in some situation
where we can neglect the effect of
nontrivial homology cycles.  We will argue in later sections
that this is indeed the case.

Now in terms of the matrix fields $A, B$, the Boltzmann factors
for the original Ising links are
related to the bare propagators as
\begin{eqnarray}
\langle AA \rangle_0=\langle BB \rangle_0= L {\rm e}^{\beta}, \\
\langle AB \rangle_0= L {\rm e}^{-\beta}
\end{eqnarray}
with $L= \sqrt{c}/(1-c^2)$, where $\langle \cdot  \rangle_0$ indicates
the expectation value with the quadratic action.
Substituting the Fourier transformation
(\ref{fouriertransf1}) and
(\ref{fouriertransf2}),
we have
\begin{eqnarray}
\langle AA \rangle_0=\langle BB \rangle_0= {1\over 2\sqrt{1-c^2}}
({\rm e}^{\tilde {\beta}} + {\rm e}^{-\tilde{\beta}}), \\
\langle AB \rangle_0 = {1\over 2\sqrt{1-c^2}}({\rm e}^{\tilde {\beta}} - {\rm
e}^{-\tilde{\beta}}).
\end{eqnarray}
This clearly shows that the dual transformation is performed simply
by changing to the new matrix fields $X$ and $Y$, which diagonalize the kinetic
term of the original action,
as
\begin{eqnarray}
X ={1\over \sqrt{2}}( A+B) \\
Y={1\over \sqrt{2}}( A-B)
\end{eqnarray}
whose propagators just give the Boltzmann factors for the
dual model,
\begin{eqnarray}
\langle XX \rangle_0 = {1\over \sqrt{1-c^2}}{\rm e}^{\tilde {\beta}} = 
{1\over 1-c},  \\
\langle YY \rangle_0 = {1\over \sqrt{1-c^2}}{\rm e}^{-\tilde {\beta}} =
{1\over 1+c}.
\end{eqnarray}
The new matrix fields $X, Y$ can be regarded as living  on the dual vertices
(surface elements or
triangles of  the {\it original} lattice) but their
propagators are associated with the dual Ising links
(dual to the edges of the dual surface elements),
having one-to-one correspondence with original Ising links.
 Thus they should not be confused with
the dual spin variables themselves
which are supposed to live on dual-spin sites, namely,
the center of the dual surface
elements corresponding to the vertices of the original random surfaces.

In terms of the new matrix fields, the partition function is
($\hat{g}=g/\sqrt{2}$)
\begin{equation}
Z= \int d^{N^2}X d^{N^2}Y \exp [-S_D(X,Y)],
\end{equation}
\begin{equation}
S_D(X,Y)= N{\rm tr}({1-c\over 2}X^2 + {1+c\over 2}Y^2
-{\hat{g}\over 3}(X^3 + 3XY^2) ).
\label{twomatrixmodel-xy}
\end{equation}
In comparison with the original representation, this representation itself
does not show any trace of possible dual symmetry,
reflecting the situation explained above.
This simply defines a mapping of  an Ising model
on random surfaces to a special case of the O(n) model as has been
discussed by several authors \cite{Kostov}.

The homological property we have discussed can now be
described by  the configurations  of closed loops formed by the
$Y$ field. The global $\mbox{\boldmath $Z$}_2$ vector fields
correspond to the closed $Y$-loops
winding around the nontrivial homological cycles of the
surfaces. The $\mbox{\boldmath $Z$}_2$ nature comes from
the fact that even number of closed $Y$-loops winding
around nontrivial homology cycles are
continuously deformed into null  closed $Y$-loop.
In the case of sphere, there is no nontrivial homology cycles
and hence we can introduce dual $\mbox{\boldmath $Z$}_2$ spin variables which
reside on the domains bounded by closed $Y$-loops.
The existence of such closed $Y$-loops winding around nontrivial cycles
forbids to define dual spin domains globally. Equivalently, the
Hamiltonian description of the time development
of the random surfaces necessarily requires us to introduce
the observables containing odd number of the $Y$-fields.

Let us next discuss the basis of string field representations
used in the present paper.
In the previous work, we used the component expansion of string fields
with respect to the number of spin domains.
In the dual formulation, the basis which can be compared
with this is the expansion with respect to the number of the
$Y$-fields. Thus the observables we consider
are of the
following type:
\begin{equation}
{1\over N}{\rm{tr}}({1\over \xi-X}) \equiv \Psi_X(\xi),
\label{dualbasis0}
\end{equation}
\begin{equation}
{1\over N}{\rm{tr}}({1\over \xi_1-X}Y) \equiv \Psi_1(\xi_1),
\label{dualbasis1}
\end{equation}
\begin{equation}
{1\over N}{\rm{tr}}({1\over \xi_1-X}Y{1\over \xi_2-X}Y)
\equiv \Psi_2(\xi_1, \xi_2),
\label{dualbasis2}
\end{equation}
\begin{equation}
{1\over N}{\rm{tr}}({1\over \xi_1-X}Y{1\over \xi_2-X}Y
{1\over \xi_3-X}Y) \equiv \Psi_3(\xi_1, \xi_2, \xi_3),
\label{dualbasis3}
\end{equation}
\[
. . . . . . {\rm{etc.}}
\]
We note that if we could have neglected the closed $Y$-loops
winding around the nontrivial homology cycles,  $\Psi$'s with odd number
of $Y$'s were not necessary.  We would then have the
interpretation of these quantities
in terms of domains with respect to dual spins,
and $\Psi_{2n}$ would correspond to the component of the
field with $n$ pairs of domains. Note that the matrix $Y$  just represents
the domain boundaries of the dual-spin interpretation.
Thus, in studying the duality symmetry, $\Psi_{2n}$ should be
substituted in  place of  the observables considered in the
previous paper, namely,
\begin{equation}
\Phi_n(\zeta_1, \sigma_1, \cdots, \zeta_n, \sigma_n)\equiv {1\over N}
{\rm{tr}}(\prod_i^n {1\over \zeta_i -A}{1\over  \sigma_i -B})
\label{originalbasis}
\end{equation}
after an appropriate symmetrization with respect to two matrices $A$ and $B$,
taking into account the fact that the $\Psi_{2n}$ does not discriminate global
dual spin directions.
A detailed discussion of the duality symmetry of the stochastic Hamiltonian
will be presented later in section VI.

      From the above discussion, it is clear at least conceptually that
the dual transformation is nothing but a change of the
basis of the string fields between (\ref{originalbasis}) and
(\ref{dualbasis0}) $\sim$ (\ref{dualbasis3}) etc. 
\footnote{
We here mention a paper \cite{KZ}
which has discussed the T-duality in the
framework of covariant string field theories
of critical bosonic string theory.
}
However, it turns out that performing this transformation
directly for the string fields in the {\it continuum} limit
is technically very difficult, because the
process of taking the continuum limit
involves an intricate operator mixing and
does not seem to be commutative with the above change of the basis.
Thus the question of the T-duality symmetry
for the Ising model on the random surface is quite nontrivial.
The approach in the present paper will therefore be
somewhat indirect such that we perform an independent study of
the scaling limit of the Schwinger-Dyson equations and the stochastic
Hamiltonian using the dual basis and compare the results with
those obtained using the original basis.
One practical virtue of this approach would  be
that the ranges of the components where  we can compute
correlation functions explicitly are different from
each other depending on different bases,
 and hence our results are useful for
showing the universal nature of the structure of, say, the stochastic
Hamiltonians, as has been emphasized in the previous work \cite{SY}.
Presenting such a  piece of evidence for the universal
structure of the string field theories is another purpose of the
present paper.

Finally, we here fix our notations in
what follows.
As introduced above, the basis operators are $\Psi_n$.
We denote the expectation values of $\Psi_n$ in the
sphere approximation by $V^{(n)}(\cdot)=\langle \Psi_n(\cdot) \rangle_0
$ with the same arguments and suffices.
We have used the notation $W^{(2n)}$ for $\langle \Phi_n \rangle$
in the case of the original spin formulation \cite{SY}.

\section{Stochastic Hamiltonian
in the Dual Formalism}

Let us first derive the stochastic Hamiltonian using the
dual basis of the observables $\Psi$'s  introduced in the previous section.
The generating functional of the dual two-matrix model is defined by
\begin{equation}
  Z[K] = \frac1{Z} \int d^{N^2}X d^{N^2}Y {\rm e}^{-S_D} {\rm e}^{K \cdot
\Psi},
\end{equation}
\begin{equation}
  Z    = \int d^{N^2}X d^{N^2}Y{\rm e}^{-S_D} ,
\end{equation}
\begin{equation}
  K \cdot \Psi =
\int \frac{d\xi }{2\pi i}K_X\left
(\xi\right)\Psi_X\left(\xi\right)
            + \sum_{n=1}^{\infty}\int
\prod_{i=1}^n \frac{d\xi_i}{2\pi  i}
K_n\left( \xi_1,\cdots ,\xi_n \right)
\Psi_n\left( \xi_1,\cdots ,\xi_n \right),
 \end{equation}
where and in what follows
the integral contour, unless specified otherwise,
runs parallel to the imaginary axis
towards the positive imaginary infinity
on the right hand side of all the  pole  singularities
 of  the observables $ \Psi_X,\Psi_n $.

As discussed in the previous paper\cite{SY},
the Stochastic Hamiltonian is derived from the identity,
\begin{equation}
0= - \int d^{N^2}Xd^{N^2}Y \,\sum_{\alpha=1}^{N^2} \left[
\frac{\partial}{\partial X_{\alpha}} {\rm e}^{- S}
\frac{\partial}{\partial X_{\alpha}}
+ \frac{\partial}{\partial Y_{\alpha}} {\rm e}^{-S}
\frac{\partial}{\partial Y_{\alpha}} \right] {\rm e}^{K\cdot \Psi}.
\label{matrixlaplacian}
\end{equation}
The notations $X_{\alpha}, Y_{\alpha}$ denote the
components of the hermitian matrix in the expansion in terms
of the basis $\{t_{\alpha}\}
 \, (\mbox{tr}(t_{\alpha}t_{\beta})=\delta_{\alpha\beta})$.
The derivative operator inside the integral may
be regarded as the Laplacian operator for the present model.
Expressed as a functional differential equation in terms
of the sources, this takes the form
\begin{equation}
0={\cal H} Z\left[ K \right],
\end{equation}
\begin{equation}
{\cal H}= -K\cdot \left({\cal K} \frac{\delta}{\delta K}\right)
-K\cdot\left(\frac{\delta}{\delta K} \vee \frac{\delta}{\delta K}\right)
 - \frac{1}{N^2} K\cdot\left( K \cdot \left( \wedge \frac{\delta}{\delta K}
\right)\right) - K\cdot T .
\label{Hamiltonian}
\end{equation}
Here the functional derivative $\frac{\delta}{\delta K_n} $ acting
on the source terms is defined as
\begin{equation}
\frac{\delta K_m \left( \xi_1^\prime ,\cdots,\xi_m^\prime\right)}
{\delta K_n \left( \xi_1,\cdots,\xi_n \right)}=\delta_{m,n}\frac{1}{n}
\left(2\pi  i \right)^n \sum_c
 \delta\left( \xi_1-{\xi^\prime}_{c(1)}\right)\cdots
\delta\left( \xi_{n}-{\xi^\prime}_{c(n)}\right).
\end{equation}
Reflecting the cyclic symmetry of $\Psi_n$ with respect
to its variables $\{ \xi_i \}$, the summation in the right hand side
is over cyclic permutations $c(i)$ of the indices $i=1,2,\ldots,n$
of $\xi'_i$.
The origin and definition of each term of the Hamiltonian (\ref{Hamiltonian})
are explained below.
\begin{enumerate}
\item The first term, ``kinetic term''
$ K\cdot \left({\cal K} \frac{\delta}{\delta K}\right) $, comes from
the  product of
the first derivatives of the source term and the action
($n=1,2,\ldots$),
\begin{eqnarray}
\left( {\cal K} \frac{\delta}{\delta K} \right)_X(\xi) & = &
 \partial_{\xi}\left[\xi(1-c-\hat{g}\xi)
\frac{\delta}{\delta K_X(\xi)}\right]
 -\hat{g}
\oint \frac{d \xi^{\prime}}{2 \pi i} \partial_\xi
\frac{\delta} {\delta K_2 (\xi ,\xi^{\prime})} ,   \\
\left({\cal K} \frac{\delta}{\delta K}\right)_n(\xi_1, \cdots, \xi_n)
 & = & \sum_{j=1}^{n} \left[-2c+
(1-c-\hat{g}\xi_j)\xi_j \partial_{\xi_j}
\right]
 \frac{\delta}{\delta K_n(\xi_1,
\cdots,\xi_n)}  \nonumber \\
& & +\sum_{j=1}^n\hat{g}\oint\frac{d\xi}{2 \pi i}\frac{\delta}{\delta K_{n+2}
(\xi_1,\cdots,\xi_{j-1},\xi_j,\xi,\xi_j,\xi_{j+1},\cdots, \xi_n)} \nonumber \\
& & -\sum_{j=1}^n  \hat{g}\oint \frac{d \xi}{2 \pi i}
\frac{\delta}{\delta K_n
(\xi_1,\cdots,\xi_{j-1},\xi,\xi_{j+1}, \cdots, \xi_n)},\label{3.15}
\end{eqnarray}
where the integral $\oint$ is over a closed curve
encircling all the pole singularities of the observables $\Psi$'s.
The structure of the kinetic term is slightly
different from that with the original
variables $A, B$. If for the moment we allow ourselves
to use the dual-spin language for convenience,
this contains a part which measures the number of dual-spin flip links,
corresponding to the $Y$-matrices,
in addition to a part measuring the length of each domain.
 The other parts
describe the infinitesimal motion of string loops,
 changing their lengths and/or dual spin directions.
\item The second term
,  $K\cdot\left(\frac{\delta}{\delta K}
\vee \frac{\delta}{\delta K}\right)$, of (\ref{Hamiltonian})
comes from
the second derivative of
the source term $K\cdot \Psi$
and represents processes where a loop
 splits into two. The symbol $\left(\frac{\delta}{\delta K}
\vee \frac{\delta}{\delta K}\right)_I $ stands for
the spin configurations of the resultant two loops
       from a single loop with spin configuration $I$.
 \begin{eqnarray*}
\left(\frac{\delta}{\delta K} \vee \frac{\delta}{\delta K}\right)_{X}(\xi)
&=& -\partial_\xi \frac{\delta^2}{{\delta K_X(\xi)}^2} , \\
\left(\frac{\delta}{\delta K} \vee \frac{\delta}{\delta K}\right)_{1}(\xi_1)
&=& -  2\frac{\delta}{{\delta K_X(\xi_1)}}
\partial_{\xi_1} \frac{\delta}{{\delta K_1(\xi_1)}},
\end{eqnarray*}
\begin{eqnarray*}
\left(\frac{\delta}{\delta K} \vee \frac{\delta}{\delta K}\right)_2
(\xi_1,\xi_2)&=& -2\sum_{j=1}^{2}\frac{\delta}
{\delta K_X(\xi_j)}
\partial_{\xi_j}
\frac{\delta}{\delta K_2(\xi_1,
\xi_2)}    \\
 & &
+2\left( D_z(\xi_1,
\xi_2)\frac{\delta}{{\delta K_1(z)}}\right)^2
+ 2\frac{\delta^2}{\delta K_X(\xi_1)
\delta K_X(\xi_2)} ,
\end{eqnarray*}
\begin{eqnarray}
\lefteqn{\left(\frac{\delta}{\delta K} \vee \frac{\delta}{\delta K}\right)_n
(\xi_1, \cdots, \xi_n)
= -2\sum_{j=1}^{n}  \frac{\delta}
{\delta K_X(\xi_j)}
\partial_{\xi_j}
\frac{\delta}{\delta K_n(\xi_1, \cdots,
\xi_n) }} \nonumber \\
 & + &  2\sum_{k<l} D_z(\xi_k,\xi_l)
 \frac{\delta}{\delta K_{l-k}(z,\xi_{k+1},\cdots, \xi_{l-1})}
 D_{\zeta}(\xi_k,\xi_l)\frac{\delta}{\delta K_{n-l+k}( \xi_1,\cdots,\xi_{k-1},
\zeta,\xi_{l+1}, \cdots, \xi_{n})}  \nonumber \\
 & - & 2\sum_{j=1}^{n}\frac{\delta}{\delta K_X(\xi_j)}
D_z(\xi_{j-1}, \xi_{j+1})
\frac{\delta}{\delta K_{n-2}(\xi_1,\cdots,\xi_{j-2},z,\xi_{j+2},\cdots,
\xi_n)}  \nonumber \\
 & +  & 2\sum_{1<l-k<n-1}  D_z(\xi_{k+1},\xi_l)
 \frac{\delta}{\delta K_{l-k-1}(z,\xi_{k+2},
\cdots, \xi_{l-1})}  \nonumber \\
 &  & ~~ \times
D_{\zeta}(\xi_k,\xi_{l+1})
\frac{\delta}{\delta K_{n-l+k-1}(\xi_1,\cdots,\xi_{k-1},
\zeta,\xi_{l+2},\cdots, \xi_n)},
\, \, \,  (n\ge 3).
\label{dualsplit}
\end{eqnarray}
Here we used the definition of the combinatorial
derivative,
$$
D_z(\xi_1,\xi_2)f(z)= \frac{f(\xi_1)-f(\xi_2)}{\xi_1-\xi_2}.
$$
Note also that the last term of (\ref{dualsplit}) is absent for $n=3$.
\item The third term  $\frac{1}{N^2} K\cdot\left( K \cdot \left(
\wedge \frac{\delta}{\delta K}\right)\right) $ comes from
the square of the first derivative of the source term,
and  represents
processes in which two loops merge into a single loop. The symbol
$ \left( \wedge \frac{\delta}{\delta K}\right)_{I,J} $ expresses
the spin configuration of the resultant single loop into which two loops
with spin configurations $I,J$  merge. The variables on
the left hand  side of a  semicolon represent the variables  conjugate
to  the lengths of domains of a loop with     configuration $I$,
while those on the right hand side
represent the variables with configuration $J$:
\begin{eqnarray*}
\left( \wedge \frac{\delta}{\delta K}\right)_{X,X}(\xi;\xi^{\prime})
&=& -\partial_{\xi} \partial_{\xi^{\prime}}
D_z(\xi,\xi^{\prime})
\frac{\delta}{\delta K_X(z)}, \\
\left( \wedge \frac{\delta}{\delta K}\right)_{n,X}(
\xi_1,\cdots, \xi_n;\xi^{\prime})
&=&\left( \wedge \frac{\delta}{\delta K}\right)_{X,n}(\xi^{\prime};
\xi_1,\cdots, \xi_n)  ~~~~(n=1,2,3,\cdots) \\
& =& -\sum_{j=1}^{n}\partial_{\xi'} \partial_{\xi_j }
D_z(\xi',\xi_j)\frac{\delta}{\delta K_n(\xi_1,\cdots,\xi_{j-1},z,
\xi_{j+1},\cdots,\xi_n)}, \\
\left( \wedge \frac{\delta}{\delta K} \right)_{1,1}
(\xi_1;\xi^{\prime}_1)
 & = &D_{\zeta}(\xi_1,\xi^{\prime}_1)D_z(\xi_1,\xi^{\prime}_1)
\frac{\delta}{\delta K_2(z,\zeta)}
-D_z(\xi_1,\xi^{\prime}_1)\frac{\delta}{\delta K_X(z)},
\end{eqnarray*}
\begin{eqnarray*}
\lefteqn{\left( \wedge \frac{\delta}{\delta K} \right)_{n,1}(
\xi_1,\cdots, \xi_n;\xi^{\prime}_1)
= \left( \wedge \frac{\delta}{\delta K}
\right)_{1,n}(\xi^{\prime}_1;
\xi_1,\cdots, \xi_n)  ~~~~(n=2,3,4,\cdots)}\\
 & = & \sum_{j=1}^nD_z(\xi^{\prime}_1,\xi_j)
D_{\zeta}(\xi^{\prime}_1,\xi_j)
\frac{\delta}{\delta K_{n+1}(\xi_1,\cdots,\xi_{j-1},
z,\zeta,\xi_{j+1},
\cdots, \xi_n)} \\
 &  & +\sum_{j=1}^{n}D_{\zeta}(\xi_j,
\xi^{\prime}_1)D_z(\xi_{j+1},\zeta)
\frac{\delta}{\delta K_{n-1}
(\xi_1,\cdots,\xi_{j-1},
z,\xi_{j+2},
\cdots, \xi_n)},
\end{eqnarray*}
\begin{eqnarray}
 \lefteqn{\left( \wedge \frac{\delta}{\delta K} \right)_{n,m}
( \xi_1,\cdots, \xi_n;{\xi^{\prime}}_1,\cdots,{\xi^{\prime}}_m)
 \quad (n,m=2,3,4,\cdots) }\nonumber \\
& =&\sum_{j=1}^{n}\sum_{k=1}^{m}D_z(\xi_j,
{\xi^{\prime}}_k)
D_{\zeta}(\xi_j,
{\xi^{\prime}}_k)  \nonumber \\
&&~~\times\frac{\delta}{\delta K_{n+m}(\xi_1,
\cdots, \xi_{j-1},z,
{\xi^{\prime}}_{k+1},\cdots,
{\xi^{\prime}}_{m},{\xi^{\prime}}_1,\cdots,{\xi^{\prime}}_{k-1},
\zeta,\xi_{j+1},\cdots,\xi_n)}  \nonumber \\
& &+\sum_{j=1}^{n}\sum_{k=1}^{m}D_z(\xi_{j}, {\xi^{\prime}}_{k+1})
D_{\zeta}( {\xi^{\prime}}_{k},
\xi_{j+1}) \nonumber \\
&& ~~\times \frac{\delta }{\delta K_{n+m-2}(\xi_1,
\cdots,\xi_{j-1}, z,
{\xi^{\prime}}_{k+2},\cdots,
{\xi^{\prime}}_m,{\xi^{\prime}}_1,\cdots,{\xi^{\prime}}_{k-1},\zeta,
\xi_{j+2},\cdots,\xi_n)} . \label{dualmerge}
\end{eqnarray}

\item The last term, a tadpole term, arises again from
the product of
the first derivative of the source term $K_X \cdot  \Psi_X $
and the action, and describes
the process of the annihilation of a {\it shortest} loop into nothing :
\begin{equation}
K\cdot T= \int \frac{d \xi}{2 \pi i} K_X(\xi) \hat{g}.
\end{equation}
\end{enumerate}
Similarly to the Hamiltonian of the original formalism,
the present dual Hamiltonian  satisfies a  locality property.
For instance,  only the loops $\Psi_X$ with
no domain boundary
 can be annihilated into nothing.
Also, only a single pair of domains or a single pair of the boundaries of
the domains can participate
in the splitting or merging processes, and other domains remain  unchanged.
One of the differences between the Hamiltonians in the original
and dual formulations arises from the derivatives
with respect to $Y$.  The $Y_{\alpha}$-derivatives produce
processes which have no counter part in the Hamiltonian
in terms of the original variables.  For instance,
the last terms of  the kinetic (\ref{3.15})
and merging terms  (for $I, J \ge 1$)
in the above expressions contain the contributions from the 
$Y_{\alpha}$-derivatives.
For the splitting interaction, the last (for $I=2,3$) or the last two terms
(for $I \ge 4$) are the contributions from the $Y_{\alpha}$-derivatives.
     From the view point of continuum theory, these are
very singular; they would be
 measure-zero contributions. In later sections, we will show
that the terms arising from the $Y_{\alpha}$-derivatives
in fact vanish in the continuum limit.

\section{Schwinger-Dyson Equations
and the Scaling Limit of the Stochastic Hamiltonian}

Now we consider the scaling limit of the Hamiltonian
(\ref{Hamiltonian}). To do this, we closely follow \cite{SY}.
Namely, we will first  identify and subtract
the non-universal parts of the disk amplitudes
and establish the operator mixing, and then
will rewrite the Hamiltonian (\ref{Hamiltonian})
in terms of the universal parts of the operators.

As we discuss  in  Appendix B,
the original operators $\Psi_{I}$ and their universal parts $\hat{\Psi}_{I}$
are related by a linear transformation of the following form:
\begin{equation}
\Psi_{I}=\sum_{J}{\cal M}_{IJ}\hat{\Psi}_{J}+\psi_{I},
                      ~~~~~~~~  I = X,1,2,\cdots,
                             \label{dualmix}
\end{equation}
where ${\cal M}_{IJ}$ is a mixing matrix of upper-triangular form,
${\cal M}_{IJ}=0$ for $I<J$,
and $\psi_{I}$ is the non-universal c-number function.
The upper-triangular  form of ${\cal M}$ comes from a property of the operator
mixing that the operators corresponding to simpler configurations
can mix as non-universal parts in taking the continuum limit,
as has been discussed in the previous work \cite{SY}.
Note that $\psi_{I}$ vanishes by itself for an arbitrary odd $I$,
because there exists the $\mbox{\boldmath $Z$}_{2}$ symmetry under
$Y \rightarrow -Y$.
 The first few components of (\ref{dualmix}) are
\begin{eqnarray*}
\Psi_{X}(\xi) & = & \hat{\Psi}_{X}(\xi)+\psi_{X}(\xi),  \\
\Psi_{1}(\xi_{1}) & = & \hat{\Psi}_{1}(\xi_{1}),   \\
\Psi_{2}(\xi_{1},\xi_{2}) & = & \frac{2}{\sqrt{5c}}(\hat{\Psi}_{X}(\xi_{1})+
   \hat{\Psi}_{X}(\xi_{2})) +\hat{\Psi}_{2}(\xi_{1},\xi_{2})
                            +\psi_{2}(\xi_{1},\xi_{2}),
                                                                            \\
\Psi_{3}(\xi_{1},\xi_{2},\xi_{3}) & = & \frac{2}{\sqrt{5c}}
                   [-D_{\xi}(\xi_{1},\xi_{2})\hat{\Psi}_{1}(\xi)
                    -D_{\xi}(\xi_{2},\xi_{3})\hat{\Psi}_{1}(\xi)
                    -D_{\xi}(\xi_{3},\xi_{1})\hat{\Psi}_{1}(\xi)]           \\
  &  &             +\hat{\Psi}_{3}(\xi_{1},\xi_{2},\xi_{3}),
                                                                             \\
\Psi_{4}(\xi_{1},\xi_{2},\xi_{3},\xi_{4}) & = & -\frac{4}{5c}
                   (D_{\xi}(\xi_{1},\xi_{3})\hat{\Psi}_{X}(\xi)
                     +D_{\xi}(\xi_{2},\xi_{4})\hat{\Psi}_{X}(\xi))      \\
  &  & -\frac{2}{\sqrt{5c}}[D_{\xi}(\xi_{1},\xi_{3})
          (\hat{\Psi}_{2}(\xi,\xi_{2})+\hat{\Psi}_{2}(\xi,\xi_{4}))\nonumber \\
  &  & \mbox{  }+D_{\xi}(\xi_{2},\xi_{4}) (\hat{\Psi}_{2}(\xi_{1},\xi)+
           \hat{\Psi}_{2}(\xi_{3},\xi))] \nonumber \\
  &  & +\hat{\Psi}_{4}(\xi_{1},\xi_{2},\xi_{3},\xi_{4})+
           \psi_{4}(\xi_{1},\xi_{2},\xi_{3},\xi_{4}),
\end{eqnarray*}
\begin{equation}
      \cdots,                \label{dualsf}
\end{equation}
where
\begin{eqnarray*}
\psi_{X}(\xi) & = & \frac{2c(c+1)}{3\hat{g}}
                     -\frac{1}{3}((5c-1)\xi+\hat{g}\xi^{2}),  \\
\psi_{2}(\xi_{1},\xi_{2}) & = & \frac{1}{5}(1+4\hat{s})
                               -2\sqrt{\frac{c}{5}}(\xi_{1}+\xi_{2}), \\
\psi_{4}(\xi_{1},\xi_{2},\xi_{3},\xi_{4}) & = & 1,
\end{eqnarray*}
\begin{equation}
                 \cdots,
\end{equation}
and $c$ is at the critical value: $c=\frac{-1+2\sqrt{7}}{27}$,
and $\hat{s}=1+\sqrt{7}$. All these results are derived explicitly
in Appendix B\footnote
{ As we comment in (\ref{V4mixing}) in Appendix B,
$\psi_{4}$ has possibly a linear term of $y$ in addition.
This term affects higher components of spin flip amplitudes
through the Schwinger-Dyson equations.
However, since it changes only the non-universal parts
of these amplitudes, it can be absorbed into
the redefinition of  the non-universal parts
and does not change the scaling limit of the Hamiltonian.}.

Denoting  the connected $k$-point
correlator for the $K=0$ background as
\begin{equation}
G^{(k)}_{I_{1},\cdots,I_{k}} = \langle\Psi_{I_{1}}\cdots\Psi_{I_{k}}\rangle,
                          ~~~~~~~~  I_{1},\cdots,I_{k}=X,1,2,\cdots,
                                 \label{5-7-1}
\end{equation}
the generating functional takes the form
\begin{equation}
Z[K]=\exp\left[K\cdot G^{(1)}+\frac{1}{2!}K\cdot (K\cdot G^{(2)})+
  \frac{1}{3!}K\cdot(K\cdot (K\cdot G^{(3)}))+\cdots\right].
\label{Z}
\end{equation}
Separation of the universal parts $\hat{G}^{(k)}_{I_{1},\cdots,I_{k}}$
amounts to a linear transformation
\begin{eqnarray*}
G^{(1)}_{I} & = & \sum_{J}{\cal M}_{IJ}\hat{G}^{(1)}_{J}+\psi_{I},  \\
G^{(k)}_{I_{1},\cdots,I_{k}} & = & \sum_{J_{1},\cdots,J_{k}}
  {\cal M}_{I_{1}J_{1}}\cdots{\cal M}_{I_{k}J_{k}}
\hat{G}^{(k)}_{J_{1},\cdots,J_{k}} ~~~
    (k\geq 2).
\end{eqnarray*}
  Thus, on introducing the transformed source $\hat{K}$ by
\begin{equation}
K_{I}=\sum_{J}\hat{K}_{J}({\cal M}^{-1})_{JI},              \label{5-10-1}
\end{equation}
the generating functional $\hat{Z}[\hat{K}]$ in the continuum theory
is obtained by  a rescaling $
Z[K]= {\rm e}^{K\cdot\psi}\hat{Z}[\hat{K}]$, giving
\begin{equation}
\hat{Z}[\hat{K}]=
\exp\left[\hat{K}\cdot \hat{G}^{(1)}+\frac{1}{2!}\hat{K}\cdot
  (\hat{K}\cdot \hat{G}^{(2)})+
  \frac{1}{3!}\hat{K}\cdot(\hat{K}\cdot (\hat{K}\cdot \hat{G}^{(3)}))
  +\cdots\right].
\label{5-7-3}
\end{equation}
The Hamiltonian acting on $\hat{Z}[\hat{K}]$ now becomes
\begin{eqnarray}
0 & = & {\cal H} \hat{Z}[\hat{K}],   \label{5-10-2} \\
{\cal H} & = & -(\hat{K}{\cal M}^{-1})\cdot
\left({\cal K}({\cal M}\frac{\delta}{\delta \hat{K}}+\psi)\right)
                   -(\hat{K}{\cal M}^{-1})\cdot T            \nonumber \\
 &  & -(\hat{K}{\cal M}^{-1})\cdot
\left( ({\cal M}\frac{\delta}{\delta \hat{K}}+\psi)\vee
         ({\cal M}\frac{\delta}{\delta \hat{K}}+\psi)\right) \nonumber \\
 &  & -\frac{1}{N^{2}}(\hat{K}{\cal M}^{-1})\cdot
\left( (\hat{K}{\cal M}^{-1})\cdot
         \left(\wedge({\cal M}\frac{\delta}{\delta
\hat{K}}+\psi)\right)\right).
                  \label{5-10-3}
\end{eqnarray}

    Next, we try to eliminate the mixing matrix
    from Eq. (\ref{5-10-3}).
We claim the validity of the following
identities:
\begin{eqnarray}
\lefteqn{-(\hat{K}{\cal M}^{-1})\cdot
\left({\cal K}({\cal M}\frac{\delta}{\delta\hat{K}}+\psi)\right)
                  -(\hat{K}{\cal M}^{-1})\cdot T } \nonumber \\
     &  &   -(\hat{K}{\cal M}^{-1})
\cdot\left(({\cal M}\frac{\delta}{\delta\hat{K}}+\psi)\vee
      ({\cal M}\frac{\delta}{\delta\hat{K}}+\psi)\right) \nonumber \\
     &  &   -\frac{1}{N^{2}}(\hat{K}{\cal M}^{-1})\cdot
\left( (\hat{K}{\cal M}^{-1})\cdot
               \left(\wedge\psi\right)\right) \nonumber \\
 & = & -\hat{K}\cdot \left({\cal F}\frac{\delta}{\delta\hat{K}}\right)
-(\hat{K}{\cal M}^{-1})\cdot
\left(\left({\cal M}\frac{\delta}{\delta\hat{K}}\right)
              \vee\left({\cal M}\frac{\delta}{\delta\hat{K}}\right)\right),
        \label{kineticterm}
\end{eqnarray}
\begin{eqnarray}
(\hat{K}{\cal M}^{-1})\cdot
\left(\left({\cal M}\frac{\delta}{\delta\hat{K}}\right)
\vee\left({\cal M}\frac{\delta}{\delta\hat{K}}\right)
\right) & = & \hat{K}\cdot\left(\frac{\delta}{\delta\hat{K}}
\vee\frac{\delta}{\delta\hat{K}}\right),   \label{splitterm} \\
(\hat{K}{\cal M}^{-1})\cdot\left((\hat{K}{\cal M}^{-1})\cdot\left(
\wedge{\cal M}\frac{\delta}{\delta\hat{K}}\right) \right)& = &
                      \hat{K}\cdot\left(\hat{K}\cdot\left(
\wedge\frac{\delta}{\delta\hat{K}}\right)\right),         \label{mergeterm}
\end{eqnarray}
where ${\cal F}$ is a part of the kinetic operator ${\cal K}$,
representing only the spin flip
processes. Note that
the term $(\hat{K}{\cal M}^{-1})\cdot( (\hat{K}{\cal M}^{-1})\cdot(
\wedge\psi ))$ does not completely vanish in the dual theory
in contrast to the corresponding identity
in the original spin formulation \cite{SY}.
These identities enable us to rewrite the
Hamiltonian as
\begin{equation}
{\cal H}=
-\hat{K}\cdot
\left({\cal F}\frac{\delta}{\delta\hat{K}}\right)-\hat{K}\cdot\left
(\frac{\delta}{\delta\hat{K}}\vee\frac{\delta}{\delta\hat{K}}\right)
 -\frac{1}{N^{2}}
\hat{K}\cdot\left(\hat{K}\cdot
\left(\wedge\frac{\delta}{\delta\hat{K}}\right)\right).
\label{hatHamiltonian}
\end{equation}

\vspace{0.5cm}
\noindent
{\bf Justification  of  the identities
(\ref{kineticterm})$\sim$(\ref{mergeterm}) }:\\
We now try to establish the above identities.

\vspace{0.3cm}

\noindent
(1) {\em Kinetic term}

In order to make the expression for
 the kinetic operator ${\cal F}$ explicit,
we first consider the spin flip process in the continuum theory.
For the universal parts of the disk
and cylinder amplitudes, the string fields with a microscopic domain
containing a single flipped spin
are obtained from the string fields without any
microscopic domain by the following rule (for derivation see Appendix C):
\begin{eqnarray}
-\partial_{\xi}\hat{V}_{2}(\xi)
   & = & -\hat{\oint}\frac{d\eta}{2\pi i}
\partial_{\xi}\hat{V}^{(2)}(\xi,\eta),
                           \label{V2flip} \\
{\hat{V}_{2}}^{(3)}(\xi_{1},\xi_{2},\xi_{3};) & = &
\hat{\oint}\frac{d\eta}{2\pi i}\hat{V}^{(4)}
        (\xi_{1},\xi_{2},\xi_{3},\eta),            \label{V4flip} \\
       &    &   \cdots ,   \nonumber \\
\hat{V}^{\mbox{\scriptsize cyl}~(1)}_{Y}(\xi)
   & = & \hat{\oint}\frac{d\eta}{2\pi i}
         \hat{V}^{\mbox{\scriptsize cyl}}_{1|1}(\left.\xi \right|\eta),
                           \label{V1flip} \\
\hat{V}^{\mbox{\scriptsize cyl}}_{2|Y}
(\xi_{1};\xi_{2}) & = &
\hat{\oint}\frac{d\eta}{2\pi i}
\hat{V}^{\mbox{\scriptsize cyl}~(3)}_{Y}(\xi_{1},\eta,\xi_{2}),
                                                          \label{V3flip} \\
       &    &   \cdots .   \nonumber
\end{eqnarray}
where the domain corresponding to the variable $\eta$ has been shrunk 
into the microscopic domain by integration.
The integral symbol $\hat{\oint}\frac{d\eta}{2\pi i}$
is used in the sense of the integral with respect to
the variable $y$ of the continuum theory ($\eta=\xi_{*}(1+ay)$)
\begin{equation}
\hat{\oint}\frac{d\eta}{2\pi i}=\xi_{*}a\int_{C}\frac{dy}{2\pi i},
        \label{5-12-3}
\end{equation}
where the contour $C$ encircles
around the negative real axis
and the singularities of the left half plane, and
$\xi_{*}=\frac{1}{5}\hat{s}c^{-1/2}$.
Here and in what follows $a$ is the lattice spacing.
Note a slight difference from the original Ising theory in the point
that there is no finite renormalization
factor denoted by $s^{-1}$ in \cite{SY}.

As already emphasized in \cite{SY}, these relations are natural
since the spin flip process occurs locally
with respect to domains;
in  (\ref{V2flip}), (\ref{V4flip}), (\ref{V1flip}) and (\ref{V3flip})
only the $\eta$-domain
is concerned and the other domains do not change at all.
Because of the locality, we can reasonably expect
that the relations such as (\ref{V2flip})$\sim$(\ref{V3flip})
should hold for any amplitudes with an arbitrary number of handles
with generic spin
configurations, although a completely general proof
for this cannot be given at present.
By assuming this, we can rewrite  the spin flip process in
$ \left({\cal K}\left({\cal M}\frac{\delta}{\delta \hat{K}}
+\psi\right)\right)_{X}(\xi)$
as,
\begin{eqnarray}
\lefteqn{-\partial_{\xi}\oint\frac{d\eta}{2\pi i}
[\left({\cal M}\frac{\delta}{\delta \hat{K}}\right)_{2}
 (\xi,\eta)+\psi_{2}(\xi,\eta)]}  \nonumber \\
 & = &
 -\hat{\oint}\frac{d\eta}{2\pi i}
\partial_{\xi}\frac{\delta}{\delta\hat{K}_{2}(\xi,\eta)}-
\partial_{\xi}\left(\left(\frac{1}{\hat{g}}
                     ((1-c-\hat{g}\xi)\xi-2\psi_{X}(\xi))\right)
 \frac{\delta}{\delta\hat{K}_{X}(\xi)}\right)  \nonumber \\
 &  &  -\frac{1}{\hat{g}}\partial_{\xi}
                [\hat{g}\xi+((1-c-\hat{g}\xi)\xi-\psi_{X}(\xi))\psi_{X}(\xi)],
\label{5-14-1}
\end{eqnarray}
where the first term is the universal part, and the others
 are the non-universal parts.
Similarly, we can derive the expressions for the spin-flip processes
in $\left({\cal K}\left({\cal M}\frac{\delta}{\delta \hat{K}}
+\psi\right)\right)_{2}
(\xi_{1},\xi_{2}).$
However, it must be noted that the cylinder amplitude
$V^{\mbox{\scriptsize cyl}~(1)}_{Y}(\xi)$ has a non-universal c-number term as
shown in (\ref{V1mixing}) in  Appendix B.
In this case, a special care must  be exercised in rewriting
$\left({\cal K}\left({\cal M}\frac{\delta}{\delta\hat{K}}
+\psi \right)\right)_{1}(\xi_{1})$.
 Namely, from (\ref{5-7-3}) and (\ref{V1mixing}),
we can obtain the following relations:
\begin{eqnarray}
\lefteqn{ \left.\frac{\delta}{\delta\hat{K}_{1}(\xi)}
\oint\frac{d\eta}{2\pi i}[\left({\cal M}
\frac{\delta}{\delta\hat{K}}\right)_{1}(\eta)
        +\psi_{1}(\eta)] \hat{Z}[\hat{K}]\right|_{\hat{K}=0}
=\langle\hat{\Psi}_{1}(\xi)\oint\frac{d\eta}{2\pi i}\Psi_{1}(\eta)\rangle }
\nonumber \\
 & & =\frac{1}{N^2}\frac{1}{\sqrt{5c}} +
      \langle\hat{\Psi}_{1}(\xi)\hat{\oint}\frac{d\eta}{2\pi i}
      \hat{\Psi}_{1}(\eta)\rangle
     =\frac{1}{N^2}\frac{1}{\sqrt{5c}} +
      \left.\frac{\delta}{\delta\hat{K}_{1}(\xi)}
      \hat{\oint}\frac{d\eta}{2\pi i}\frac{\delta}{\delta\hat{K}_{1}(\eta)}
      \hat{Z}[\hat{K}]\right|_{\hat{K}=0},
\label{5-13-1}
\end{eqnarray}
and therefore the expression requires a slight modification as follows:
\begin{equation}
\oint\frac{d\eta}{2\pi i}
[\left({\cal M}\frac{\delta}{\delta\hat{K}}\right)_{1}(\eta)
        +\psi_{1}(\eta)]
=\int \frac{d\eta}{2\pi i}\hat{K}_{1}(\eta)\frac{1}{N^{2}}\frac{1}{\sqrt{5c}}
       +\hat{\oint}\frac{d\eta}{2\pi i}\frac{\delta}{\delta\hat{K}_{1}(\eta)}.
\label{modifiedspinflip}
\end{equation}
The first term in (\ref{modifiedspinflip}) leads to the non-universal
c-number term
$\frac{1}{\sqrt{5c}}$
in the cylinder amplitude $V^{\mbox{\scriptsize cyl}~(1)}_{Y}(\xi)$.
The appearance of  such a  term may be
interpreted as arising from the difference
between the two methods of regularizations
for the $\eta$-integral, namely,
the lattice regularization (matrix model) and the
Beta-function regularization.  For higher components,
we expect that there is no such modification coming from the terms
containing $\hat{K}_{n}$ ($n\geq 2$),
because $V^{\mbox{\scriptsize cyl}~(3)}_{Y}$
and $V^{\mbox{\scriptsize cyl}~(5)}_{Y}$ have no non-universal c-number terms
and the higher cylinder amplitudes scale with negative powers of $a$,
according to the results of the cylinder amplitudes
(\ref{V3mixing}) and (\ref{V5mixing}) in  Appendix B.

  Concerning the third term in the l.h.s. of (\ref{kineticterm}),
$(\wedge\psi)_{I,J}$ vanishes, as we will show below,
except for the $I=J=1$ component:
\begin{equation}
(\wedge\psi)_{I,J}(\xi_{1},\cdots,\xi_{I};\xi'_{1},\cdots,\xi'_{J})
= \left\{ \begin{array}{ll}
             \frac{1}{3}(\hat{g}(\xi_{1}+\xi'_{1})+5c-1)
                                                 & \mbox{$I=J=1$} \nonumber \\
             0                                   & \mbox{otherwise}.
          \end{array}
  \right.
\label{c-nomerge}
\end{equation}

   Using these results,  we arrive at
Eq. (\ref{kineticterm}) with
\begin{eqnarray}
\hat{K}\cdot \left({\cal F}\frac{\delta}{\delta\hat{K}}\right)
 & = &
\int \frac{d\xi}{2\pi i}\hat{K}_{X}(\xi)\hat{g}(-\partial_{\xi})\hat{\oint}
\frac{d\eta}{2\pi i}\frac{\delta}
         {\delta\hat{K}_{2}(\xi,\eta)} \nonumber \\
 &  &
+\int \frac{d\xi_{1}}{2\pi i}\hat{K}_{1}(\xi_{1})\hat{g}
 \hat{\oint}\frac{d\eta}{2\pi i}
            \frac{\delta}{\delta\hat{K}_{3}(\xi_{1},\eta,\xi_{1})}
+\cdots \nonumber \\
 &  &
+\int \frac{d\xi_{1}}{2\pi i}\frac{d\xi_{2}}{2\pi i}
      \hat{K}_{2}(\xi_{1},\xi_{2})\hat{g}\hat{\oint}
\frac{d\eta}{2\pi i}
\left(\frac{\delta}{\delta\hat{K}_{4}(\xi_{1},\eta,\xi_{1},\xi_{2})}
     +\frac{\delta}{\delta\hat{K}_{4}(\xi_{1},\xi_{2},\eta,\xi_{2})}
                    \right)
+\cdots  \nonumber \\
 &  & +\cdots,  \label{5-15-1}
\end{eqnarray}
where the ellipsis stands for the subleading terms and the terms containing
the higher components $\hat{K}_{n}$ ($n\geq 3$).

It is noted that the tadpole term is canceled
with a contribution of the same form from the kinetic term,
as in  the original Ising theory \cite{SY},  and that
$\psi$-dependent terms are absent because their contributions
   from kinetic and splitting terms cancel each other.
In particular
the leading contribution of $(\wedge\psi)_{1,1}$ is precisely canceled
with the non-universal c-number term of (\ref{modifiedspinflip}).
  Although we do not elaborate further on determining
the explicit continuum
limits for higher components, it is natural, because of the
local nature of the spin-flip processes, to suppose that
the above expression already indicates the generic structure of
the kinetic term, namely the flipping of a single spin on a loop with
general spin configurations,  and the absence of the tadpole terms and
the terms without spin flipping.

\vspace{0.3cm}
\noindent
(2) {\em Interaction term}

   Next, we consider
the splitting and merging processes, Eqs. (\ref{splitterm}) and
(\ref{mergeterm}).
In the next section, we will argue the validity of
the following important relations,
showing the commutativity of the splitting and merging
processes with the operator mixing,
\begin{equation}
\left(\left({\cal M}\frac{\delta}{\delta\hat{K}}\right)
\vee\left({\cal M}\frac{\delta}{\delta\hat{K}}\right)\right)_{I}=
\left({\cal M}\left(\frac{\delta}{\delta\hat{K}}
\vee\frac{\delta}{\delta\hat{K}}\right)\right)_{I},
\label{splitmix}
\end{equation}
\begin{equation}
\left(\wedge\left({\cal M}\frac{\delta}{\delta\hat{K}}\right)\right)_{I,J}=
\sum_{K,L}{\cal M}_{IK}{\cal M}_{JL}\left(\wedge\frac{\delta}{\delta\hat{K}}
\right)_{K,L},
\label{mergemix}
\end{equation}
for completely general case, provided the mixing obeys some rules
which can be explicitly confirmed for several lower nontrivial
components.
\footnote{
The proof of the next section is applicable both for the
original and dual theories, generalizing the results for the
original theory given in the previous work \cite{SY}.
}
This commutativity property
ensures that the general structure
of the splitting and the merging in the continuum limit are
essentially the same as
in the lattice theory.

\vspace{0.3cm}
\noindent
(3) {\em C-number term}

   Finally, let us turn to (\ref{c-nomerge}). We can
check its validity
 for the components
$ (I,J) = (X,X),(X,2),(X,2k+1),(1,1),(1,3),(1,2k),(2,2),(2,2k+1)$
($k$ is integer) by direct calculation. For higher components,
we give a general proof of (\ref{c-nomerge}) in the same way as
\cite{SY}. First we show that $\psi_{2k}$ must be polynomial for
every $k$. For this purpose, we start with the Schwinger-Dyson equations
and derive the dual version of Staudacher's recursion equations
\cite{St}. They relate $V^{(k)}$ to the amplitudes
$V^{(l)}$ ($l < k$) as
\begin{eqnarray}
\lefteqn{V^{(2k)}(\xi_{1},\cdots,\xi_{2k})
     = \frac{1}{1+c-\hat{g}(\xi_{1}+\xi_{2})}} \nonumber \\
     & & \times \left\{V(\xi_{1})(-D_{\xi}(\xi_{2},\xi_{2k}))
               V^{(2k-2)}(\xi_{1},\xi_{3},\cdots,\xi_{2k-1})
                                                               \right.
             +V(\xi_{2})(-D_{\xi}(\xi_{1},\xi_{3}))
               V^{(2k-2)}(\xi,\xi_{4},\cdots,\xi_{2k}) \nonumber \\
     & & +\sum_{l=2}^{k-1} \left[D_{\xi}(\xi_{2},\xi_{2l})
               V^{(2l-2)}(\xi,\xi_{3},\cdots,\xi_{2l-1})\right]
               \left[D_{\xi}(\xi_{1},\xi_{2l+1})
               V^{(2k-2l)}(\xi,\xi_{2l+2},\cdots,\xi_{2k})\right] \nonumber \\
      & & -\left.\hat{g}V^{(2k-1)}_{2}(\xi_{2},\cdots,\xi_{2k};)
                    -\hat{g}V^{(2k-1)}_{2}
                          (\xi_{3},\cdots,\xi_{2k},\xi_{1};)\right\}, \\
\lefteqn{\hat{g}V^{(2k-1)}_{2}(\xi_{2},\cdots,\xi_{2k};)} \nonumber \\
        & & = (-1)^{k+1}\hat{g}V_{2}(\xi_{k+1})
                + \sum_{p=0}^{k-2}(-1)^{k+p+1}
                   D_{\xi}(\xi_{k-p},\xi_{k+p+2}) \nonumber \\
        & &\times \left[\left((1-c)\xi-\hat{g}\xi^{2}-V(\xi_{k-p})
                                  -V(\xi_{k+p+2})\right)\right.
                \left.V^{(2p+2)}
                          (\xi,\xi_{k-p+1},\cdots,\xi_{k+p+1})\right]
\nonumber \\
        & & + \sum_{p=0}^{k-2}\sum_{l=1}^{p}(-1)^{k+p+1}
                \left[D_{\xi}(\xi_{k-p},\xi_{k-p+2l})
               V^{(2l)}(\xi,\xi_{k-p+1},\cdots,\xi_{k-p+2l-1})\right]
\nonumber \\
        & & \times \left.\left[D_{\xi}(\xi_{k-p+2l},\xi_{k+p+2})
          V^{(2p-2l+2)}(\xi,\xi_{k-p+2l+1},\cdots,\xi_{k+p+1})
                            \right]\right\},
\end{eqnarray}
and $\hat{g}V^{(2k-1)}_{2}(\xi_{3},\cdots,\xi_{2k},\xi_{1};)$
has a  similar form. Suppose that $\psi_{2k}$'s are polynomials up to some $k$.
Then using above equations we see
that the part of the numerator for $V^{(2k)}$
consisting only of $\psi$ is a polynomial,
because in general the combinatorial derivative of a polynomial
is also a polynomial and c-number function $\psi$ in
$V_{2}(\xi_{k+1})$ is a polynomial. The denominator,
on the other hand, behaves in the scaling limit as
$$
1+c-\hat{g}(\xi_{1}+\xi_{2}) = -ac\hat{s}(y_{1}+y_{2}) + O(a^{2}).
$$
Thus, by using the scaled variables
 $\xi_{i}=\xi_{*}(1+ay_{i}),$
$\psi_{2k}$ can be written as
$$
\psi_{2k}=\frac{\mbox{Polynomial of }(y_{1},y_{2},\cdots,y_{2k})}
                 {y_{1}+y_{2}}.
$$
However, from the cyclic symmetry with respect to $\xi_i$'s in
$V^{(2k)}(\xi_{1},\xi_{2},\cdots, \xi_{2k})$,
the denominator $y_{1}+y_{2}$ must be
cancelled with the numerator, and thus $\psi_{2k}$ should have the form:
$$
\psi_{2k}=\mbox{Polynomial of } (y_{1},y_{2},\cdots,y_{2k}),
$$
where the polynomial has the same symmetry as $V^{(2k)}$.
Thus, by induction, the $\psi_{2k}$ must be a polynomial for general $k$.

     Now from the scaling behavior of the universal part $\hat{V}^{(2k)}$
$$
\hat{V}^{(2k)}(\xi_{1},\cdots,\xi_{2k}) =
\langle \hat{\Psi}_{2k}(\xi_1,\cdots,\xi_{2k})\rangle_0=
a^{\frac{7}{3}-\frac{2}{3}k}v ^{(2k)}(y_{1},y_{2},\cdots,y_{2k})
$$
derived in  Appendix B,
we expect that the relevant part of $\psi_{2k}$
takes the form
\begin{equation}
\psi_{2k}=\left\{\begin{array}{ll}\mbox{const.} &  k=3 \\

0 & k\geq 4 .     \end{array}  \right.
\label{phik}
\end{equation}
Since every component of $\wedge\psi$ contains the derivative or
the combinatorial derivative, Eq. (\ref{phik}) leads to (\ref{c-nomerge}).

We are now ready to take the continuum limit of the stochastic
Hamiltonian (\ref{hatHamiltonian}). The scaling property of
the string fields
presented in Appendix B enables us to
introduce the variables in the continuum
theory
as follows:
$$
\hat{g} =
\hat{g}_{*}(1-a^{2}\frac{\hat{s}^{2}}{10}\hat{T}),~~~\xi=\xi_{*}(1+ay),
          ~~\frac{1}{N} =a^{7/3}g_{\mbox{st}},
$$
\begin{eqnarray}
\frac{\delta}{\delta\hat{K}_X(\xi)} & = &
a^{4/3}\xi_{*}^{-1}\frac{\delta}{\delta\tilde{K}_X(y)},
{}~~~\hat{K}_X(\xi)=a^{-7/3}\tilde{K}_X(y),\nonumber \\
\frac{\delta}{\delta\hat{K}_n(\xi_{1},\cdots,\xi_{n})}
& = & a^{(7-n)/3}\xi_{*}^{-n}
 \frac{\delta}{\delta\tilde{K}_{n}(y_{1},\cdots,y_{n})}, \nonumber \\
 &  & \hat{K}_n(\xi_{1},\cdots,\xi_{n})=a^{-(7+2n)/3}\tilde{K}_{n}
       (y_{1},\cdots,y_{n}), \nonumber \\
 &  & (n=1,2,3,\cdots).
\end{eqnarray}

 In the limit $a\rightarrow 0$, the leading contributions in
(\ref{hatHamiltonian})
start with
$O(a^{1/3})$, which come from the $X_{\alpha}$-derivatives in
(\ref{matrixlaplacian}). On the other hand, the contributions from the
$Y_{\alpha}$-derivatives are subleading and do not survive in the
continuum limit. After the finite rescaling
$$
\tilde{K}_{I}\rightarrow (\xi_*\hat{g}_{*})^{-1}\tilde{K}_{I},
{}~~~\frac{\delta}{\delta\tilde{K}_{I}}\rightarrow
\xi_*\hat{g}_{*}\frac{\delta}{\delta\tilde{K}_{I}},
{}~~~g_{\mbox{st}}^{2}\rightarrow (\xi_*\hat{g}_{*})^{2}g_{\mbox{st}}^{2},
$$
and absorbing the overall factor $a^{1/3}\xi_*^{-1}\hat{g}_*$ into the
renormalization of the fictitious time, we have the final result for the
continuum stochastic Hamiltonian:
\begin{equation}
{\cal H}_D=-\tilde{K}\cdot\left({\cal F}\frac{\delta}{\delta\tilde{K}}\right)
-\tilde{K}\cdot
\left(\frac{\delta}{\delta\tilde{K}}\vee\frac{\delta}{\delta\tilde{K}}\right)-
   g_{\mbox{st}}^{2}\tilde{K}\cdot
\left(\tilde{K}\cdot\left(\wedge\frac{\delta}{\delta\tilde{K}}\right)\right),
\label{continuumH}
\end{equation}
where the inner product is defined by
\begin{equation}
f\cdot g \equiv \int_{-i\infty}^{i\infty}\frac{dy}{2\pi i} f_{X}(y)g_{X}(y)
  +\sum_{n=1}^{\infty} \int_{-i\infty}^{i\infty}\prod_{i=1}^{n}
     \frac{dy_{i}}{2\pi i} f_{n}(y_{1},\cdots,y_{n})
            g_{n}(y_{1},\cdots,y_{n}).
\end{equation}
Each term in (\ref{continuumH}) has the following structure.
The first term is the kinetic term consisting only of the spin-flip
processes:
\begin{eqnarray}
\left({\cal F}\frac{\delta}{\delta\tilde{K}}\right)_X(y) & = &
-\partial_{y}\int_{C}\frac{dx}{2\pi i}
       \frac{\delta}{\delta\tilde{K}_2(y,x)}, \nonumber \\
\left({\cal F}\frac{\delta}{\delta\tilde{K}}\right)_1(y_1) & = &
\int_{C}\frac{dx}{2\pi i}
       \frac{\delta}{\delta\tilde{K}_3(y_1,x,y_1)}, \nonumber\\
\left({\cal F}\frac{\delta}{\delta\tilde{K}}\right)_2(y_{1},y_2) & = &
 \int_{C}\frac{dx}{2\pi i}
\left[\frac{\delta}{\delta\tilde{K}_4(y_{1},x,y_{1},y_2)}+
 \frac{\delta}{\delta\tilde{K}_4(y_{1},y_2,x,y_2)}\right], \nonumber \\
   &  & \cdots,  \nonumber \\
\left({\cal F}\frac{\delta}{\delta\tilde{K}}\right)_n(y_{1},\cdots,y_n)
& = & \sum_{j=1}^n\int_{C}\frac{dx}{2\pi i}
\frac{\delta}{\delta\tilde{K}_{n+2}(y_{1},\cdots,y_j,x,y_j,\cdots,y_n)},
\nonumber \\
   & & \cdots.
\end{eqnarray}

  The second and third terms represent the splitting and merging processes,
respectively. Due to the commutativity of
these processes with the operator
mixing, they have the same structure as the contributions from the
$X_{\alpha}$-
derivatives in the lattice theory. The components of the splitting term are
given as
\begin{eqnarray*}
\left(\frac{\delta}{\delta\tilde{K}}
\vee\frac{\delta}{\delta\tilde{K}}\right)_X(y) & = &
-\partial_{y}\frac{\delta^{2}}
   {\delta \tilde{K}_X(y)^{2}},   \\
\left(\frac{\delta}{\delta\tilde{K}}
\vee\frac{\delta}{\delta\tilde{K}}\right)_1(y_1) & = &
-2\frac{\delta}{\delta \tilde{K}_X(y_1)}
    \partial_{y_1}\frac{\delta}{\delta \tilde{K}_1(y_1)},   \\
\left(\frac{\delta}{\delta\tilde{K}}
\vee\frac{\delta}{\delta\tilde{K}}\right)_2(y_1,y_2) & = & -2\sum_{j=1}^2
    \frac{\delta}{\delta\tilde{K}_X(y_j)}\partial_{y_j}
    \frac{\delta}{\delta\tilde{K}_2(y_1,y_2)}
    +2\left(D_z(y_1,y_2)\frac{\delta}{\delta\tilde{K}_1(z)}\right)^2,
\end{eqnarray*}
$$ \cdots, $$
\begin{eqnarray*}
\lefteqn{\left(\frac{\delta}{\delta\tilde{K}}
\vee\frac{\delta}{\delta\tilde{K}}\right)_n(y_{1},\cdots,y_n)
  =  -2\sum_{j=1}^n\frac{\delta}{\delta\tilde{K}_X(y_{j})}\partial_{y_{j}}
          \frac{\delta}{\delta\tilde{K}_n(y_{1},\cdots,y_n)}}   \\
 &+  & 2\sum_{k<l}D_z(y_k,y_l)
          \frac{\delta}{\delta\tilde{K}_{l-k}(z,y_{k+1},\cdots,y_{l-1})}
       D_w(y_k,y_l) \frac{\delta}{\delta\tilde{K}_{n-l+k}
          (y_1,\cdots,y_{k-1},w,y_{l+1},\cdots,y_n)}
\end{eqnarray*}
\begin{equation}
 \cdots.
\end{equation}
Also, the components of the merging term are
$$
\left(\wedge \frac{\delta}{\delta \tilde{K}}\right)_{X,X}(y; y') =
 -\partial_{y}\partial_{y'}D_z(y,y')\frac{\delta}{\delta \tilde{K}_X(z)},
$$
 \begin{eqnarray*}
\lefteqn{\left(\wedge \frac{\delta}{\delta \tilde{K}}\right)_{n,X}
(y_1,\cdots,y_n; y') =
\left(\wedge \frac{\delta}{\delta \tilde{K}}\right)_{X,n}
(y'; y_1,\cdots, y_n)} \\
 & = & -\partial_{y'}\sum_{j=1}^{n}\partial_{y_j}D_z(y_j, y')
        \frac{\delta}{\delta \tilde{K}_n(y_1,\cdots, y_{j-1},z,y_{j+1},
\cdots, y_n)} \qquad (n\geq 1),
\end{eqnarray*}
$$
\left(\wedge \frac{\delta}{\delta \tilde{K}}\right)_{1,1}(y_1; y'_1)
=D_z(y_1,y'_1)D_w(y_1,y'_1)
\frac{\delta}{\delta\tilde{K}_2(z,w)},
$$
\begin{eqnarray*}
\lefteqn{\left(\wedge \frac{\delta}{\delta \tilde{K}}\right)_{n,1}
(y_1,\cdots,y_n;y'_1) =
\left(\wedge \frac{\delta}{\delta \tilde{K}}\right)_{1,n}
(y'_1; y_1,\cdots, y_n)} \\
 & = & \sum_{j=1}^n   D_z(y_j,y'_1)D_w(y_j,y'_1)
\frac{\delta}{\delta \tilde{K}_{n+1}
(y_1,\cdots,y_{j-1},z,w,y_{j+1}, \cdots,y_n)}
\qquad (n \geq 2),
\end{eqnarray*}
\begin{eqnarray}
\lefteqn{\left(\wedge \frac{\delta}{\delta \tilde{K}}\right)_{n,m}
(y_1,\cdots,y_n;y'_1, \cdots,y'_m) =
   \sum_{j=1}^n\sum_{k=1}^mD_z(y_j,y'_k)D_w(y_j,y'_k)} \nonumber\\
 & \times &
\frac{\delta}{\delta\tilde{K}_{n+m}(y_1,\cdots,y_{j-1},z,y'_{k+1},\cdots,y'_m,
   y'_1,\cdots,y'_{k-1},w,y_{j+1},\cdots,y_n)} \nonumber \\
   &  &  \qquad (n,m \geq 2).
\end{eqnarray}
As we have promised earlier, the terms of the
Hamiltonian arising from the $Y_{\alpha}$-derivatives
do not contribute in the continuum limit.
Thus it should be noted that the symbols
$\vee$ and $\wedge$ used here do not represent completely identical
objects with $\vee$ and $\wedge$ in eq.
(\ref{hatHamiltonian}), because the latter contains terms
vanishing in the continuum limit
which originate from
the $Y_{\alpha}$-derivatives in Eq. (\ref{matrixlaplacian}).

  In this Hamiltonian, the interactions including the string fields
with
odd domains survive after the continuum limit.
This is important for discussing the duality symmetry.
Remember that the string fields with
even domains can be interpreted in terms of dual spins,
 while those with odd domains cannot be. If we allow  only the
strings with even domains in the initial state
and consider its time
evolution
by this Hamiltonian {\it within} the tree approximation,  the strings with
odd domains never contribute,
because the vacuum expectation value of a single
string field with
odd domains vanishes owing to the $\mbox{\boldmath $Z$}_2$-symmetry.
In general, however, the strings with odd domains necessarily
contribute in the
intermediate states, since there exists such a process
that a string
with even domains splits into two strings with odd domains
and they merge to
a single string again.
It represents the excitation of
odd $Y$-loops
along
the non-trivial cycles of the handles.
In section VI, we will further examine the correspondence between the
original
and dual theories beyond the level of the disk and cylinder amplitudes
presented in Appendix B.

\section{Algebra of Splitting and Merging Interactions}

We saw that the commutativity of the mixing matrix with
splitting and merging processes plays
a crucial role in obtaining the continuum
Hamiltonian in both the original and the dual models.
In view of its potential importance
for the development of string field theories,
we devote the present section to its
derivation and further elucidation.

The commutativity was previously confirmed explicitly in
several simple (but
nontrivial) configurations
and conjectured for general configurations.
Although we cannot still give a complete
proof, we shall improve the situation by proposing
the  general form of ${\cal M}_{IJ}$, and then
proving the commutativity  for generic case on the basis of this
assumption.

The general forms of the mixing matrices are
summarized as:
\begin{enumerate}
\item ${\cal M}$ can be expressed in terms of a matrix ${\cal R}$ as
\begin{equation}
{\cal M}_{IJ} = ( e^{\alpha{\cal R}} )_{IJ},  \label{rule1}
\end{equation}
where $\alpha$ is a constant whose value is $\alpha = \sqrt{10c}$
for the original model and $\alpha = \frac2{\sqrt{5c}}$ for the dual
one.
\item The definition of ${\cal R}$.
\begin{enumerate}
\item The original model:
\begin{eqnarray}
&&({\cal R} f)_A (\zeta) = 0, \quad
  ({\cal R} f)_B (\sigma) = 0, \quad
  ({\cal R} f)_1 (\zeta, \sigma) = f_A (\zeta) + f_B (\sigma), \nonumber\\
&&({\cal R} f)_n (\zeta_1, \sigma_1, \cdots, \zeta_n, \sigma_n)
\quad ( n \ge 2) \nonumber\\
&& ~~~~ = - \sum_{j=1}^n D_\zeta (\zeta_{j}, \zeta_{j+1}) f_{n-1}
(\zeta_1, \cdots, \sigma_{j-1}, \zeta, \sigma_{j+1}, \cdots, \sigma_n)
\nonumber\\
&& ~~~~~~~~ - \sum_{j=1}^n D_\sigma (\sigma_{j-1}, \sigma_j) f_{n-1}
(\zeta_1, \cdots, \zeta_{j-1}, \sigma, \zeta_{j+1}, \cdots, \sigma_n)
\label{rule3} \\
&& ~~~~ \quad ( \zeta_0 \equiv \zeta_n, \; \zeta_{n+1} \equiv \zeta_1, \;
\sigma_0 \equiv \sigma_n, \; \sigma_{n+1} \equiv \sigma_1 ). \nonumber
\end{eqnarray}
\item The dual model:
\begin{eqnarray}
&&({\cal R} f)_X (\xi) = 0, \quad
  ({\cal R} f)_1 (\xi_1) = 0, \quad
  ({\cal R} f)_2 (\xi_1, \xi_2) = f_X (\xi_1) + f_X (\xi_2), \nonumber\\
&&({\cal R} f)_n (\xi_1, \cdots, \xi_n)
\quad ( n \ge 3) \nonumber\\
&& ~~~~ = - \sum_{j=1}^n D_z (\xi_{j-1}, \xi_{j+1}) f_{n-2}
(\xi_1, \cdots, \xi_{j-2}, z, \xi_{j+2}, \cdots, \xi_n)
\label{rule5} \\
&& ~~~~ \quad ( \xi_0 \equiv \xi_n, \; \xi_{n+1} \equiv \xi_1 ).
\nonumber
\end{eqnarray}
\end{enumerate}
\end{enumerate}
The validity of these rules is explicitly
confirmed for the original model up
to the $I=A,B,1,2$ components of ${\cal M}_{IJ}$
as given in Eq. (99) of \cite{SY}, and
for the dual model up to the $I=X,1,2,3,4,5$
components of ${\cal M}_{IJ}$ as given
in Eqs. (\ref{dualsf}) and (\ref{V5mixing}).

The physical meaning of the operator ${\cal R}$ is
reducing one domain of spins.
Reduction of one domain in generic spin configurations causes
merging of two domains which have been separated by the reduced
domain. This explains the appearance of the
 combinatorial derivatives in ${\cal R}$.

Now assuming the above general structure (\ref{rule1}) for ${\cal M}_{IJ}$,
 the proof
of the commutativity Eqs. (\ref{splitmix}) and (\ref{mergemix})
is reduced to establish the
following equations (\ref{splitR}) and (\ref{mergeR}), respectively:
\begin{eqnarray}
\left( {\cal R} ( f \vee g ) \right)_I
= \left( ({\cal R} f) \vee g \right)_I
+ \left( f \vee ({\cal R} g) \right)_I, \label{splitR} \\
\left( \wedge ({\cal R} f) \right)_{I,J}
= \sum_{K} {\cal R}_{IK} ( \wedge f)_{K,J}
+ \sum_{L} {\cal R}_{JL} ( \wedge f)_{I,L}, \label{mergeR}
\end{eqnarray}
where we introduced two independent
arbitrary vectors $f_I$ and $g_I$ to emphasize
that the above equations should be valid irrespectively of the particular
structure of $\left( \frac{\delta}{\delta\hat{J}} \right)_I$ or $\left(
\frac{\delta}{\delta\hat{K}} \right)_I$.
The relations (\ref{splitR}) and (\ref{mergeR})
say that ${\cal R}$ acts like a derivation on the string
fields with respect to the rule of their multiplications
defined by the merging-splitting interactions.
Note also the possibility of interpreting  the operator
${\cal R}$ as a conserved charge on the world sheet.

The definitions of $\vee$ and $\wedge$ for lower components in
the original model are given in (95) and (96) of \cite{SY}.
For attempting the proof of commutativity,
we need more precise definitions for general configurations.
\begin{eqnarray}
& &(f \vee g)_A (\zeta) =
- f_A (\zeta) \partial_\zeta g_A (\zeta) -g_A(\zeta) \partial_\zeta f_A(\zeta),
\nonumber\\
& &(f \vee g)_B (\sigma) =
- f_B (\sigma) \partial_\sigma g_B (\sigma) -g_B(\sigma)
\partial_\sigma f_B(\sigma), \nonumber\\
& &(f \vee g)_n (\zeta_1, \sigma_1, \cdots, \zeta_n, \sigma_n) \nonumber\\
& & =
- \sum_{k=1}^n f_A(\zeta_k) \partial_{\zeta_k}
g_n(\zeta_1, \sigma_1, \cdots, \zeta_n, \sigma_n)
- \sum_{k=1}^n f_B(\sigma_k) \partial_{\sigma_k}
g_n(\zeta_1, \sigma_1, \cdots, \zeta_n, \sigma_n) \nonumber\\
& & ~~~~ +
\sum_{k<l}
D_\zeta(\zeta_k,\zeta_l) f_{l-k}(\zeta, \sigma_k, \cdots,\zeta_{l-1},
\sigma_{l-1})
D_{\zeta'}(\zeta_k,\zeta_l) g_{n-l+k}
(\zeta_1, \cdots, \zeta_{k-1}, \sigma_{k-1}, \zeta', \sigma_l, \cdots,
\sigma_n)
\nonumber\\
& & ~~~~ +
\sum_{k<l}
D_\sigma(\sigma_k,\sigma_l) f_{l-k}(\zeta_{k+1}, \sigma_{k+1},
\cdots,\zeta_l, \sigma)
D_{\sigma'}(\sigma_k,\sigma_l) g_{n-l+k}
(\zeta_1, \cdots, \zeta_k, \sigma', \zeta_{l+1}, \sigma_{l+1},
\cdots, \sigma_n)
\nonumber\\
& & ~~~~
- \sum_{k=1}^n g_A(\zeta_k) \partial_{\zeta_k}
f_n(\zeta_1, \sigma_1, \cdots, \zeta_n, \sigma_n)
- \sum_{k=1}^n g_B(\sigma_k) \partial_{\sigma_k}
f_n(\zeta_1, \sigma_1, \cdots, \zeta_n, \sigma_n) \nonumber\\
& & ~~~~ +
\sum_{k<l}
D_\zeta(\zeta_k,\zeta_l) g_{l-k}(\zeta, \sigma_k, \cdots,\zeta_{l-1},
\sigma_{l-1})
D_{\zeta'}(\zeta_k,\zeta_l) f_{n-l+k}
(\zeta_1, \cdots, \zeta_{k-1}, \sigma_{k-1}, \zeta', \sigma_l, \cdots,
\sigma_n)
\nonumber\\
& & ~~~~ +
\sum_{k<l}
D_\sigma(\sigma_k,\sigma_l) g_{l-k}(\zeta_{k+1}, \sigma_{k+1},
\cdots,\zeta_l, \sigma)
D_{\sigma'}(\sigma_k,\sigma_l) f_{n-l+k}
(\zeta_1, \cdots, \zeta_k, \sigma', \zeta_{l+1}, \sigma_{l+1},
\cdots, \sigma_n),
\nonumber\\
& &
\end{eqnarray}
\begin{eqnarray}
& & (\wedge f)_{A,A}(\zeta;\zeta') =
-\partial_\zeta \partial_{\zeta'} D_z(\zeta,\zeta')f_A(z), \nonumber\\
& & (\wedge f)_{A,B}(\zeta;\sigma) =
(\wedge f)_{B,A}(\sigma;\zeta) = 0, \nonumber\\
& & (\wedge f)_{B,B}(\sigma;\sigma') =
-\partial_\sigma \partial_{\sigma'} D_s(\sigma,\sigma') f_B(s), \nonumber\\
& & (\wedge f)_{n,A}(\zeta_1,\sigma_1,\cdots,\zeta_n,\sigma_n;\zeta') =
(\wedge f)_{A,n}(\zeta';\zeta_1,\sigma_1,\cdots,\zeta_n,\sigma_n) \nonumber\\
& & ~~~~ =
- \partial_{\zeta'} \sum_{k=1}^n \partial_{\zeta_k} D_z(\zeta_k,\zeta')
f_n(\zeta_1,\cdots,\zeta_{k-1},\sigma_{k-1},z,\sigma_k,\cdots,\sigma_n),
\nonumber\\
& & (\wedge f)_{n,B}(\zeta_1,\sigma_1,\cdots,\zeta_n,\sigma_n;\sigma') =
(\wedge f)_{B,n}(\sigma';\zeta_1,\sigma_1,\cdots,\zeta_n,\sigma_n) \nonumber\\
& & ~~~~ =
- \partial_{\sigma'} \sum_{k=1}^n \partial_{\sigma_k} D_s(\sigma_k,\sigma')
f_n(\zeta_1,\cdots,\zeta_k,s,\zeta_{k+1},\sigma_{k+1},\cdots,\sigma_n),
\nonumber\\
& & (\wedge f)_{n,m}
(\zeta_1,\sigma_1,\cdots,\zeta_n,\sigma_n;\zeta'_1,\sigma'_1,
\cdots,\zeta'_m,\sigma'_m)
\nonumber\\
& & =
\sum_{k=1}^n \sum_{l=1}^m D_z(\zeta_k,\zeta'_l) D_w(\zeta_k,\zeta'_l)
f_{n+m}(\zeta_1,\sigma_1,\cdots,\sigma_{k-1},z,\sigma'_l,\cdots,\sigma'_m,
\zeta'_1,\cdots,\sigma'_{l-1},w,\sigma_k,\cdots,\sigma_n) \nonumber\\
& & +
\sum_{k=1}^n \sum_{l=1}^m D_s(\sigma_k,\sigma'_l) D_t(\sigma_k,\sigma'_l)
f_{n+m}(\zeta_1,\cdots,\zeta_k,s,\zeta'_{l+1},\cdots,\zeta'_m,\sigma'_m
\zeta'_1,\cdots,\zeta'_l,t,\zeta_{k+1},\cdots,\zeta_n,\sigma_n). \nonumber\\
& &
\end{eqnarray}
Note that we have slightly generalized
the original definition of $\vee$  by introducing
two different vectors
$f_I$ and $g_I$.

For the dual model, we have already
presented the definitions of $\vee$ and
$\wedge$ in (\ref{dualsplit}) and (\ref{dualmerge}) in section III.
Again, we make a slight generalization
of the definition of $\vee$ as above.
The prescription in this generalization is the same as
above. It is sufficient to
explain it using a simple example.
For instance, the $I=1$ component for the dual model is defined in
(\ref{dualsplit}) as
\begin{eqnarray}
\left( \frac{\delta}{\delta K} \vee \frac{\delta}{\delta K} \right)_1 (\xi_1)
=  -2\frac{\delta}{\delta K_{X}(\xi_{1})}\partial_{\xi_{1}}
      \frac{\delta}{\delta K_{1}(\xi_{1})}. \nonumber
\end{eqnarray}
Correspondingly, we generalize the definition as
\begin{equation}
( f \vee g)_1(\xi_1) =
-  f_X(\xi_1) \partial_{\xi_1} g_1(\xi_1)
-  g_X(\xi_1) \partial_{\xi_1} f_1(\xi_1).
\end{equation}
The prescriptions for generic cases can be easily deduced from this
example.\\

Let us now proceed to prove Eqs. (\ref{splitR}) and
(\ref{mergeR}) for both original and dual models on
the basis of the rules (\ref{rule1}) $\sim$ (\ref{rule5}).
Most parts of the proof consist of
straightforward calculations after plugging
all these definitions into the formulas. However, we need several non-trivial
identities for some cases. The set of such identities and the
cases to use them are the same for both original and dual models,
signaling their universal nature.

For Eq. (\ref{splitR}), we use the following two identities valid
for arbitrary functions $F(\zeta)$, $G(\zeta)$ and $H(z,w)$
:
\begin{eqnarray}
\lefteqn{
D_\zeta(\zeta_1,\zeta_2) F(\zeta) \partial_\zeta G(\zeta) =
\sum_{k=1}^2 F(\zeta_k) \partial_{\zeta_k} D_{\zeta}(\zeta_1,\zeta_2)
G(\zeta) + D_{\zeta}(\zeta_1,\zeta_2) F(\zeta) D_{\zeta'}(\zeta_1,\zeta_2)
G(\zeta'),
} \label{E6}\\
\lefteqn{ D_\zeta(\zeta_1,\zeta_2) D_{\zeta'}(\zeta,\zeta_k) F(\zeta')
 D_{\zeta''}(\zeta,\zeta_k)G(\zeta'') } \nonumber\\
& & = D_\zeta(\zeta_1,\zeta_k) D_{\zeta'}(\zeta,\zeta_2) F(\zeta')
D_{\zeta''}(\zeta_1,\zeta_k) G(\zeta'')
+ D_{\zeta'}(\zeta_2,\zeta_k) F(\zeta')
D_\zeta(\zeta_2,\zeta_k) D_{\zeta''}(\zeta_1,\zeta) G(\zeta''). \nonumber\\
& & \label{eq12-1}
\end{eqnarray}
For Eq. (\ref{mergeR}):
\begin{equation}
D_\zeta(\zeta_1,\zeta_2) \partial_\zeta D_z(\zeta,\zeta') F(z) =
\partial_{\zeta_1} D_z(\zeta',\zeta_1) D_\zeta (z,\zeta_2) F(\zeta)
+ \partial_{\zeta_2} D_z(\zeta',\zeta_2) D_\zeta (\zeta_1, z) F(\zeta),
\label{E10}
\end{equation}
\begin{equation}
\partial_\zeta \partial_{\zeta'} D_z(\zeta,\zeta') F(z)
= D_z(\zeta,\zeta') D_w(\zeta,\zeta') D_\zeta(z,w) F(\zeta), \label{E11}
\end{equation}
\begin{eqnarray}
& & D_\zeta(\zeta_1,\zeta_2) D_z(\zeta,\zeta') D_w(\zeta,\zeta') H(z,w)
\nonumber\\
&& ~~~~ =
D_\zeta(\zeta_2,\zeta') D_z(\zeta,\zeta_1) D_w(\zeta_2,\zeta') H(z,w)
\nonumber\\
&& ~~~~~~~~ +
D_\zeta(\zeta_1,\zeta') D_z(\zeta_1,\zeta') D_w(\zeta,\zeta_2) H(z,w).
\label{eq11-1}
\end{eqnarray}

The nature of these identities can be seen pictorially
in Figure 1 $\sim$ Figure 5.\footnote{
The notations for the original model are used in these figures,
but the argument
is also valid for the dual one, after changing
the variables $\zeta_1, \sigma, \zeta_2$
into $\xi_1, \xi_2, \xi_3$.}
For convenience of the reader, we
have included a Table,
indicating precisely where
these identities should be used.
Identities (\ref{E6}), (\ref{E10}) and (\ref{E11}) have already
appeared in Appendix E of \cite{SY} as Eqs. (E6), (E10) and (E11),
respectively.
On the other hand, identities (\ref{eq12-1}) and (\ref{eq11-1})
are not mentioned in \cite{SY} where simpler spin
configurations are dealt with.

Finally, we will explain a property of the combinatorial derivative
which is useful in the proof.
Obviously, the combinatorial derivative is symmetric under
interchange of its arguments:
\begin{equation}
D_z(\zeta_1,\zeta_2) F(z) =
\frac{F(\zeta_1)}{\zeta_1-\zeta_2} + \frac{F(\zeta_2)}{\zeta_2-\zeta_1}.
\end{equation}
For two or three successive combinatorial derivatives, we have
\begin{eqnarray}
&& D_z(\zeta_1,\zeta_2) D_{z'}(z,\zeta_3) F(z') \nonumber\\
&& ~~~~ = \frac{F(\zeta_1)}{(\zeta_1-\zeta_2)(\zeta_1-\zeta_3)}
+ \frac{F(\zeta_2)}{(\zeta_2-\zeta_3)(\zeta_2-\zeta_1)}
+ \frac{F(\zeta_3)}{(\zeta_3-\zeta_1)(\zeta_3-\zeta_2)}, \\
&& \nonumber\\
&& D_z(\zeta_1,\zeta_2) D_{z'}(z,\zeta_3) D_{z''}(z',\zeta_4) F(z'')
= D_z(\zeta_1,\zeta_2) D_{z'}(\zeta_3,\zeta_4) D_{z''}(z,z') F(z'')
\nonumber\\
&& ~~~~ =
\frac{F(\zeta_1)}{(\zeta_1-\zeta_2)(\zeta_1-\zeta_3)(\zeta_1-\zeta_4)} +
\frac{F(\zeta_2)}{(\zeta_2-\zeta_3)(\zeta_2-\zeta_4)(\zeta_2-\zeta_1)}
\nonumber\\
&& ~~~~~ +
\frac{F(\zeta_3)}{(\zeta_3-\zeta_4)(\zeta_3-\zeta_1)(\zeta_3-\zeta_2)} +
\frac{F(\zeta_4)}{(\zeta_4-\zeta_1)(\zeta_4-\zeta_2)(\zeta_4-\zeta_3)},
\end{eqnarray}
which are symmetric under interchange among $\zeta_1,\zeta_2,\zeta_3$ or
$\zeta_1,\zeta_2,\zeta_3,\zeta_4$, respectively.
Physically, these properties imply that the order of merging of three
or four domains is irrelevant.\\

The commutativity we have just proven
on the basis of the rules (\ref{rule1})$\sim$(\ref{rule5})
 means that the interaction terms in the
Hamiltonian in terms of the transformed fields
$\left( \frac{\delta}{\delta\hat{J}} \right)_I$ or
$\left(\frac{\delta}{\delta\hat{K}} \right)_I$ take the same form as those in
the bare Hamiltonian in terms of
$\left( \frac{\delta}{\delta J} \right)_I$ or
$\left(\frac{\delta}{\delta K} \right)_I$,
with a proviso that some of the terms present before the
scaling limit can vanish after the scaling limit, as occurring in
the $Y_{\alpha}$-derivative terms of our Hamiltonian.
 Since our proof is based on
a general structure of the mixing matrix, being  the exponential of a
derivation operator ${\cal R}$,
we strongly suspect that the commutativity
is a universal property of the matrix models
and their string field theories.
Furthermore, we might think a transformation such as (\ref{dualmix}) as a
special case of more general symmetry transformations of the string fields:
If a linear transformation of string fields
obeys the general rules (\ref{rule1}) $\sim$ (\ref{rule5}),
 the interaction terms of the string field theory
are invariant under the transformation, leaving aside the
question of generating kinetic terms
in attempting a background independent formulation of the string field theory
as discussed in
\cite{SY}.

\vspace{3cm}
\begin{center}
\begin{tabular}{|c|l|l|c|} \hline
Process   & Original model & Dual model & Necessary Identities \\
\hline\hline
Splitting & I=n (n $\geq$ 2)
          & I=n (n $\geq$ 3) & (\ref{E6}) \\
\cline{2-4}
          & I=n (n $\geq$ 3)
          & I=n (n $\geq$ 4) & (\ref{eq12-1}) \\
\hline
Merging   & (I,J)=(n,A),(n,B) (n $\geq$ 2)
          & (I,J)=(n,X) (n $\geq$ 3) & (\ref{E10})\\
\cline{2-4}
 				     & $\qquad\quad$ ---------
          & (I,J)=(2,1) & (\ref{E11})\\
\cline{2-4}
          & $\qquad\quad$ ---------
          & (I,J)=(n,1) (n $\geq$ 3) & (\ref{eq11-1})\\
\cline{2-4}
          & (I,J)=(1,1)
     					& (I,J)=(2,2) & (\ref{E11})\\
\cline{2-4}
          & (I,J)=(n,1) (n $\geq$ 2)
          & (I,J)=(n,2) (n $\geq$ 3) & (\ref{E11}) , (\ref{eq11-1})\\
\cline{2-4}
          & (I,J)=(n,m) (n,m $\geq$ 2)
          & (I,J)=(n,m) (n,m $\geq$ 3) & (\ref{eq11-1}) \\ \hline
\end{tabular}

\vspace{0.3cm}
\noindent
Table: Appearance of non-trivial identities.
\end{center}

\vspace{2cm}

\begin{center}
\begin{picture}(245,225)
\put(0,0){ \epsfxsize 275pt \epsfbox{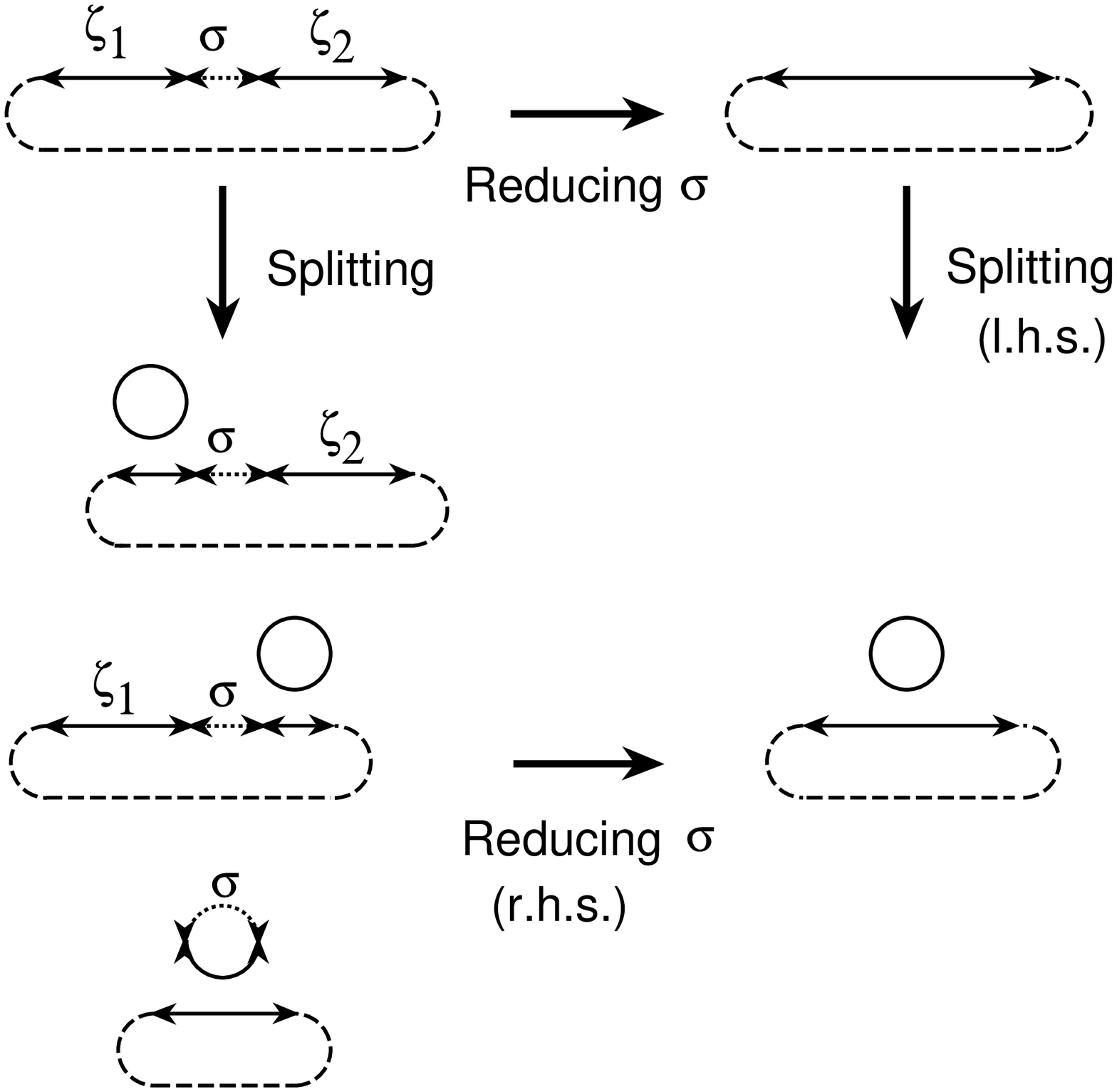}}
\end{picture}\\
\begin{quote}
\small
Figure 1: The pictorial explanation of the identity (\ref{E6}).
This is required for the following cases.
Suppose there are three successive domains with $\zeta_1, \sigma, \zeta_2$.
The identity shows that the reduction of the domain with $\sigma$
commutes with the particular splitting processes which occur in
the $\zeta_1$ domain or in the $\zeta_2$ domain.
\end{quote}
\end{center}

\vspace{2cm}

\begin{center}
\begin{picture}(233,199)
\put(0,0){ \epsfxsize 261pt \epsfbox{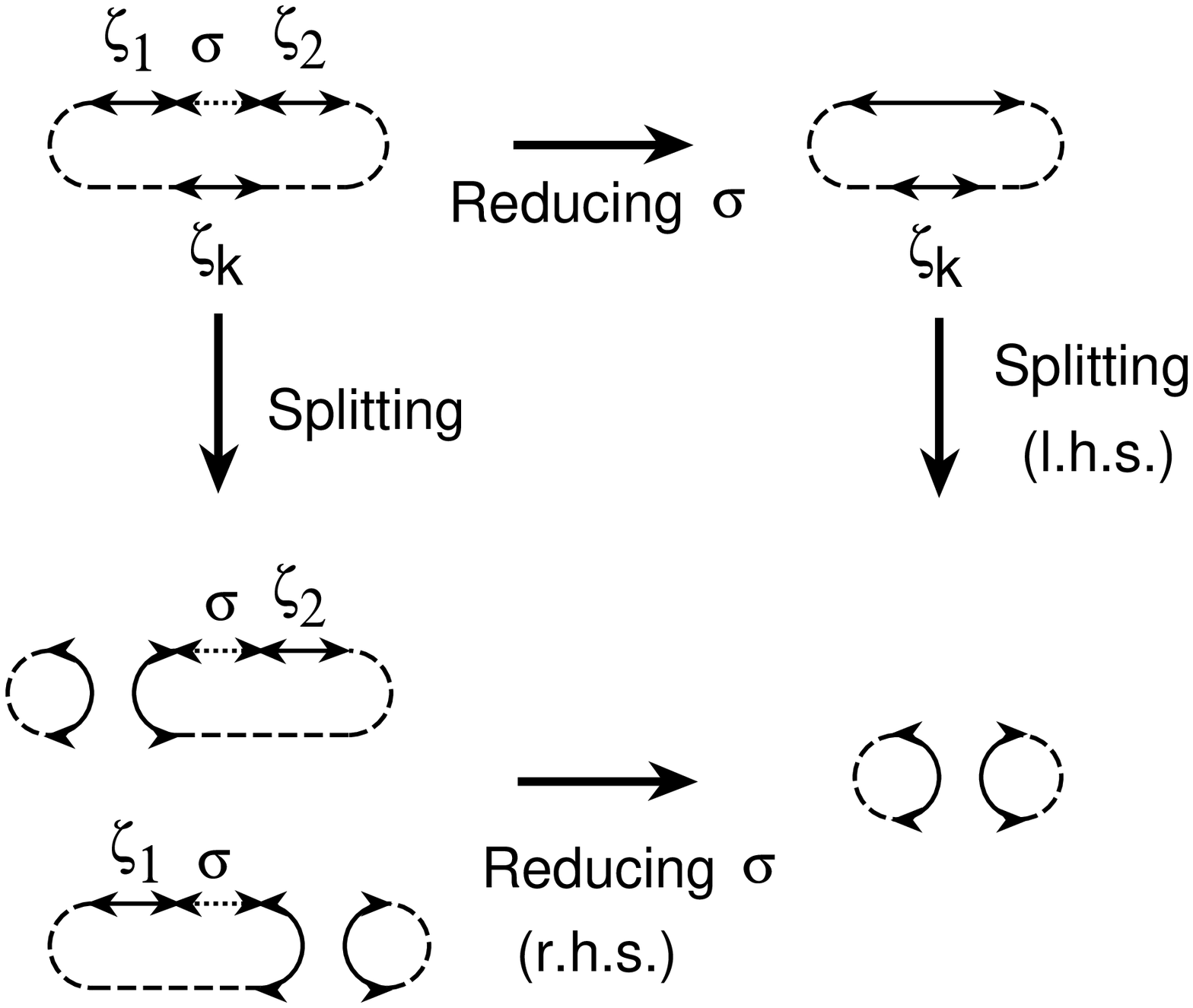}}
\end{picture}\\
\begin{quote}
\small
Figure 2: The pictorial explanation of the identity (\ref{eq12-1}).
Consider the three successive domains with $\zeta_1, \sigma, \zeta_2$ as
in Figure 1.
The identity shows that the reduction of the domain with $\sigma$
commutes with a splitting process between the domain with $\zeta_1$
and another domain with $\zeta_k$ and one between the domain with
$\zeta_2$ and the domain with $\zeta_k$.
\end{quote}
\end{center}

\vspace{1cm}

\begin{center}
\begin{picture}(248,174)
\put(0,0){ \epsfxsize 279pt \epsfbox{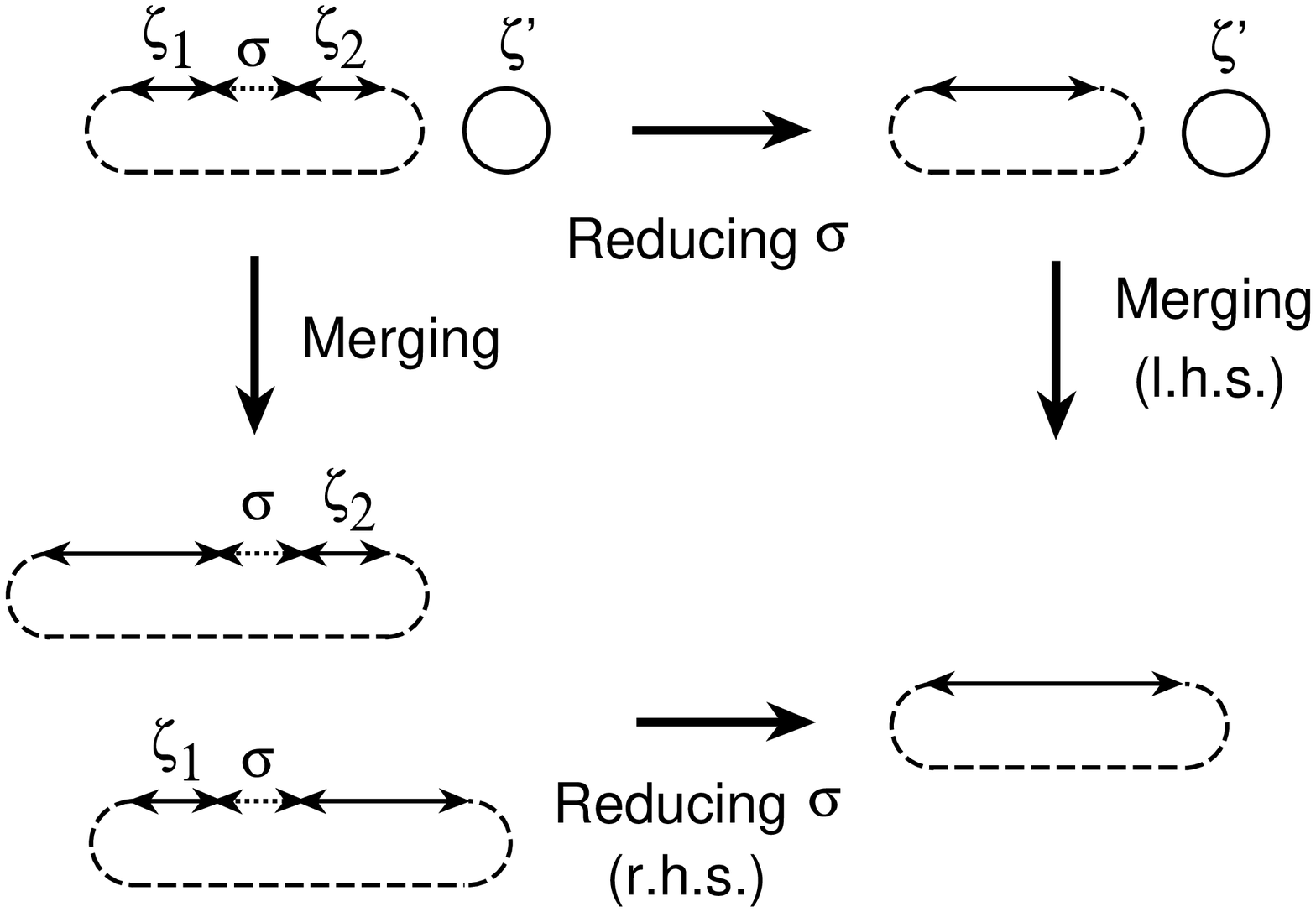}}
\end{picture}\\
\begin{quote}
\small
Figure 3: The pictorial explanation of the identity (\ref{E10}).
This identity is required when one of two strings
consists of only one domain with $\zeta'$.
(By ``only one domain", we mean $\Psi_X$ (not $\Psi_1$) for the dual
model.)
Consider the three successive domains with $\zeta_1, \sigma, \zeta_2$ in
the other string.
Merging processes that the string with one domain merges into the
$\zeta_1$ domain or the $\zeta_2$ domain commute with the reduction of
the $\sigma$ domain.
\end{quote}
\end{center}

\begin{center}
\begin{picture}(218,187)
\put(0,0){ \epsfxsize 244pt \epsfbox{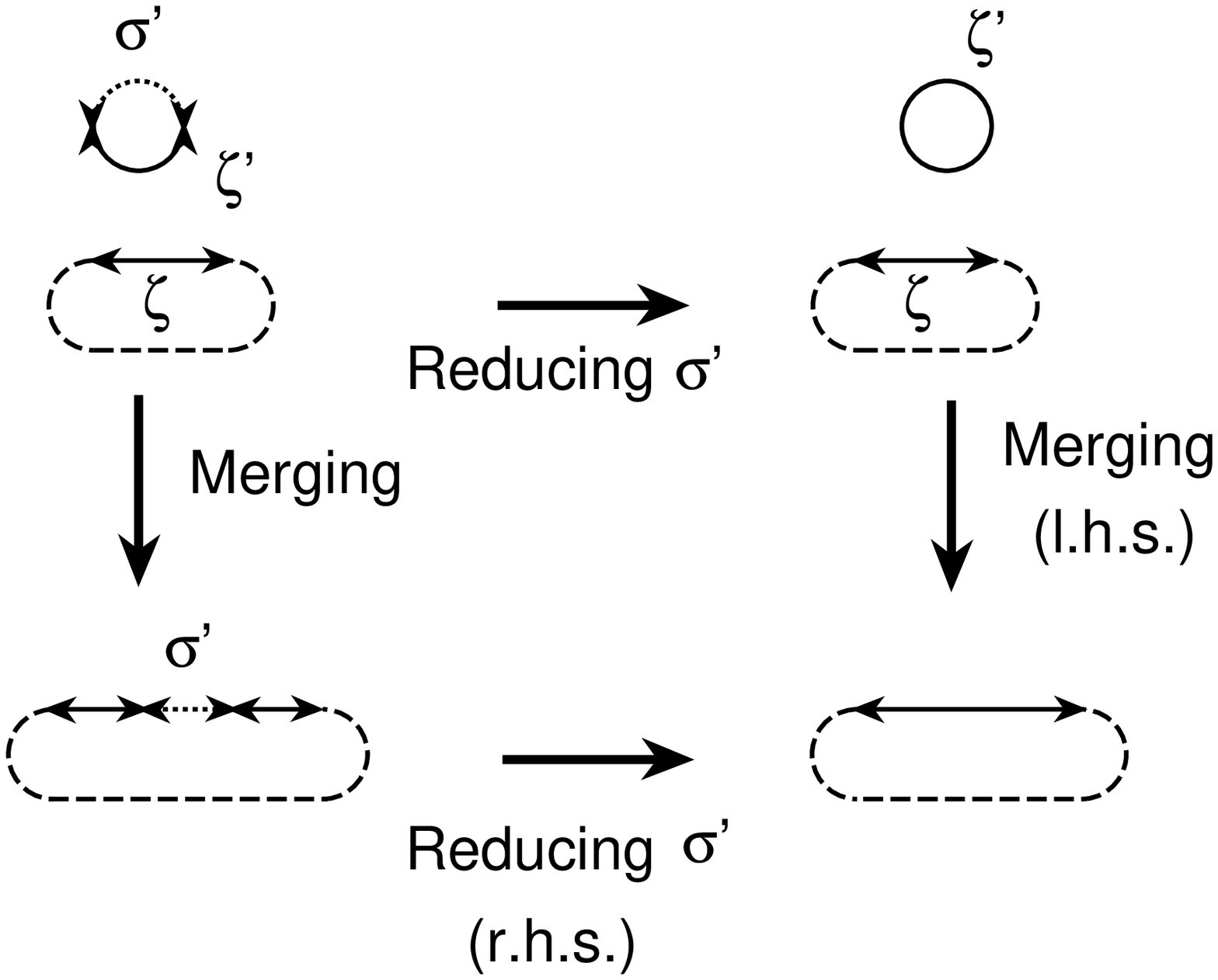}}
\end{picture}\\
\begin{quote}
\small
Figure 4: The pictorial explanation of the identity (\ref{E11}).
When one of two strings consists of two domains with $\zeta'$ and
$\sigma'$ and we consider the reduction of the $\sigma'$ domain, we need
this identity.
We consider a merging process involving the $\zeta'$ domain
and a $\zeta$ domain in the other string.
The identity shows that it commutes with the reduction of the domain
with $\sigma'$.
\end{quote}
\end{center}

\vspace{1cm}

\begin{center}
\begin{picture}(270,199)
\put(0,0){ \epsfxsize 303pt \epsfbox{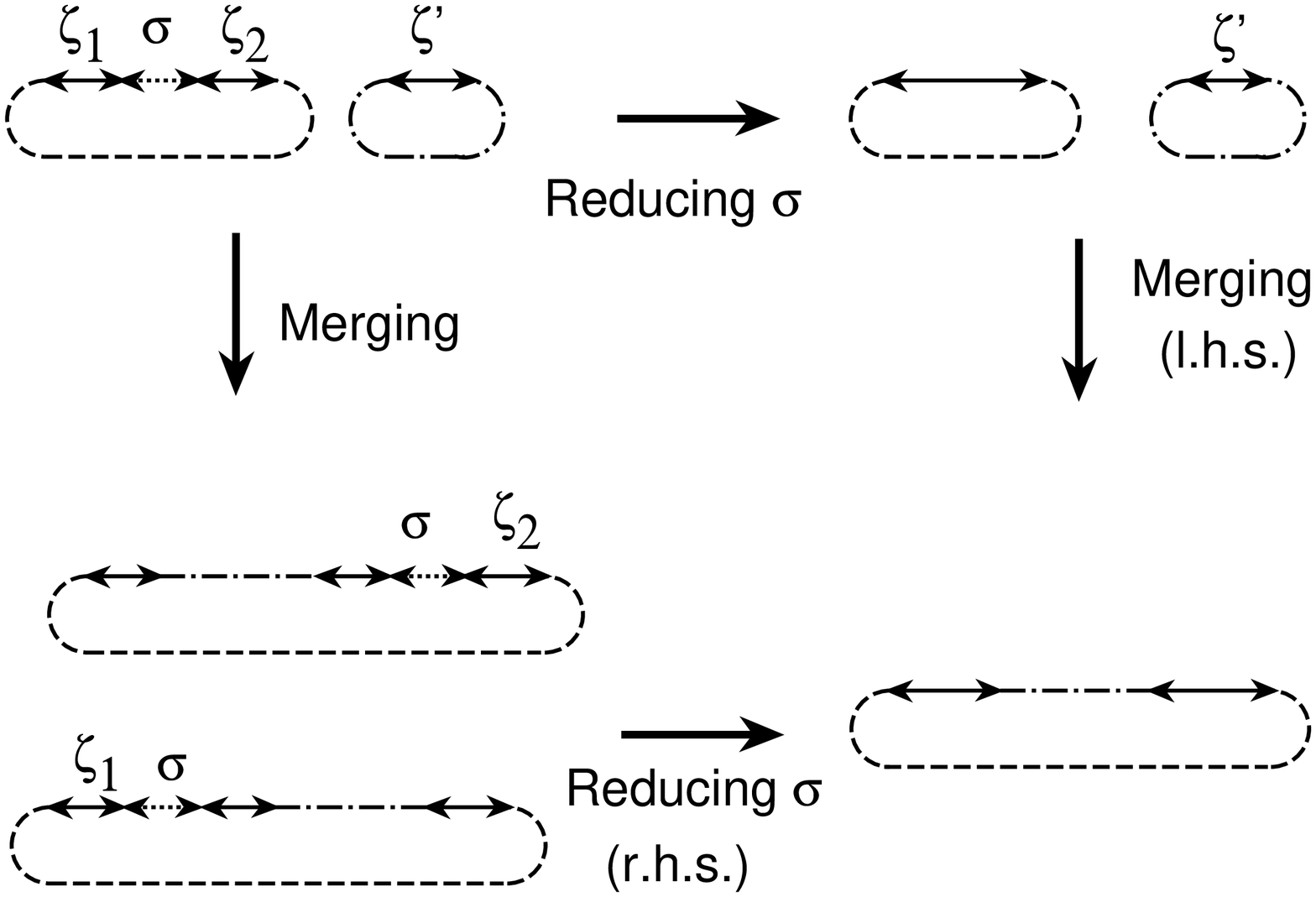}}
\end{picture}\\
\begin{quote}
\small
Figure 5: The pictorial explanation of the identity (\ref{eq11-1}).
We focus on the three successive domains with $\zeta_1, \sigma,
\zeta_2$ in a string. We consider a merging process between the
$\zeta_1$ domain and a $\zeta'$ domain in the other string and one
between the $\zeta_2$ domain and the $\zeta'$ domain.
This identity represents the commutativity of those
merging processes with the reduction of the $\sigma$ domain.
\end{quote}
\end{center}

\section{Discussion of Duality}
\setcounter{subsection}{0}
\setcounter{equation}{0}

  In section IV, we have derived the string field Hamiltonian in
the dual formalism.  The result shows that
the T-duality symmetry would be broken if the topological excitation of the
$Y$-loops at the quantum level could not be neglected.
Here, in order to check this, we shall calculate explicitly
the disk amplitudes with one handle in both the original and dual
theories.  We will find that the topological $Y$-loops indeed
contribute to the amplitude in the continuum theory. Also, a further
correspondence between the original and dual Hamiltonians will be discussed.

\subsection{Amplitudes for cylinder and disk with one handle
in the original theory}

   We first derive the amplitude for disk with one handle
in the original theory by taking
the continuum limit of the Schwinger-Dyson equation.
To do this, it is necessary to know the form of the cylinder amplitude
$ \left\langle\Phi_A(\zeta)\Phi_A(\zeta')\right\rangle_0$.
The Schwinger-Dyson equation
determining
this is
\begin{eqnarray}
0 & = & \left[3 \left\langle\Phi_A(\zeta)\right\rangle_0^2+
       2(-2\zeta+2g\zeta^2+\frac{c}{g}) \left\langle\Phi_A(\zeta)
\right\rangle_0-
       g \left\langle\frac{1}{N}\mbox{tr} A \right\rangle_0 \right.
\nonumber \\
 & & ~~\left.+1-g\zeta+(\zeta-g\zeta^2)(\zeta-g\zeta^2-\frac{c}{g})
+\frac{c^3}{g}\zeta\right]
            \left\langle\Phi_A(\zeta')\Phi_A(\zeta) \right\rangle_0
\nonumber \\
 & & +\left[-g \left\langle\Phi_A(\zeta) \right\rangle_0
+g(\zeta-g\zeta^2-\frac{c}{g})
-1+g\zeta\right]
      \left\langle\Phi_A(\zeta')\frac{1}{N}\mbox{tr} A \right\rangle_0
\nonumber \\
 & & +g(2-g\zeta)
      \left\langle\Phi_A(\zeta')\frac{1}{N}\mbox{tr} A^2 \right\rangle_0
            -g^2 \left\langle\Phi_A(\zeta')\frac{1}{N}\mbox{tr} A^3
\right\rangle_0 \nonumber \\
 & & \left. +\frac{1}{N^2}\partial_{\zeta'}D_z(\zeta,\zeta')
\right[ \left\langle\Phi_A(z) \right\rangle_0^2+
       \left\langle\Phi_A(\zeta) \right\rangle_0 \left\langle\Phi_A(z)
\right\rangle_0 \nonumber \\
 & &~~ \left.+(-2z+2gz^2+\frac{c}{g}) \left\langle\Phi_A(z)
\right\rangle_0\right],
                               \label{cylinderSD}
\end{eqnarray}
where $\Phi_A(\zeta)=\frac{1}{N}\mbox{tr}\frac{1}{\zeta-A}$, and
the symbol $ \left\langle\cdots \right\rangle_h$ represents a connected
amplitude with $h$ handles.
This is derived using a similar method as in the closed cubic equation for the
disk amplitude $W(\zeta)$ \cite{SY}.
Eq. (\ref{cylinderSD}) can be solved by using the following two facts.
One is that the factor
in front of
$ \left\langle\Phi_A(\zeta')\Phi_A(\zeta) \right\rangle_0$ becomes
\begin{equation}
a^{8/3}\frac{3cs^{8/3}}{40\cdot 2^{2/3}}(w(y)^2-T^{4/3})+O(a^3)
                     \label{coeff}
\end{equation}
in the continuum limit, where for the continuum limit in the original
theory, we use the notation in Appendix C of Ref. \cite{SY}.
For instance,
$$
g=g_*(1-a^2\frac{s^2}{20}T),~~~\zeta=\frac{1+3c}{2g_*}(1+ay),~~~
s=2+\sqrt{7},
$$
$$
  \left\langle\hat{\Phi}_A(\zeta) \right\rangle_0 =
              a^{4/3}\frac{c^{1/2}s^{4/3}}{\sqrt{10}\cdot 2^{4/3}}
\hat{w}(y),
$$
and $w(y)$ is the leading term of $\hat{w}(y)$:
$$
w(y)=(y+\sqrt{y^2-T})^{4/3}+(y-\sqrt{y^2-T})^{4/3}.
$$
The other is that the leading term of (\ref{coeff}) vanishes at $y=0,
\pm\sqrt{\frac{T}{2}}$.\footnote
{The Riemann surface of the disk $w(y)$
consists of a 3-sheeted plane. The first sheet has one cut in the region
$y<-\sqrt{T}$ on the real axis, the second one has the two cuts
$y<-\sqrt{T}$ and $y>\sqrt{T}$, and the third has the one cut
$y>\sqrt{T}$.
In this statement, it should be considered that $y=0$ and
$\sqrt{\frac{T}{2}}$
are points
on the first sheet but $y=-\sqrt{\frac{T}{2}}$ on the second sheet.
Here, we derive
the cylinder amplitude by assuming that it is regular at these
three points.}
Using the scaling of the cylinder $ \left\langle\Phi_A(\zeta')\Phi_A(\zeta)
\right\rangle_0
=O(\frac{1}{N^2}a^{-2})$ and the above properties, we can easily see
\begin{equation}
 \left\langle\Phi_A(\zeta)\frac{1}{N}\mbox{tr} A^n \right\rangle_0
=O(\frac{1}{N^2}a^{-2/3})
\end{equation}
for $n=1,2,3$. Then, note that Eq. (\ref{cylinderSD}) can be written as
\begin{eqnarray}
0 & = & a^{8/3}\frac{3cs^{8/3}}{40\cdot 2^{2/3}}(w(y)^2-T^{4/3})
          \left\langle\Phi_A(\zeta')\Phi_A(\zeta) \right\rangle_0 \nonumber \\
 & &-a^{4/3}\frac{c^2s^{4/3}}{2^{4/3}}w(y)
             \left\langle\Phi_A(\zeta')\frac{1}{N}\mbox{tr} A \right\rangle_0
+ ac^2syg_2(y')+c^2g_3(y') \nonumber \\
 & & +\frac{1}{N^2}a^{2/3}\frac{c^2s^{2/3}}{4\cdot2^{2/3}}\partial_{y'}\left[
     \frac{w(y)-w(y')}{y-y'}(2w(y)+w(y'))\right]
+O(\frac{1}{N^2}a),      \label{cylinderSD2}
\end{eqnarray}
where
\begin{eqnarray*}
g_2(y') & = & 2\sqrt{7} \left\langle\Phi_A(\zeta')\frac{1}{N}\mbox{tr} A
\right\rangle_0
   -\sqrt{10c} \left\langle\Phi_A(\zeta')\frac{1}{N}\mbox{tr} A^2
\right\rangle_0 \\
g_3(y') & = &  -4 \left\langle\Phi_A(\zeta')\frac{1}{N}\mbox{tr} A
\right\rangle_0
    +\sqrt{70c} \left\langle\Phi_A(\zeta')\frac{1}{N}\mbox{tr} A^2
\right\rangle_0
    -\frac{10}{3}c \left\langle\Phi_A(\zeta')\frac{1}{N}\mbox{tr} A^3
\right\rangle_0 \\
 &  &    -\frac{1}{N^2}\frac{2}{3}
        +a^{1/3}\frac{1}{N^2}\frac{s^{1/3}}{3\cdot 2^{1/3}}
               \partial_y\hat{w}(y).
\end{eqnarray*}
     From the consistency of Eq. (\ref{cylinderSD2}), the nontrivial
orders of $g_2$ and $g_3$
must be at most
\begin{equation}
g_2(y)=O(\frac{1}{N^2}a^{-1/3}), ~~~g_3(y)=O(\frac{1}{N^2}a^{2/3}).
\end{equation}
Now,  using the second of the above two facts,
solving  the equation (\ref{cylinderSD2}) to the leading order
is an easy task.
We arrive at
\begin{eqnarray}
 \left\langle\Phi_A(\zeta)\frac{1}{N}\mbox{tr} A \right\rangle_0 & = &
    \left. \frac{1}{N^2}a^{-2/3}2^{-4/3}s^{-2/3}\partial_y
     \right[y^{-1}w(y) \nonumber \\
 & & \left.+\frac{T^{1/3}}{4(y^2-T/2)y}(w(y)^2+T^{2/3}w(y)-2T^{4/3})
\right]
                   +O(\frac{1}{N^2}a^{-1/3}),
\label{6-6} \\
g_2(y)
 & = & -\frac{1}{N^2}a^{-1/3}2^{-8/3}s^{-1/3}\partial_y
\left[\frac{1}{y^2-T/2}
       (w(y)^2+T^{2/3}w(y)-2T^{4/3})\right] \nonumber \\
 &   & +O(\frac{1}{N^2}a^0), \\
g_3(y)
 & = & \left.-\frac{1}{N^2}a^{2/3}\frac{s^{2/3}}{12\cdot 2^{2/3}}
           \partial_y\right[y^{-1}(w(y)^2-2T^{4/3})  \nonumber \\
 &   & ~\left.+\frac{T}{4(y^2-T/2)y}(w(y)^2+T^{2/3}w(y)-2T^{4/3})\right]
+O(\frac{1}{N^2}a),         \label{eight}
\end{eqnarray}
\begin{eqnarray}
 \left\langle\Phi_A(\zeta)\Phi_A(\zeta') \right\rangle_0 & = &
\frac{1}{N^2}a^{-2}\frac{10c}{s^2}w(y,y')
+O(\frac{1}{N^2}a^{-5/3}),                    \nonumber \\
w(y,y')& = & \frac{4}{9}\frac{1}{f(y,y')g(y,y')}
  \frac{z_+(y)^{1/3}}{z_+(y)^{2/3}+z_-(y)^{2/3}+T^{1/3}} \nonumber \\
 & & \times
  \frac{z_+(y')^{1/3}}{z_+(y')^{2/3}+z_-(y')^{2/3}+T^{1/3}}
 \left[1+\frac{3T^{1/3}}{f(y,y')}+3\frac{z_+(y)z_+(y')}{g(y,y')}\right],
\end{eqnarray}
where
\begin{eqnarray*}
z_{\pm}(y) & = & y \pm \sqrt{y^2-T}, \\
f(y,y') & = & z_+(y)^{1/3}z_+(y')^{1/3}+z_-(y)^{1/3}z_-(y')^{1/3}
+T^{1/3}, \\
g(y,y') & = & z_+(y)^{2/3}+z_+(y')^{2/3}+z_+(y)^{1/3}z_+(y')^{1/3}.
\end{eqnarray*}

  Next, let us proceed to the amplitude for disk with one handle.
The Schwinger-Dyson equation we consider is
\begin{eqnarray}
0 & = & \left[3 \left\langle\Phi_A(\zeta) \right\rangle_0^2+
       2(-2\zeta+2g\zeta^2+\frac{c}{g}) \left\langle\Phi_A(\zeta)
\right\rangle_0-
       g \left\langle\frac{1}{N}\mbox{tr} A \right\rangle_0\right. \nonumber \\
 & & ~~\left.+1-g\zeta+(\zeta-g\zeta^2)(\zeta-g\zeta^2-\frac{c}{g})
               +\frac{c^3}{g}\zeta\right]
              \left\langle\Phi_A(\zeta) \right\rangle_1 \nonumber \\
 & & +[3 \left\langle\Phi_A(\zeta)
\right\rangle_0+\frac{c}{g}-2\zeta+2g\zeta^2]
 \left\langle\Phi_A(\zeta)^2 \right\rangle_0
                                                \nonumber \\
 & & -g \left\langle\Phi_A(\zeta)\frac{1}{N}\mbox{tr} A \right\rangle_0
      -g \left\langle\Phi_A(\zeta) \right\rangle_0
\left\langle\frac{1}{N}\mbox{tr} A \right\rangle_1
+\frac{1}{N^2}\frac{1}{2}\partial_{\zeta}^2 \left\langle\Phi_A(\zeta)
\right\rangle_0
\nonumber \\
 & & +[g(\zeta-g\zeta^2-\frac{c}{g})-c(1-g\zeta)+1-c]
\left\langle\frac{1}{N}\mbox{tr} A \right\rangle_1
          -g^2 \left\langle\frac{1}{N}\mbox{tr} A^3 \right\rangle_1.
\label{hdiskSD}
\end{eqnarray}
Note that the factor in front of $ \left\langle\Phi_A(\zeta)
\right\rangle_1$ is the same as the
first term in Eq. (\ref{cylinderSD}).
This equation can then be solved by the
same technique as before. The result is
\begin{equation}
 \left\langle\frac{1}{N}\mbox{tr} A \right\rangle_1=
  \frac{1}{N^2}a^{-2}\frac{\sqrt{10}}{24c^{1/2}s^2}T^{-1}
  +O(\frac{1}{N^2}a^{-4/3}),
\end{equation}
\begin{equation}
 \left\langle\frac{1}{N}\mbox{tr} A \right\rangle_1-\frac{c}{3}
\left\langle\frac{1}{N}\mbox{tr} A^3 \right\rangle_1
     =-\frac{1}{N^2}a^{-2/3}\frac{\sqrt{10}}{216\cdot 2^{7/3}c^{1/2}
s^{2/3}}
        T^{-1/3}+O(\frac{1}{N^2}a^0),
\end{equation}
\begin{equation}
 \left\langle\Phi_A(\zeta) \right\rangle_1 = \frac{1}{N^2}a^{-10/3}
    \frac{40\sqrt{10c}}{2^{2/3}s^{10/3}}w_{(1)}(y)
     +O(\frac{1}{N^2}a^{-3}),
\end{equation}
where
\begin{eqnarray}
w_{(1)}(y) & = & \frac{1}{216T}
\frac{1}{(z_+(y)^{2/3}+z_-(y)^{2/3}+T^{1/3})^5}
                         [12y^2+77T       \nonumber \\
  &   & ~~+15T^{1/3}(z_+(y)^{4/3}+z_-(y)^{4/3})+
                 50T^{2/3}(z_+(y)^{2/3}+z_-(y)^{2/3})].
\end{eqnarray}

\subsection{Amplitudes for cylinder and disk with one handle
in the dual theory}

  On the other hand, in the dual theory, the Schwinger-Dyson equation
for the cylinder is
\begin{eqnarray}
0 & = & [-3\hat{g} \left\langle\Psi_X(\xi) \right\rangle_0^2+2(2c(1+c)
-\hat{g}^2\xi^2+(1-5c)\hat{g}\xi)
         \left\langle\Psi_X(\xi) \right\rangle_0    \nonumber \\
  &   & ~~+2\hat{g}^2 \left\langle\frac{1}{N}\mbox{tr} X \right\rangle_0-
          \hat{g} (1-3c)-2c\xi (1+c-2\hat{g}\xi)(1-c-\hat{g}\xi)]
\nonumber \\
  &   & ~\times  \left\langle\Psi_X(\xi')\Psi_X(\xi) \right\rangle_0
\nonumber \\
  &   & +\hat{g} [2\hat{g} \left\langle\Psi_X(\xi) \right\rangle_0-2c(3-c)
+4c\hat{g}\xi]
           \left\langle\Psi_X(\xi')\frac{1}{N}\mbox{tr} X \right\rangle_0
\nonumber \\
  &   & +4c\hat{g}^2 \left\langle\Psi_X(\xi')\frac{1}{N}\mbox{tr} X^2
\right\rangle_0
+\hat{g}^2\frac{1}{N^2}\partial_{\xi'}
          \left\langle\frac{1}{N}\mbox{tr} \left(\frac{1}{\xi'-X}Y
                               \frac{1}{\xi-X}Y\right) \right\rangle_0
\nonumber \\
  &   & +\frac{1}{N^2}\partial_{\xi'}D_z(\xi,\xi')[-\hat{g}
\left\langle\Psi_X(z) \right\rangle_0^2
                 -\hat{g} \left\langle\Psi_X(\xi) \right\rangle_0
\left\langle\Psi_X(z) \right\rangle_0   \nonumber \\
   &    & ~~+ (2c(1+c)-\hat{g}^2z^2+(1-5c)\hat{g} z)
\left\langle\Psi_X(z) \right\rangle_0],
\label{dualcylinderSD}
\end{eqnarray}
which is also derived in a similar way as obtaining Eq. (\ref{diskSD1}).
Note that the term
$$
\hat{g}^2\frac{1}{N^2}\partial_{\xi'}
          \left\langle\frac{1}{N}\mbox{tr}\left(\frac{1}{\xi'-X}Y
\frac{1}{\xi-X}Y\right) \right\rangle_0
$$
comes from the following Schwinger-Dyson equation
\begin{equation}
0=\int d^{N^2}Xd^{N^2}Y\sum_{\alpha}\frac{\partial}{\partial X_{\alpha}}\left[
    \mbox{tr}\frac{1}{\xi'-X}\mbox{tr}\left(\frac{1}{\xi-X}Yt^{\alpha}Y\right)
{\rm e}^{-S_D}\right],
\end{equation}
which reflects the fact that the dual theory does not distinguish
overall spin directions. When both domains of $\xi$
and $\xi'$ are of
spin-up
(spin-down), such a term does not appear in the original theory.
As we will see, this survives in the scaling limit
and leads to the difference
between the cylinder amplitudes.

Since the scaling limit of the factor in front of
$ \left\langle\Psi_X(\xi')\Psi_X(\xi) \right\rangle_0$ becomes
$$ a^{8/3}\frac{3c^{5/2}\hat{s}^{8/3}}{2^{4/3}\sqrt{5}}
   (\hat{T}^{4/3}-v(y)^2)+O(a^3),  $$
Eq. (\ref{dualcylinderSD}) can be solved by the same argument as before.
We have the following result:
\begin{eqnarray}
 \left\langle\Psi_X(\xi)\frac{1}{N}\mbox{tr} X \right\rangle_0 & = &
\frac{1}{N^2}a^{-2/3}2^{-2/3}\hat{s}^{-2/3}\partial_y[y^{-1}v(y) \nonumber \\
 & & ~~+\frac{\hat{T}^{1/3}}{4(y^2-\hat{T}/2)y}(v(y)^2+\hat{T}^{2/3}v(y)
-2\hat{T}^{4/3})]
        \nonumber \\
 & & +O(\frac{1}{N^2}a^{-1/3}),
\end{eqnarray}
\begin{eqnarray}
\lefteqn{ -108c \left\langle\Psi_X(\xi)\frac{1}{N}\mbox{tr} X \right\rangle_0
     +324\sqrt{5c} \left\langle\Psi_X(\xi)\frac{1}{N}\mbox{tr} X^2
\right\rangle_0
     +\frac{4}{N^2} }  \nonumber \\
 & = & \frac{1}{N^2}a^{2/3}\hat{s}^{2/3}2^{-1/3}\partial_y\left[
       v(y)^2-2\hat{T}^{4/3}
     +\frac{\hat{T}}{4(y^2-\hat{T}/2)}(v(y)^2+\hat{T}^{2/3}v(y)
-2\hat{T}^{4/3})\right]
                                \nonumber \\
 &  & +O(\frac{1}{N^2}a),
\end{eqnarray}
\begin{equation}
 \left\langle\Psi_X(\xi')\Psi_X(\xi) \right\rangle_0=
  \frac{1}{N^2}a^{-2}\hat{s}^{-2}5c[2w_D(y,y')+u_D(y,y')]
   +O(\frac{1}{N^2}a^{-5/3}),
\end{equation}
where $w_D(y,y')$ is the result of the replacement $T\rightarrow\hat{T}$ in
$w(y,y')$,
and $u_D(y, y')$ represents the difference of the functional forms
between the cylinder amplitudes:
\begin{eqnarray}
u_D(y,y') & = & -\frac{4}{9}\frac{y}{v(y)+\hat{T}^{2/3}}
               \frac{y'}{v(y')+\hat{T}^{2/3}}\frac{1}{(y^2-y'^2)^2} \nonumber
\\
 & & \times [\hat{z}_{+}(y)^{8/3}+\hat{z}_{-}(y)^{8/3}+\hat{z}_{+}(y')^{8/3}
+\hat{z}_{-}(y')^{8/3}
\nonumber \\
 & & ~-6(\hat{z}_{+}(y)^{4/3}+\hat{z}_{-}(y)^{4/3})(\hat{z}_{+}(y')^{4/3}
+\hat{z}_{-}(y')^{4/3})
\nonumber \\
 & & ~-5\hat{T}^{2/3}(\hat{z}_{+}(y)^{4/3}+\hat{z}_{-}(y)^{4/3}
+\hat{z}_{+}(y')^{4/3}
+\hat{z}_{-}(y')^{4/3})
                          \nonumber \\
 & & ~+6\hat{T}^{2/3}(\hat{z}_{+}(y)^{2/3}
+\hat{z}_{-}(y)^{2/3})(\hat{z}_{+}(y')^{2/3}
+\hat{z}_{-}(y')^{2/3})
                           \nonumber \\
 & & ~+4(2y^2-\hat{T})(\hat{z}_{+}(y')^{2/3}
+\hat{z}_{-}(y')^{2/3}) \nonumber \\
 & & ~+4(\hat{z}_+(y)^{2/3}+\hat{z}_-(y)^{2/3})(2y'^2-\hat{T})] \nonumber \\
 & & +\frac{1}{(y+y')^2},
\end{eqnarray}
where $\hat{z}_{\pm}(y)$ is obtained
by replacing $T$ to $\hat{T}$ in $z_{\pm}(y)$.
As will be mentioned later, the difference is interpreted
as coming from the odd $Y$-loop excitations
on a nontrivial homology cycle of the cylinder.

  Also, for the disk with one handle, we can solve the Schwinger-Dyson
equation:
\begin{eqnarray}
0 & = & [-3\hat{g} \left\langle\Psi_X(\xi) \right\rangle_0^2+2(2c(1+c)
-\hat{g}^2\xi^2+(1-5c)\hat{g}\xi)
         \left\langle\Psi_X(\xi) \right\rangle_0    \nonumber \\
  &   & ~~+2\hat{g}^2 \left\langle\frac{1}{N}\mbox{tr} X \right\rangle_0-
          \hat{g} (1-3c)-2c\xi (1+c-2\hat{g}\xi)(1-c-\hat{g}\xi)]
            \left\langle\Psi_X(\xi) \right\rangle_1 \nonumber \\
  &   & +[-3\hat{g} \left\langle\Psi_X(\xi) \right\rangle_0+2c(1+c)
-\hat{g}^2\xi^2+(1-5c)\hat{g}\xi]
            \left\langle\Psi_X(\xi)^2 \right\rangle_0              \nonumber \\
  &   & +2\hat{g}^2 \left\langle\Psi_X(\xi)\frac{1}{N}
\mbox{tr} X \right\rangle_0
          -\hat{g}^2 \left\langle(\frac{1}{N}\mbox{tr}\frac{1}{\xi-X}Y)^2
\right\rangle_0
-\hat{g}\frac{1}{N^2}\frac{1}{2}\partial_{\xi}^2 \left\langle\Psi_X(\xi)
\right\rangle_0
                                         \nonumber \\
  &   & +[2\hat{g}^2 \left\langle\Psi_X(\xi) \right\rangle_0
+2c\hat{g}(-3+c+2\hat{g}\xi)]
                  \left\langle\frac{1}{N}\mbox{tr} X \right\rangle_1
            +4c\hat{g}^2 \left\langle\frac{1}{N}\mbox{tr} X^2 \right\rangle_1.
    \label{dualhdiskSD}
\end{eqnarray}
Note the appearance of the term
$-\hat{g}^2 \left\langle(\frac{1}{N}\mbox{tr}\frac{1}{\xi-X}Y)^2
\right\rangle_0 $,
which does not allow
the interpretation in terms of  dual spins.

 The solution is as follows:
$$
 \left\langle\frac{1}{N}\mbox{tr} X \right\rangle_1=
   \frac{1}{N^2}a^{-2}\frac{\sqrt{5}}{24c^{1/2}\hat{s}^2}\hat{T}^{-1}
    +O(\frac{1}{N^2}a^{-4/3}),
$$
\begin{eqnarray}
\lefteqn{-108\sqrt{5}c^{3/2} \left\langle\frac{1}{N}\mbox{tr} X \right\rangle_1
      +20c \left\langle\frac{1}{N}\mbox{tr} X^2 \right\rangle_1 }  \nonumber \\
 & = & -\frac{1}{N^2}a^{-2/3}\hat{s}^{-2/3}\frac{25}{36\cdot 2^{2/3}}
       \hat{T}^{-1/3} +O(\frac{1}{N^2}a^0),
\end{eqnarray}
\begin{equation}
 \left\langle\Psi_X(\xi) \right\rangle_1 =
\frac{1}{N^2}a^{-10/3}\hat{s}^{-10/3}
   5^{3/2}\cdot 2^{2/3}c^{1/2}[2w_{(1)D}(y)+u_{(1)}(y)]
                +O(\frac{1}{N^2}a^{-3}),
\end{equation}
where $w_{(1)D}(y)$ is obtained by replacing $T$ to $\hat{T}$ in $w_{(1)}(y)$,
and the difference from the original theory is
\begin{equation}
u_{(1)}(y) =-\frac{1}{54\hat{T}^{1/3}}
  \frac{5(\hat{z}_{+}(y)^{2/3}+\hat{z}_{-}(y)^{2/3})+11\hat{T}^{1/3}}
  {(\hat{z}_{+}(y)^{2/3}+\hat{z}_{-}(y)^{2/3}+\hat{T}^{1/3})^5}.
\end{equation}

\subsection{Comparison of the amplitudes}

  We have obtained the amplitudes for cylinder and disk with one handle
both in
the original and dual theories. The result of the disk
with one handle
illustrates that the T-duality symmetry does not hold
in the higher genus
amplitudes as is expected from the discussion in section II.
On the other hand, the difference
between the cylinder amplitudes seems to be puzzling.
This is resolved as
follows.
Since  the dual theory can not discriminate overall dual spin
directions, it is necessary to take account of
the cylinder amplitude
$ \left\langle\Phi_B(\zeta')\Phi_A(\zeta) \right\rangle_0 $, in comparing
correlators with the original theory.
The Schwinger-Dyson equation needed to obtain this is
\begin{eqnarray}
0 & = & \left[3 \left\langle\Phi_A(\zeta) \right\rangle_0^2+
       2(-2\zeta+2g\zeta^2+\frac{c}{g}) \left\langle\Phi_A(\zeta)
\right\rangle_0-
       g \left\langle\frac{1}{N}\mbox{tr} A \right\rangle_0\right.  \nonumber
\\
 & & ~~\left.+1-g\zeta+(\zeta-g\zeta^2)(\zeta-g\zeta^2-\frac{c}{g})
          +\frac{c^3}{g}\zeta\right]
            \left\langle\Phi_B(\zeta')\Phi_A(\zeta) \right\rangle_0 \nonumber
\\
 & & +[-g \left\langle\Phi_A(\zeta) \right\rangle_0+g(\zeta-g\zeta^2
-\frac{c}{g})-1+g\zeta]
      \left\langle\Phi_B(\zeta')\frac{1}{N}\mbox{tr} A \right\rangle_0
\nonumber \\
 & & +g(2-g\zeta)
      \left\langle\Phi_B(\zeta')\frac{1}{N}\mbox{tr} A^2 \right\rangle_0
       -g^2 \left\langle\Phi_B(\zeta')\frac{1}{N}\mbox{tr} A^3 \right\rangle_0
\nonumber \\
 & & -\frac{c^2}{g}\frac{1}{N^2}\partial_{\zeta'}
       \left\langle\frac{1}{N}\mbox{tr}
\left(\frac{1}{\zeta-A}\frac{1}{\zeta'-B}\right) \right\rangle_0.
 \label{cylinderABSD}
\end{eqnarray}
The way to solve this is the same as before. We obtain
\begin{eqnarray}
 \left\langle\Phi_B(\zeta)\frac{1}{N}\mbox{tr} A \right\rangle_0 & = &
     \frac{1}{N^2}a^{-2/3}2^{-4/3}s^{-2/3}\partial_y[y^{-1}w(y) \nonumber \\
 & & +\frac{T^{1/3}}{4(y^2-T/2)y}(w(y)^2+T^{2/3}w(y)-2T^{4/3})] \nonumber \\
 & & +O(\frac{1}{N^2}a^{-1/3}),     \label{twentysix}
\end{eqnarray}
\begin{eqnarray}
\lefteqn{2\sqrt{7} \left\langle\Phi_B(\zeta)\frac{1}{N}\mbox{tr} A
\right\rangle_0
   -\sqrt{10c}  \left\langle\Phi_B(\zeta)\frac{1}{N}\mbox{tr} A^2
\right\rangle_0}  \nonumber\\
 & = & -\frac{1}{N^2}a^{-1/3}2^{-8/3}s^{-1/3}\partial_y
\left[\frac{1}{y^2-T/2}
       (w(y)^2+T^{2/3}w(y)-2T^{4/3})\right] \nonumber\\
 &   & +O(\frac{1}{N^2}a^0),
\end{eqnarray}
\begin{eqnarray}
\lefteqn{-4 \left\langle\Phi_B(\zeta)\frac{1}{N}\mbox{tr} A \right\rangle_0
    +\sqrt{70c} \left\langle\Phi_B(\zeta)\frac{1}{N}\mbox{tr} A^2
\right\rangle_0
    -\frac{10}{3}c \left\langle\Phi_B(\zeta)\frac{1}{N}\mbox{tr} A^3
\right\rangle_0} \nonumber \\
 &  &    -\frac{1}{N^2}\frac{1}{3}
        -a^{1/3}\frac{1}{N^2}\frac{s^{1/3}}{3\cdot 2^{4/3}}
\partial_y\hat{w}(y)
     \nonumber \\
 & = & -\frac{1}{N^2}a^{2/3}\frac{s^{2/3}}{12\cdot 2^{2/3}}\partial_y\left[
      y^{-1}(w(y)^2-2T^{4/3})
       +\frac{T}{4(y^2-T/2)y}(w(y)^2+T^{2/3}w(y)-2T^{4/3})\right]  \nonumber \\
 &  & +O(\frac{1}{N^2}a),
\label{6-28}
\end{eqnarray}
and at the leading order the Schwinger-Dyson equation for
$ \left\langle\Phi_B(\zeta')\Phi_A(\zeta) \right\rangle_0
+ \left\langle\Phi_A(\zeta')\Phi_A(\zeta) \right\rangle_0$
becomes the same form as that for $ \left\langle\Psi_X(\zeta')\Psi_X(\zeta)
\right\rangle_0$.
Thus, we have
\begin{equation}
 \left\langle\Phi_B(\zeta')\Phi_A(\zeta) \right\rangle_0
+ \left\langle\Phi_A(\zeta')\Phi_A(\zeta) \right\rangle_0
=\frac{1}{N^2}a^{-2}\frac{10c}{s^2}[2w(y,y')+u(y,y')] +
O(\frac{1}{N^2}a^{-5/3}),
\end{equation}
where $u(y,y')$ is obtained by replacing $\hat{T}$ to $T$ in $u_D(y,y')$.

It can now be seen that the mixed cylinder amplitude
$$ \left\langle(\Phi_A(\zeta)+\Phi_B(\zeta))(\Phi_A(\zeta')+\Phi_B(\zeta'))
\right\rangle_0$$
of  the original
theory has the same functional form as $ \left\langle\Psi_X(\xi)\Psi_X(\xi')
\right\rangle_0$
of the dual theory, which means the duality symmetry in the cylinder level.
On the
cylinder in the dual theory,  $Y$-loops along a
nontrivial homology
cycle can exit.  A natural interpretation is that the amplitude
 with the even number
$Y$-loops
corresponds to $ \left\langle\Phi_A\Phi_A \right\rangle_0
+ \left\langle\Phi_B\Phi_B \right\rangle_0$ in the original
theory, and that the amplitude with the odd number ones to the cross term
$ \left\langle\Phi_A\Phi_B \right\rangle_0+ \left\langle\Phi_B\Phi_A
\right\rangle_0$. This is consistent with the
picture of the dual spins.

We can thus conclude that, to the leading order
of the universal parts of the disk and cylinder amplitudes,
the following identification between
the original and dual theories holds:
$$T\Leftrightarrow \hat{T} ,$$
$$\frac{s}{\sqrt{2}}\Leftrightarrow \hat{s} ,$$
$$\frac{1}{\sqrt{2}}(\Phi_A(\zeta)+\Phi_B(\zeta))
\Leftrightarrow\Psi_X(\xi), $$
$$\frac{1}{\sqrt{2}}\mbox{tr} (A+B)\Leftrightarrow \mbox{tr} X.$$
This can be checked for all the disk and cylinder amplitudes
we have obtained.

\subsection{Comparison of the string field Hamiltonians}

  Until now, after a long analysis of the Schwinger-Dyson
equations,  we have shown that the T-duality
symmetry holds in some disk and cylinder amplitudes without handles,
and that it is
violated in the higher genus amplitude. Here, we make sure that
in general the
T-duality is a symmetry  in the planar approximation, by comparing
the forms
of the string field Hamiltonians.

  We start with the Hamiltonian of the original theory, where the
numerical
constant $cs^{-1}$ in the kinetic term is dropped,
since it can be absorbed
into the finite renormalization of the fictitious time
by rescaling as
$$
\tilde{J}_I\rightarrow \frac{1}{cs^{-1}}\tilde{J}_I,~~~
  \frac{\delta}{\delta\tilde{J}_I}
\rightarrow cs^{-1}\frac{\delta}{\delta\tilde{J}_I}, ~~~
   g_{\mbox{st}}\rightarrow cs^{-1}g_{\mbox{st}}.
$$
We introduce the string fields which have a definite parity
with respect to the reversal $(A\leftrightarrow B)$ of spin directions:
\begin{eqnarray}
\tilde{J}^{(\pm)}_X(y) & = & \tilde{J}_A(y)\pm\tilde{J}_B(y), \nonumber \\
\tilde{J}^{(\pm)}_{2n}(y_1,y_2,\cdots, y_{2n}) & = & \frac{1}{2}(
   \tilde{J}_{n}(y_1,y_2,\cdots, y_{2n})
\pm\tilde{J}_{n}(y_2,\cdots, y_{2n},y_1)),
\nonumber \\
\frac{\delta}{\delta\tilde{J}^{(\pm)}_X(y)} & = & \frac{1}{2}\left(
   \frac{\delta}{\delta\tilde{J}_A(y)}\pm\frac{\delta}{\delta\tilde{J}_B(y)}
\right), \nonumber \\
\frac{\delta}{\delta\tilde{J}^{(\pm)}_{2n}(y_1,y_2,\cdots, y_{2n})} & = &
   \frac{1}{2}\left[\frac{\delta}{\delta\tilde{J}_{n}(y_1,y_2,\cdots, y_{2n})}
\pm
      \frac{\delta}{\delta\tilde{J}_{n}(y_2,\cdots, y_{2n},y_1)}\right]
\nonumber \\
  &  &  ~~~~~~~~~(n=1,2,3,\cdots).
\end{eqnarray}
Recalling that the variable $\tilde{J}_n(y_1,\cdots,y_{2n})$ in the original
theory have the cyclic symmetry with respect to the
permutations of $n$-pairs $(y_1,y_2),(y_3,y_4),
\cdots,
(y_{2n-1},y_{2n})$, we see that the new variables $\tilde{J}^{(\pm)}$'s and
the
derivatives with respect to them
have a definite symmetric or anti-symmetric property
under the cyclic permutations of the $2n$ variables $y_1,y_2,\cdots,y_{2n}$.
Note that
the symmetric property of $\tilde{J}^{(+)}$'s is identical with that of the
string fields in the dual theory.

  Then the functional derivatives of the new variables become
\begin{eqnarray}
\frac{\delta\tilde{J}^{(P)}_X(y)}{\delta\tilde{J}^{(Q)}_X(y')} & = &
2\pi i\delta_{P,Q}
\delta(y-y'),               \nonumber \\
\frac{\delta\tilde{J}^{(+)}_{2n}(y_1,\cdots,y_{2n})}{\delta\tilde{J}^{(+)}_{2n}
    (y'_1,\cdots,y'_{2n})} & = & \frac{1}{2n}(2\pi i)^{2n}\sum_c
    \delta(y_1-y'_{c(1)})\dots\delta(y_{2n}-y'_{c(2n)}), \nonumber \\
\frac{\delta\tilde{J}^{(-)}_{2n}(y_1,\cdots,y_{2n})}{\delta\tilde{J}^{(-)}_{2n}
    (y'_1,\cdots,y'_{2n})} & = & \frac{1}{2n}(2\pi i)^{2n}
\sum_c(-1)^{|c|}
    \delta(y_1-y'_{c(1)})\dots\delta(y_{2n}-y'_{c(2n)}),
\end{eqnarray}
and the others vanish, where $P$ and $Q$ take $+$ or $-$, and
$c$ represents the cyclic
permutations of $1,\cdots, 2n$.
The notation $(-1)^{|c|}$ denotes its sign, namely,  $(+1)$ for even cyclic
permutations or $(-1)$ for odd cyclic permutations.

  By using these variables, the Hamiltonian is written in the following
form:
\begin{eqnarray}
{\cal H} & = & {\cal H}^{(+)}+{\cal H}^{(-)}, \nonumber \\
{\cal H}^{(+)} & = & -\tilde{J}^{(+)}
\cdot\left(F\frac{\delta}{\delta\tilde{J}^{(+)}}\right)
      -\tilde{J}^{(+)}\cdot
  \left(\frac{\delta}{\delta\tilde{J}^{(+)}}
\vee\frac{\delta}{\delta\tilde{J}^{(+)}}\right)-
   \frac{1}{2}g_{\mbox{st}}^{2}\tilde{J}^{(+)}\cdot\left(\tilde{J}^{(+)}\cdot
     \left(\wedge\frac{\delta}{\delta\tilde{J}^{(+)}}\right)\right),
\nonumber \\
{\cal H}^{(-)} & = &  -\tilde{J}^{(-)}\cdot
\left(F\frac{\delta}{\delta\tilde{J}^{(-)}}\right)
\nonumber \\
 &  &-\tilde{J}^{(+)}\cdot
  \left(\frac{\delta}{\delta\tilde{J}^{(-)}}
\vee\frac{\delta}{\delta\tilde{J}^{(-)}}\right)
  -\tilde{J}^{(-)}\cdot
  \left(\frac{\delta}{\delta\tilde{J}^{(+)}}
\vee\frac{\delta}{\delta\tilde{J}^{(-)}}
    +((+)\leftrightarrow (-))\right) \nonumber \\
 &  &-\frac{1}{2}g_{\mbox{st}}^{2}\tilde{J}^{(-)}
\cdot\left(\tilde{J}^{(-)}\cdot
     \left(\wedge\frac{\delta}{\delta\tilde{J}^{(+)}}\right)\right)
\nonumber \\
 &  &-\frac{1}{2}g_{\mbox{st}}^{2}(\tilde{J}^{(+)}\cdot\tilde{J}^{(-)}
+((+)\leftrightarrow (-)))\cdot
     \left(\wedge\frac{\delta}{\delta\tilde{J}^{(-)}}\right),
\end{eqnarray}
where only even
number of the $(-)$-fields appear because of
the $\mbox{\boldmath $Z$}_2$-symmetry of the
original Hamiltonian.

  The above expression shows that if the coupling is rescaled as
$g_{\mbox{st}}^2\rightarrow 2g_{\mbox{st}}^2$, the form of ${\cal H}^{(+)}$
becomes completely
identical
with that of the dual Hamiltonian ${\cal H}_D$ (\ref{continuumH}) after
setting
the fields with odd domains to zero. There still remain the portions
that do not match
between the original and dual theories. They are ${\cal H}^{(-)}$
in the original theory and the terms containing the
fields with odd domains in the dual theory. These contributions give no
effects in the tree amplitudes
when the initial state does not contain the $(-)$-fields, in
the original theory, because the vacuum expectation
value
of a single $(-)$-field vanishes due to the $\mbox{\boldmath $Z$}_2$-symmetry.
Similarly, in the dual theory,
the string fields with odd domains never appear
for the tree amplitudes when the initial state consists only of the fields
with even
domains.
Thus we can conclude that at the level of the tree amplitudes the
T-duality
symmetry holds, and that $\tilde{J}^{(+)}_{2n}$ in the original theory are
identified with $\tilde{K}_{2n}$ in the dual theory.

\section{Concluding remarks}

We have presented a detailed study of the Schwinger-Dyson equations
and the stochastic Hamiltonians for the two-matrix model formulated in both
the
dual and original  variables.
We have first performed the duality transformation
at the discrete level and then taking the
scaling limit separately in both the original and the dual formulations.
To deepen our
understanding on duality and also to
extend our results to more general cases, it is
desirable to find a direct transformation
of the string fields under the T-dual transformation
in the continuum limit.

Our two main results were followings:
\begin{enumerate}
\item The theory in the scaling limit is duality symmetric only in the sphere
approximation. The absence  of the duality symmetry for higher genus
originates from the existence of nontrivial
global $\mbox{\boldmath $Z$}_2$ vector fields and is explicitly confirmed for
the disk amplitude with one handle.
\item We extended the previous discussion on the
remarkable commutativity property between the
mixing of the string fields in the process of taking the scaling limit
and the merging-splitting interactions of the string fields,
by proposing a general rule of the string-field mixing.
\end{enumerate}

The first result, which is a consequence of the absence
of the momentum-winding symmetry, might be important in thinking about
T-duality transformation in string theory for general manifolds
without simple target-space isometry.
If we want to treat the dual transformed theories
in equal footing with the original theories  as
required once we introduce D-branes,
this would suggest that the
fundamental variables are worldsheet-vector fields or currents instead of
the target space coordinates. Also,
if we reformulate the original spin model
    from the beginning in terms of the
link variables instead of the site variables,
it may be possible to construct
completely duality symmetric models.
In fact, it is easy to define such a self-dual lattice model using
the same method as  a self-dual
$XY$ model as a lattice-regularized version of  a
circle-compactified string theory. This is discussed in Appendix A.
A difficulty is that such a model in general requires us
to introduce both the site and link variables
simultaneously
\footnote{
For an example of such a model, see \cite{kaz}.
} in order to include
the global winding-like modes before  the dual transformation
and cannot be described in terms of simply solvable matrix models.

The second result concerning on the
universal structure of  string field theories might be useful in discussing
possible symmetry structure and the background independence
of string field
theories in general, quite independently of the
problem of the operator mixing.

      From the viewpoint of our original motivation
to the present work, a crucial question is whether
this kind of the formulation of string field theory
is really  possible for the higher-dimensional models,
especially,
including supersymmetry. If we can write down a single
Hamiltonian which is supposed to govern all-genus behaviors of the
theory at once ,
it would be a promising starting point
for studying the non-perturbative symmetries
 including S-duality structure.

\acknowledgements
The work of T. Y. is partially supported by a US-Japan Collaborative
Program from the  Japan Society for the Promotion of Science.
The work of Y. O. is partially supported by a Predoctoral
Research Fellowship from the Japan Society for the
Promotion of Science.

\appendix
\renewcommand{\theequation}{\Alph{section}\arabic{equation}}

\noindent
\section{String T-Duality as Kramers-Wannier Duality}
In this Appendix, we briefly discuss how the T-duality
symmetry of circle-compactified
string theories is reformulated as a special case of the
Kramers-Wannier duality, namely that  of an XY model with a Villain-type
action on a general lattice. For convenience, we will use the
familiar notation of square lattice,
with $\triangle_{\mu}$ being the directional difference
between nearest neighbor sites along the direction $\mu$.
Note however that the arguments below
are valid for arbitrary lattices
with nearest neighbor interactions.

The partition function of the model for a
lattice on an arbitrary two-dimensional surface is
defined as
\begin{equation}
Z=\oint [dX]\bigl(
\prod_{s, \mu} \sum_{m_{s,\mu}}\bigr)
\prod_s \delta(\triangle_{\mu} m_{s, \nu}-\triangle_{\nu}m_{s, \mu})
\exp \bigl[-{R^2 \over  4\pi}\sum_{s, \mu}
(\triangle_{\mu}X_s - 2\pi m_{s, \mu})^2
\bigr]
\label{orgpartition}
\end{equation}
where the constraint
$\triangle_{\mu} m_{s, \nu}-\triangle_{\nu}m_{s, \mu}=0$
 is imposed in order to suppress the {\it local} vortex excitations
which are forbidden in the
usual continuum formulation of circle-compactified
string theory. The target space coordinate $X$
of the compactification circle is
normalized such that
the periodicity is $2\pi$ and the symbol $\oint [dX]$
denotes the path integral with respect to
$X$ over a single period.
Correspondingly, the variable $m_{s, \mu}$
takes integer values, and the radius parameter $R$
appears in front of the
world-sheet action.  Thus, the $R^2$ just plays the role
of the inverse temperature.  Note also that, after solving the
constraint, the integer-valued vector field $m_{s, \mu}$ just
represents the winding modes along nontrivial homology cycles
for an arbitrary surface of generic genus, since
the pure gauge mode of the form $m_{s, \mu}=
\triangle_{\mu}m_s$  is absorbed into a shift
of the target coordinate
$X_s \rightarrow X_s + 2\pi m_s$, which in turn ensures the
periodicity of the integrand of the partition function
after the summation over $m_s$.

Now performing the Fourier transformation and
introducing auxiliary variable $\tilde X_{\tilde s}$
 to exponentiate the constraint,
we have
\[
Z\propto \oint [dX]
\bigl(\prod_{s, \mu}\sum_{m_{s,\mu}}\bigr)
\oint [d\tilde{X}]
\int \bigl(\prod_{\tilde{s}, \mu}d\psi_{\tilde{s},\mu}\bigr)
\exp
\sum_{\tilde s}[i\tilde X_{\tilde{s}}{\epsilon_{\mu\nu}\over 2}
(\triangle_{\mu}m_{s, {\nu}}
-\triangle_{\nu}m_{s, {\mu}})
\]
\[+
i {1\over 2\pi}\epsilon_{\mu\nu}\psi_{\tilde{s},\mu}(\triangle_{\nu}X_s
- 2\pi m_{s, \nu})
-{1 \over 4\pi R^2}\psi_{\tilde{s}, \mu}^2]
\]
where the integrals for $\tilde{X}_{\tilde{s}}$ is again over
a single
period $2\pi$, while the integral for $\psi_{\tilde{s},\mu}$
is the whole real axis. Note also that all the new
variables are introduced on the dual lattice;
namely, $\tilde{X}_{\tilde{s}}$ on dual sites,
$\psi_{\tilde{s}, \mu}$ on dual links, and the
correspondence between the original site
$s$ and the dual site $\tilde{s}$ is
fixed uniquely in the first and the second terms
on the exponential of the integrand.
The summation over $m_{s,\mu}$ leads to the constraint
\begin{equation}
\psi_{\tilde{s}, \mu} =\triangle_{\mu}\tilde{X}_{\tilde{s}} -
2\pi n_{\tilde{s}, \mu}
\end{equation}
where $n_{\tilde{s}, \mu}$ defined on each dual link
takes arbitrary integer values.
After substituting this result and performing the $X$ integral,
we obtain the constraint for $n_{\tilde{s}, \mu}$
\begin{equation}
\triangle_{\mu}n_{\tilde{s}, \nu}-\triangle_{\nu}n_{\tilde{s}, \mu}=0
\end{equation}
whose solutions are again the global integer-valued
vector fields
 along nontrivial
homology cycles. Thus, the partition function now
takes the form
\begin{equation}
Z\propto \oint [d\tilde{X}]\bigl(\prod_{\tilde{s}, \mu}\sum_{n_{\tilde{s},
\mu}}\bigr)
\prod_{\tilde s} \delta(\triangle_{\mu} n_{\tilde{s}, \nu}
-\triangle_{\nu}n_{\tilde{s}, \mu})
\exp\bigl[-{1 \over  4\pi R^2}\sum_{s, \mu}
(\triangle_{\mu}\tilde{X}_{\tilde{s }}-
2\pi n_{\tilde{s}, \mu})^2\bigr],
\end{equation}
which is identical to the
original form (\ref{orgpartition}) with the
correspondence $R\leftrightarrow 1/R$. Note that the
integer-valued vector field
$n_{\tilde{s},\mu}$, appearing as the
winding mode of the dual coordinate
$\tilde{X}_{\tilde{s}}$ can be interpreted
as the momentum mode
for the original coordinate $X_s$.
The symmetrical role of the winding ($m_{\mu}$) and the
momentum modes ($n_{\mu}$) for arbitrary genus is precisely the
structure of the usual T-duality symmetry.

Obviously, the above model can be easily
generalized for $Z(N)$ target space preserving
the self-dual structure. However, it is
not easy to construct a simply solvable
matrix model corresponding to such a lattice model
since it requires us to introduce both link and site
variables simultaneously.

\section{Disk and Cylinder Amplitudes in Dual Two-Matrix Model}
\renewcommand{\thesubsection}{\Alph{section}.\arabic{subsection}}
\setcounter{subsection}{0}
\setcounter{equation}{0}

Here, we obtain various disk and cylinder amplitudes in the dual two-matrix
model,
by taking the continuum limit of the corresponding Schwinger-Dyson
equations. Some of our results have already been derived in Ref. \cite{COT}.
However, for making this paper self-contained, we will give a brief mention
also for those results.

  At first, we introduce the following notations for the disk amplitudes:
\begin{eqnarray*}
V_{X^nY^m} & = &  \left\langle\frac{1}{N}\mbox{tr} X^{n}Y^{m} \right\rangle_0,
\\
V(\xi) & = & \left\langle\frac{1}{N}\mbox{tr}\frac{1}{\xi-X} \right\rangle_0,
\\
V_{j}(\xi) & = &  \left\langle\frac{1}{N}\mbox{tr}\frac{1}{\xi-X}Y^{j}
\right\rangle_0, \\
V^{(m)}(\xi_{1},\cdots,\xi_{m}) & = &  \left\langle\frac{1}{N}\mbox{tr}
  \frac{1}{\xi_{1}-X}Y\cdots
  Y\frac{1}{\xi_{m}-X}Y \right\rangle_0, \\
V^{(m)}_j(\xi_{1};\xi_2 \cdots,\xi_{m}) & = &  \left\langle\frac{1}{N}\mbox{tr}
  \frac{1}{\xi_{1}-X}Y^j\frac{1}{\xi_2-X}Y\cdots
  Y\frac{1}{\xi_{m}-X}Y \right\rangle_0,
\end{eqnarray*}
and for the cylinder amplitudes:
\begin{eqnarray*}
V^{\mbox{\scriptsize cyl}}_{Y|Y} & = &  \left\langle\mbox{tr} Y \mbox{tr} Y
\right\rangle_0, \\
V^{\mbox{\scriptsize cyl}}_{n|m}(\xi|\eta) & = &
\left\langle\mbox{tr}\frac{1}{\xi-X}Y^n
                          \mbox{tr}\frac{1}{\eta-X}Y^m \right\rangle_0, \\
V^{\mbox{\scriptsize cyl}}_{n|1}(\xi_1;\xi_2,\cdots,\xi_m|\eta)
& = &
    \left\langle\mbox{tr}\left(\frac{1}{\xi_1-X}Y^n\frac{1}{\xi_2-X}Y\cdots Y
      \frac{1}{\xi_m-X}Y\right)\mbox{tr}\frac{1}{\eta-X}Y \right\rangle_0, \\
V^{\mbox{\scriptsize cyl}}_{n|Y^j}(\xi_1;\xi_2,\cdots,\xi_m)
& = &
    \left\langle\mbox{tr}\left(\frac{1}{\xi_1-X}Y^n\frac{1}{\xi_2-X}Y\cdots Y
      \frac{1}{\xi_m-X}Y\right)\mbox{tr} Y^j \right\rangle_0, \\
V^{\mbox{\scriptsize cyl}~(m|1)}(\xi_1,\cdots,\xi_m|\eta_1)
& = &
    \left\langle\mbox{tr}\left(\frac{1}{\xi_1-X}Y\cdots Y\frac{1}{\xi_m-X}Y
\right)
        \mbox{tr}\frac{1}{\eta_1-X}Y \right\rangle_0, \\
V^{\mbox{\scriptsize cyl}~(m)}_{Y^n}(\xi_1,\cdots,\xi_m) & = &
   \left\langle\mbox{tr}\left(\frac{1}{\xi_1-X}Y\cdots
Y\frac{1}{\xi_m-X}Y\right)
       \mbox{tr} Y^n \right\rangle_0,
\end{eqnarray*}
where $ \left\langle\cdots \right\rangle_0$ stands for the connected amplitude
on the surface
with no handle.

\subsection{$V(\xi)$, $V_2(\xi)$}

 As in Ref. \cite{COT}, the closed equation for $V(\xi)$ is derived by
combining the five Schwinger-Dyson equations obtained from the following
identities in the planar limit:
\begin{eqnarray}
0 & = & \int d^{N^2}Xd^{N^2}Y \sum_{\alpha}\frac{\partial}{\partial X_{\alpha}}
  \mbox{tr}\left(\frac{1}{\xi-X}t^{\alpha}\right){\rm e}^{-S_D}, \nonumber \\
0 & = & \int d^{N^2}Xd^{N^2}Y \sum_{\alpha}\frac{\partial}{\partial Y_{\alpha}}
  \mbox{tr}\left(\frac{1}{\xi-X}Yt^{\alpha}\right){\rm e}^{-S_D}, \nonumber \\
0 & = & \int d^{N^2}Xd^{N^2}Y \sum_{\alpha}\frac{\partial}{\partial Y_{\alpha}}
  \mbox{tr}\left(\frac{1}{\xi-X}YXt^{\alpha}\right){\rm e}^{-S_D}, \nonumber \\
0 & = & \int d^{N^2}Xd^{N^2}Y \sum_{\alpha}\frac{\partial}{\partial X_{\alpha}}
  \mbox{tr}\left(\frac{1}{\xi-X}Yt^{\alpha}Y\right){\rm e}^{-S_D}, \nonumber \\
0 & = & \int d^{N^2}Xd^{N^2}Y \sum_{\alpha}\frac{\partial}{\partial X_{\alpha}}
  \mbox{tr}\left(\frac{1}{\xi-X}Y^2t^{\alpha}\right){\rm e}^{-S_D}.
\end{eqnarray}
The result is
\begin{equation}
\hat{g} V(\xi)^3+f_2  V(\xi)^2+f_1 V(\xi)+f_0=0,  \label{diskSD1}
\end{equation}
where
\begin{eqnarray*}
f_2 & = & \hat{g}^2\xi^2+(5c-1)\hat{g}\xi-2c(1+c), \\
f_1 & = & 4c\hat{g}^2\xi^3-(6c-2c^2)\hat{g}\xi^2+2c(1-c^2)\xi
-2\hat{g}^2 V_X+\hat{g}(1-3c),
 \\
f_0 & = & (6c-2c^2-4c\hat{g}\xi)\hat{g} V_X-4c\hat{g}^2 V_{X^2}
-4c\hat{g}^2\xi^2
          +(6c-2c^2)\hat{g}\xi \\
    &   & ~-2c(1-c^2)-\hat{g}^2.
\end{eqnarray*}
Since the original and dual theories are connected by the transformation
$X\leftrightarrow (A+B)/\sqrt{2}$, $Y\leftrightarrow (A-B)/\sqrt{2}$,
$\hat{g}\leftrightarrow g/\sqrt{2}$, the partition functions of both theories
are identical. So the critical points of the couplings $c, g$ are also
identical:
$c_*=\frac{-1+2\sqrt{7}}{27}$, $g_*=\sqrt{10c_*^3}$,
and $ V_X$ and $ V_{X^2}$ can be written by the amplitudes in the original
theory $W_1= \left\langle\frac{1}{N}\mbox{tr} A \right\rangle_0$ and $W_3
= \left\langle\frac{1}{N}\mbox{tr} A^3 \right\rangle_0$
whose forms in the
continuum limit are already
known in Ref. \cite{SY}:
\begin{equation}
 V_X=\sqrt{2}W_1,~~~ V_{X^2}=\frac{1-c^2}{cg}W_1-\frac{g}{c}W_3
-\frac{1}{c}.
  \label{A-4-1}
\end{equation}
The critical point $\xi_*$ is determined by a similar way as $P_*$
in the original theory,
$$  \xi_*=\frac{\hat{s}}{\sqrt{5c_*}}, ~~~\hat{s}=1+\sqrt{7}.
$$
Introducing the lattice spacing $a$ and the variables
in the continuum theory as
$$
\hat{g}=\hat{g}_*(1-a^2\frac{\hat{s}^2}{10}\hat{T}), ~~~\xi=\xi_*(1+ay),
$$
the continuum limit of the solution of (\ref{diskSD1}) is given by
\begin{eqnarray}
 V(\xi) & = &  V^{\mbox{\scriptsize non}}(\xi)+\hat{V}(\xi), \nonumber \\
 V^{\mbox{\scriptsize non}}(\xi) & = & -\frac{f_2}{3\hat{g}}, \nonumber \\
\hat{V}(\xi) & = & a^{4/3}\frac{c^{1/2}\hat{s}^{4/3}}{2^{2/3}\sqrt{5}}
      [(y+\sqrt{y^2-\hat{T}})^{4/3}+(y-\sqrt{y^2-\hat{T}})^{4/3}]+O(a^{5/3})
\nonumber \\
   & \equiv & a^{4/3}\frac{c^{1/2}\hat{s}^{4/3}}{2^{2/3}\sqrt{5}}v(y)
+O(a^{5/3}),
\end{eqnarray}
where $ V^{\mbox{\scriptsize non}}$ and $\hat{V}$ denote the non-universal and
universal parts, respectively. Here and below, $c$ appearing in the formulas
in the continuum limit is understood to be fixed at the critical point
$c_*$.

  Also, $ V_2(\xi)$ can be derived in the same way as in the original
theory:
\begin{eqnarray}
 V_2(\xi) & = &  V_2^{\mbox{\scriptsize non}}(\xi)+\hat{V}_2(\xi), \nonumber \\
 V^{\mbox{\scriptsize non}}_2(\xi) & = & \frac{1}{\hat{g}}[((1-c-\hat{g}\xi)\xi
- V^{\mbox{\scriptsize non}}(\xi))
      V^{\mbox{\scriptsize non}}(\xi) +\hat{g} V^{\mbox{\scriptsize non}}_X-1+c
+\hat{g}\xi  \nonumber \\
     &    & ~+((1-c-\hat{g}\xi)\xi-2 V^{\mbox{\scriptsize non}}(\xi))
\hat{V}(\xi)], \nonumber \\
\hat{V}_2(\xi) & = & -\frac{1}{\hat{g}}\hat{V}(\xi)^2+\hat{V}_X
      \equiv  a^{8/3}\frac{\hat{s}^{8/3}}{2^{4/3}\cdot 5 c^{1/2}}v_2(y)+O(a^3),
                  \nonumber \\
v_2(y) & =& -v(y)^2+\frac{3}{2}\hat{T}^{4/3},
\label{spinflipmixing}
\end{eqnarray}
where the non-universal and universal parts of $ V_X$:
$ V^{\mbox{\scriptsize non}}_X$ and
$\hat{V}_X$ are known from those of $W_1$ via (\ref{A-4-1}).
The $y$-independent
constant in $\hat{V}_2(\xi)$ is different from that in the corresponding
amplitude
($\hat{W}_1(\zeta)$) of the original theory. However, in the rule for
identifying
the non-universal and universal parts, the term $\hat{V}_X$,
which can be seen as
an amplitude with a simpler spin configuration than $\hat{V}_2(\xi)$,
is allowed to be absorbed in the
non-universal part. After adopting such a convention for both
original and dual theories, there are no real differences.
In fact, since the operator corresponding to this amplitude
is always accompanied by $\xi$-derivative in the leading contribution
of the Hamiltonian, this change does not affect the continuum
Hamiltonian.

\subsection{$ V^{(2)}(\xi_1,\xi_2)$, $ V^{(4)}(\xi_1,\xi_2,\xi_3,\xi_4)$}

  For the higher amplitude $ V^{(2k)}$, we can derive the dual version of
Staudacher's recursion relation (Eq. (20) in Ref. \cite{St}).
Here, we consider the continuum limit for the cases $k=1$ and 2. The
corresponding
Schwinger-Dyson equations are
\begin{equation}
 V^{(2)}(\xi_1,\xi_2)=\frac{ V(\xi_1) V(\xi_2)
   -\hat{g} (V_2(\xi_1)+V_2(\xi_2))}{1+c-\hat{g}(\xi_1+\xi_2)},
\end{equation}
\begin{eqnarray}
\lefteqn{ V^{(4)}(\xi_1,\xi_2,\xi_3,\xi_4)=\frac{1}{1+c-\hat{g}(\xi_1+\xi_4)}}
                              \nonumber \\
 &  \times &[((1-c-\hat{g}\xi_1)\xi_1- V(\xi_1)- V(\xi_3)- V(\xi_4))
                      D_z(\xi_1,\xi_3) V^{(2)}(z,\xi_2) \nonumber \\
 &         & ~+((1-c-\hat{g}\xi_2)\xi_2- V(\xi_1)- V(\xi_2)- V(\xi_4))
                      D_z(\xi_2,\xi_4) V^{(2)}(z,\xi_3) \nonumber \\
 &         & ~+(1-c-\hat{g} (\xi_1+\xi_3)) V^{(2)}(\xi_2,\xi_3)
                +(1-c-\hat{g} (\xi_2+\xi_4)) V^{(2)}(\xi_3,\xi_4)  \nonumber \\
 &         & ~+\hat{g} V_2(\xi_2)+\hat{g} V_2(\xi_3)].
\end{eqnarray}
Putting $\xi_i=\xi_*(1+ay_i)$ and expanding with respect to $a$,
the result of
$ V^{(2)}$ becomes
\begin{eqnarray}
 V^{(2)}(\xi_1,\xi_2) & = &  V^{(2)~\mbox{\scriptsize non}}(\xi_1,\xi_2)
                                +\hat{V}^{(2)}(\xi_1,\xi_2), \nonumber \\
 V^{(2)~\mbox{\scriptsize non}}(\xi_1,\xi_2) & = & \frac{1}{5}(1+4\hat{s})
    -2\sqrt{\frac{c}{5}}(\xi_1+\xi_2)+\frac{2}{\sqrt{5c}}
     (\hat{V}(\xi_1)+\hat{V}(\xi_2)),
                         \nonumber \\
\hat{V}^{(2)}(\xi_1,\xi_2) & = & a^{5/3}\frac{\hat{s}^{5/3}}{10\cdot 2^{1/3}}
     \frac{-\hat{v}(y_1)^2-\hat{v}(y_2)^2-\hat{v}(y_1)\hat{v}(y_2)
+3\hat{T}^{4/3}}{y_1+y_2}
                          \nonumber \\
  &  & +a^2\frac{2\hat{s}}{75}(18\hat{T}+10\hat{s}(y_1^2+y_2^2)
-15\hat{s} y_1y_2)
+O(a^{7/3})
                          \nonumber \\
  & \equiv & a^{5/3}\frac{\hat{s}^{5/3}}{10\cdot 2^{1/3}}v^{(2)}(y_1,y_2)
                        +O(a^2),
\label{V2mixing}
\end{eqnarray}
where $\hat{v}(y)$ is defined by the rescaling of $\hat{V}(\xi)$:
$$
\hat{V}(\xi)=a^{4/3}\frac{c^{1/2}\hat{s}^{4/3}}{2^{2/3}\sqrt{5}}\hat{v}(y),
$$
and
$$
v^{(2)}(y_1,y_2)=
\frac{-v(y_1)^2-v(y_2)^2-v(y_1)v(y_2)+3\hat{T}^{4/3}}{y_1+y_2}.
$$

Using this and repeating the same procedure for $k=2$, we have
\begin{eqnarray}
 V^{(4)}(\xi_1,\xi_2,\xi_3,\xi_4) & = &
                      V^{(4)~\mbox{\scriptsize non}}(\xi_1,\xi_2,\xi_3,\xi_4)
  +\hat{V}^{(4)}(\xi_1,\xi_2,\xi_3,\xi_4), \nonumber \\
 V^{(4)~\mbox{\scriptsize non}}(\xi_1,\xi_2,\xi_3,\xi_4) & = &
1 -\frac{4}{5c}[D_z(\xi_1,\xi_3)\hat{V}(z)+D_z(\xi_2,\xi_4)\hat{V}(z)]
\nonumber \\
  &  & -\frac{2}{\sqrt{5c}}[D_z(\xi_1,\xi_3)(\hat{V}^{(2)}(z,\xi_2)
             +\hat{V}^{(2)}(z,\xi_4)) \nonumber \\
  &  & ~+D_z(\xi_2,\xi_4)(\hat{V}^{(2)}(\xi_1,z)+\hat{V}^{(2)}(\xi_3,z))],
\nonumber \\
\hat{V}^{(4)}(\xi_1,\xi_2,\xi_3,\xi_4) &= & a\frac{\hat{s}}{20}
  v^{(4)}(y_1,y_2,y_3,y_4)+O(a^{4/3}) \nonumber \\
 & = & a\frac{\hat{s}}{20}\{-\frac{16}{3}(y_1+y_2+y_3+y_4)
        +\frac{1}{2}\frac{1}{y_1-y_3}\frac{1}{y_2-y_4}  \nonumber \\
 &   & \times [
        -v^{(2)}(y_1,y_2)(v(y_1)+v(y_2)+2v(y_3)+2v(y_4)) \nonumber \\
 &   & ~+v^{(2)}(y_1,y_4)(v(y_1)+2v(y_2)+2v(y_3)+v(y_4)) \nonumber \\
 &   & ~+v^{(2)}(y_2,y_3)(2v(y_1)+v(y_2)+v(y_3)+2v(y_4)) \nonumber \\
 &   & ~-v^{(2)}(y_3,y_4)(2v(y_1)+2v(y_2)+v(y_3)+v(y_4))]\} \nonumber \\
 &   & +O(a^{4/3}).
\label{V4mixing}
\end{eqnarray}

 $\hat{V}^{(2)}(\xi_1,\xi_2)$ has the same form as the corresponding< amplitude
$\hat{W}^{(2)}(\zeta,\sigma)$ in the original theory,
which has been pointed out in
Ref. \cite{COT}. Also, $\hat{V}^{(4)}(\xi_1,\xi_2,\xi_3,\xi_4)$ is the same as
$\hat{W}^{(4)}(\zeta_1,\sigma_1,\zeta_2,\sigma_2)$,
except the linear term of $y_i$.
Since this term is analytic with respect to $y_i$'s,
it is allowed to be absorbed to
the non-universal part. So, the difference can be regarded
as non-universal.

\subsection{$ V^{(3)}_{2}(\xi_{1},\xi_{2},\xi_{3};)$}

   This amplitude can be obtained from the Schwinger-Dyson equation:
$$
0=\int d^{N^2}Xd^{N^2}Y\sum_{\alpha}\frac{\partial}{\partial X_{\alpha}}
\mbox{tr}\left(\frac{1}{\xi_1-X}Y
  \frac{1}{\xi_2-X}Y\frac{1}{\xi_3-X}t^{\alpha}\right){\rm e}^{-S_D}.
$$
A similar calculation as before leads to the result
\begin{eqnarray}
 V^{(3)}_2(\xi_1,\xi_2,\xi_3;) & = &  V^{(3)~\mbox{\scriptsize non}}_2
(\xi_1,\xi_2,\xi_3;)
                   +\hat{V}^{(3)}_2(\xi_1,\xi_2,\xi_3;) , \nonumber \\
 V^{(3)~\mbox{\scriptsize non}}_2(\xi_1,\xi_2,\xi_3;) & = &
    \frac{1}{\hat{g}}( V^{\mbox{\scriptsize non}}(\xi_1)
+ V^{\mbox{\scriptsize non}}(\xi_3)-(1-c-\hat{g}\xi_3)\xi_3)
       D_z(\xi_1,\xi_3) V^{(2)}(z,\xi_2) \nonumber \\
  &  & -2\sqrt{\frac{c}{5}}\frac{1}{\hat{g}}(\hat{V}(\xi_1)+\hat{V}(\xi_3))
        -\frac{2}{\sqrt{5c}}D_z(\xi_1,\xi_3)\hat{V}_2(z) \nonumber \\
  &  & - V_2(\xi_2)-\frac{1}{\hat{g}}(1-c-\hat{g}(\xi_1+\xi_3))
V^{(2)}(\xi_1,\xi_2),
                    \nonumber \\
\hat{V}^{(3)}_2(\xi_1,\xi_2,\xi_3;) & = & a^2\frac{\hat{s}^2}{20\sqrt{5c}}
  (v(y_1)+v(y_3))D_z(y_1,y_3)v^{(2)}(z,y_2)+O(a^{7/3}) \nonumber \\
  & \equiv & a^2\frac{\hat{s}^2}{20\sqrt{5c}}v^{(3)}_2(y_1,y_2,y_3;)
+O(a^{7/3}).
\end{eqnarray}

\subsection{$ V^{{\rm cyl}}_{Y|Y}$, $ V^{{\rm cyl}~(1)}_{Y}(\xi)$,
$ V^{{\rm cyl}}_{1|1}(\xi|\xi')$}

  Let us next consider the cylinder amplitudes.
  As presented in \cite{COT}, $ V^{\mbox{\scriptsize cyl}}_{Y|Y}$ and
$V^{\mbox{\scriptsize cyl}~(1)}_Y(\xi)$
can be obtained from the single Schwinger-Dyson equation
\begin{equation}
 V(\xi)-(1+c-2\hat{g}\xi) V^{\mbox{\scriptsize cyl}~(1)}_{Y}(\xi)
-2\hat{g} V^{\mbox{\scriptsize cyl}}_{Y|Y}=0
\end{equation}
by using the fact that the coefficient of
$ V^{\mbox{\scriptsize cyl}~(1)}_{Y}(\xi)$ vanishes
at $\xi=\frac{1+c}{2\hat{g}}=\xi_*+O(a^2)$. The results are
\begin{eqnarray}
 V^{\mbox{\scriptsize cyl}}_{Y|Y} & = & \frac{1}{2\hat{g}}
V(\frac{1+c}{2\hat{g}}) \nonumber \\
      & = & \frac{1}{5c}-a^{4/3}\frac{\hat{s}^{4/3}}{5\cdot 2^{5/3}c}
\hat{T}^{2/3}
            +O(a^2),  \\
 V^{\mbox{\scriptsize cyl}~(1)}_Y(\xi) & = & \frac{1}{\sqrt{5c}}
  -a^{1/3}\frac{\hat{s}^{1/3}}{2^{5/3}\sqrt{5c}}\frac{\hat{v}(y)
+\hat{T}^{2/3}}{y}
+O(a)
                   \nonumber \\
    & \equiv & \frac{1}{\sqrt{5c}}+\hat{V}^{\mbox{\scriptsize cyl}~(1)}_Y(\xi).
     \label{V1mixing}
\end{eqnarray}

  For $ V^{\mbox{\scriptsize cyl}}_{1|1}(\xi|\xi')$,
considering the Schwinger-Dyson
equation
obtained from the identity
$$
0=\int d^{N^2}Xd^{N^2}Y\sum_{\alpha}\frac{\partial}{\partial Y_{\alpha}}
\left(\mbox{tr}\left(
  \frac{1}{\xi-X}t^{\alpha}\right)\mbox{tr}
\left(\frac{1}{\xi'-X}Y\right){\rm e}^{-S_D}
  \right),
$$
we have
\begin{eqnarray}
 V^{\mbox{\scriptsize cyl}}_{1|1}(\xi|\xi') & = & \frac{-D_z(\xi,\xi') V(z)
-2\hat{g} V^{\mbox{\scriptsize cyl}~(1)}_Y(\xi')}
{1+c-2\hat{g}\xi}  \nonumber \\
 & = & a^{-2/3}\frac{1}{2^{5/3}\hat{s}^{2/3}}\frac{1}{yy'}\left(
      \frac{y'v(y)-yv(y')}{y-y'}-\hat{T}^{2/3}\right)+O(a^0).
\end{eqnarray}

\subsection{$ V^{{\rm cyl}}_{2|Y}(\xi_1;\xi_2)$,
$ V^{{\rm cyl}~(3)}_Y(\xi_1,\xi_2,\xi_3)$}

We start with the following two identities:
\begin{eqnarray*}
0 & = & \int d^{N^2}Xd^{N^2}Y\sum_{\alpha}\frac{\partial}{\partial X_{\alpha}}
  \mbox{tr}\left(\frac{1}{\xi_1-X}t^{\alpha}\frac{1}{\xi_2-X}Y\right)
\mbox{tr} Y{\rm e}^{-S_D},
 \\
0 & = & \int d^{N^2}Xd^{N^2}Y\sum_{\alpha}
\frac{\partial}{\partial Y_{\alpha}}
\mbox{tr}\left(\frac{1}{\xi_1-X}Y\frac{1}{\xi_2-X}Y\frac{1}{\xi_3-X}t^{\alpha}
    \right)\mbox{tr} Y{\rm e}^{-S_D}.
\end{eqnarray*}
The amplitudes $ V^{\mbox{\scriptsize cyl}}_{2|Y}(\xi_1;\xi_2)$ and
$ V^{\mbox{\scriptsize cyl}~(3)}_Y(\xi_1,\xi_2,\xi_3)$ can be obtained by
using the
Schwinger-Dyson equations derived from the above identities.

  The results are as follows:
\begin{eqnarray}
 V^{\mbox{\scriptsize cyl}}_{2|Y}(\xi_1;\xi_2) & = &
V^{\mbox{\scriptsize cyl}~\mbox{\scriptsize non}}_{2|Y}(\xi_1;\xi_2)
    +\hat{V}^{\mbox{\scriptsize cyl}}_{2|Y}(\xi_1;\xi_2), \nonumber \\
 V^{\mbox{\scriptsize cyl}~\mbox{\scriptsize non}}_{2|Y}(\xi_1;\xi_2)
& = & \frac{1}{\hat{g}}( V^{\mbox{\scriptsize non}}(\xi_1)
  + V^{\mbox{\scriptsize non}}(\xi_2)-(1-c-\hat{g}\xi_1)\xi_1)D_z(\xi_1,\xi_2)
V^{\mbox{\scriptsize cyl}~(1)}_Y(z)
                                     \nonumber \\
  &   &  - V^{\mbox{\scriptsize cyl}}_{Y|Y}
-\frac{1}{\hat{g}}(1-c-\hat{g}(\xi_1+\xi_2))
V^{\mbox{\scriptsize cyl}~(1)}_Y(\xi_2),
                                      \nonumber \\
\hat{V}^{\mbox{\scriptsize cyl}}_{2|Y}(\xi_1;\xi_2) & = & -a^{2/3}
\frac{\hat{s}^{2/3}}{2^{7/3}\cdot 5c}
   (v(y_1)+v(y_2))D_z(y_1,y_2)\frac{v(z)+\hat{T}^{2/3}}{z} \nonumber \\
   &   & +O(a),
\end{eqnarray}
\begin{eqnarray}
 V^{\mbox{\scriptsize cyl}~(3)}_Y(\xi_1,\xi_2,\xi_3) & = &
     V^{\mbox{\scriptsize cyl}~(3)~\mbox{\scriptsize
non}}_Y(\xi_1,\xi_2,\xi_3)+
     \hat{V}^{\mbox{\scriptsize cyl}~(3)}_Y(\xi_1,\xi_2,\xi_3), \nonumber \\
 V^{\mbox{\scriptsize cyl}~(3)~\mbox{\scriptsize non}}_Y(\xi_1,\xi_2,\xi_3)
& = & \frac{2}{\sqrt{5c}}[
  -D_z(\xi_1,\xi_2)-D_z(\xi_2,\xi_3)-D_z(\xi_3,\xi_1)]
\hat{V}^{\mbox{\scriptsize cyl}~(1)}_Y(z),
\nonumber \\
\hat{V}^{\mbox{\scriptsize cyl}~(3)}_Y(\xi_1,\xi_2,\xi_3) & = & a^{-1/3}
  \frac{\hat{s}^{-1/3}}{4\cdot 2^{1/3}\sqrt{5c}}[
  \frac{\hat{T}^{2/3}}{y_1y_2y_3}v(y_1)-
   \frac{2\hat{T}^{4/3}}{(y_1+y_2)(y_2+y_3)(y_3+y_1)}  \nonumber \\
 &  & ~-v(y_1)v(y_2)\left(\frac{1}{y_1(y_1-y_3)}+\frac{1}{y_2(y_2-y_3)}
\right)
        \frac{1}{y_1+y_2} \nonumber \\
 &  & ~-v(y_1)^2
\frac{3y_1^2-y_1y_2-y_2y_3-y_3y_1}{y_1(y_1^2-y_2^2)(y_1^2-y_3^2)}
                           \nonumber \\
 &  & ~+\mbox{cyclic permutations of $y_1, y_2, y_3$}]
         +O(a^0).                   \label{V3mixing}
\end{eqnarray}

\subsection{$ V^{{\rm cyl}~(5)}_Y(\xi_1,\xi_2,\xi_3,\xi_4,\xi_5)$}

  Here, we show only the result of  the non-universal part of
this amplitude
in order to
examine the mixing among the operators with odd domains.
It is necessary to
consider the two Schwinger-Dyson equations originating from
\begin{eqnarray*}
0 & = & \int d^{N^2}Xd^{N^2}Y \sum_{\alpha}\frac{\partial}{\partial Y_{\alpha}}
\left[\mbox{tr}\left(
\frac{1}{\xi_1-X}Y\frac{1}{\xi_2-X}Y\frac{1}{\xi_3-X}Y\frac{1}{\xi_4-X}Y
\frac{1}{\xi_5-X}t^{\alpha}\right)\mbox{tr} Y{\rm e}^{-S_D}\right], \\
0 & = & \int d^{N^2}Xd^{N^2}Y \sum_{\alpha}\frac{\partial}{\partial X_{\alpha}}
\left[\mbox{tr}\left(
\frac{1}{\xi_1-X}Y\frac{1}{\xi_2-X}Y\frac{1}{\xi_3-X}Y\frac{1}{\xi_4-X}
t^{\alpha}
\right)\mbox{tr} Y{\rm e}^{-S_D}\right].
\end{eqnarray*}
Combining these, we obtain
\begin{eqnarray}
\lefteqn{ V^{\mbox{\scriptsize cyl}~(5)}_Y(\xi_1,\cdots,\xi_5)=
\frac{1}{1+c-\hat{g}(\xi_1+\xi_5)}}
  \nonumber \\
 & \times & [((1-c-\hat{g}\xi_5)\xi_5- V(\xi_1)- V(\xi_2)- V(\xi_5))
            D_z(\xi_2,\xi_5) V^{\mbox{\scriptsize cyl}~(3)}_Y(z,\xi_3,\xi_4)
\nonumber \\
 &   & ~+((1-c-\hat{g}\xi_4)\xi_4- V(\xi_1)- V(\xi_4)- V(\xi_5))
            D_z(\xi_1,\xi_4) V^{\mbox{\scriptsize cyl}~(3)}_Y(z,\xi_2,\xi_3)
\nonumber \\
 &  & ~+(D_z(\xi_3,\xi_5) V^{(2)}(z,\xi_4)+
                D_z(\xi_2,\xi_4) V^{(2)}(z,\xi_3))D_w(\xi_1,\xi_2)
               V^{\mbox{\scriptsize cyl}~(1)}_Y(w)  \nonumber \\
 &  & ~+(D_z(\xi_1,\xi_3) V^{(2)}(z,\xi_2)+
                D_z(\xi_2,\xi_4) V^{(2)}(z,\xi_3))D_w(\xi_4,\xi_5)
               V^{\mbox{\scriptsize cyl}~(1)}_Y(w)  \nonumber \\
 &  & ~+(D_z(\xi_1,\xi_3) V^{(2)}(z,\xi_2)+ V(\xi_3)+ V(\xi_4)
           -(1-c-\hat{g}\xi_4)\xi_4)D_w(\xi_3,\xi_4)
V^{\mbox{\scriptsize cyl}~(1)}_Y(w)  \nonumber \\
 &  & ~+(D_z(\xi_3,\xi_5) V^{(2)}(z,\xi_4)+ V(\xi_2)+ V(\xi_3)
           -(1-c-\hat{g}\xi_3)\xi_3)D_w(\xi_2,\xi_3)
V^{\mbox{\scriptsize cyl}~(1)}_Y(w)  \nonumber \\
 &  & ~+(1-c-\hat{g}(\xi_1+\xi_4))
V^{\mbox{\scriptsize cyl}~(3)}_Y(\xi_1,\xi_2,\xi_3)
       +(1-c-\hat{g}(\xi_2+\xi_5))
V^{\mbox{\scriptsize cyl}~(3)}_Y(\xi_2,\xi_3,\xi_4)   \nonumber \\
 &  & ~-(1-c-\hat{g}(\xi_2+\xi_3))
V^{\mbox{\scriptsize cyl}~(1)}_Y(\xi_2)
       -(1-c-\hat{g}(\xi_3+\xi_4))
V^{\mbox{\scriptsize cyl}~(1)}_Y(\xi_3)    \nonumber \\
 &  & ~-D_z(\xi_1,\xi_5) V^{(4)}(z,\xi_2,\xi_3,\xi_4)
-2\hat{g} V^{\mbox{\scriptsize cyl}}_{Y|Y}].
\end{eqnarray}

  In the continuum limit, the non-universal part
$ V^{\mbox{\scriptsize cyl}~(5)~\mbox{\scriptsize non}}_Y$
becomes
\begin{eqnarray}
 V^{\mbox{\scriptsize cyl}~(5)}_Y(\xi_1,\cdots,\xi_5)  & = &
 V^{\mbox{\scriptsize cyl}~(5)~\mbox{\scriptsize non}}_Y(\xi_1,\cdots,\xi_5)
+\hat{V}^{\mbox{\scriptsize cyl}~(5)}_Y(\xi_1,\cdots,\xi_5)
   \nonumber \\
 V^{\mbox{\scriptsize cyl}~(5)~\mbox{\scriptsize non}}_Y(\xi_1,\cdots,\xi_5)
& = & \frac{4}{5c}[
  D_z(\xi_1,\xi_3)D_w(z,\xi_5)+D_z(\xi_1,\xi_3)D_w(z,\xi_4) \nonumber \\
 &  & ~+D_z(\xi_2,\xi_4)D_w(z,\xi_5)+D_z(\xi_1,\xi_2)D_w(z,\xi_4) \nonumber \\
 &  & ~+D_z(\xi_2,\xi_3)D_w(z,\xi_5)]\hat{V}^{\mbox{\scriptsize cyl}~(1)}_Y(w)
\nonumber \\
 &  & -\frac{2}{\sqrt{5c}}[D_z(\xi_1,\xi_3)
\hat{V}^{\mbox{\scriptsize cyl}~(3)}_Y(z,\xi_4,\xi_5)
           +D_z(\xi_2,\xi_4)
\hat{V}^{\mbox{\scriptsize cyl}~(3)}_Y(\xi_1,z,\xi_5) \nonumber \\
 &  & ~+D_z(\xi_3,\xi_5)\hat{V}^{\mbox{\scriptsize cyl}~(3)}_Y(\xi_1,\xi_2,z)
           +D_z(\xi_1,\xi_4)
\hat{V}^{\mbox{\scriptsize cyl}~(3)}_Y(z,\xi_2,\xi_3) \nonumber \\
 &  & ~+D_z(\xi_2,\xi_5)
\hat{V}^{\mbox{\scriptsize cyl}~(3)}_Y(z,\xi_3,\xi_4)], \nonumber \\
\hat{V}^{\mbox{\scriptsize cyl}~(5)}_Y(\xi_1,\cdots,\xi_5) & = & O(a^{-1}).
\label{V5mixing}
\end{eqnarray}

     From the results until now, we can read off the scaling property of the
universal parts of the string fields
as follows. First, the universal part of $\mbox{tr} Y$ scales as $O(a^{2/3})$.
Since
the partition functions both in the original and dual theories are
identical,
the double scaling limits are also ($\frac{1}{N}=O(a^{7/3})$ in both theories).
Using this, it can be seen that the string fields scale as
$\hat{\Psi}_X=O(a^{4/3})$,
$\hat{\Psi}_n=O(a^{7/3-n/3})$, where $n=1,2,3,\cdots$.
This coincides with the
interpretation of the boundary conformal field theory \cite{Cardy}
as in the original theory.

\vspace{0.5cm}

\section{Continuum Spin-Flip Operator in the Dual Formalism}
\setcounter{equation}{0}

  Similarly as in the original theory, a microscopic domain consisting
only of a single flipped spin can be obtained by
an  integration $\int_C\frac{dy}{2\pi i}$ with respect to $y$
of the macroscopic domain.
We will check this  in the case of  the
disk and cylinder amplitudes.

 First, for the disk amplitudes the following relations can be verified
by direct calculations using the results in Appendix B:
\begin{eqnarray}
\partial_{y_2}\int_C\frac{dy_1}{2\pi i}v^{(2)}(y_1,y_2) & = &
\partial_{y_2}v_2(y_2),
           \nonumber \\
\int_C\frac{dy_1}{2\pi i}v^{(4)}(y_1,y_2,y_3,y_4) & = &
          v^{(3)}_2(y_2,y_3,y_4;),   \label{spin-flip1}
\end{eqnarray}
where the rule of the integral is identical with that
in the original theory.
The contour
$C$ wraps around the negative real axis and the singularities in the left
half plane, and the unintegrated variables, for example $y_2, y_3, y_4$
in the above, are understood to be outside the contour, while
$-y_2, -y_3, -y_4$ to be inside the contour.

  By including the overall normalizations,
(\ref{spin-flip1}) can be written as
\begin{eqnarray}
\partial_{\xi_2}\hat{\oint}\frac{d\xi_1}{2\pi i}\hat{V}^{(2)}(\xi_1,\xi_2) & =
&
\partial_{\xi_2}\hat{V}_2(\xi_2)+O(a^2), \nonumber \\
\hat{\oint}\frac{d\xi_1}{2\pi i}\hat{V}^{(4)}(\xi_1,\xi_2,\xi_3,\xi_4) & = &
  \hat{V}^{(3)}_2(\xi_2,\xi_3,\xi_4;)+O(a^{7/3}),
\end{eqnarray}
where the integral symbol $\hat{\oint}\frac{d\xi}{2\pi i}$ is
used in the sense
of
$$
\hat{\oint}\frac{d\xi}{2\pi i}=a\xi_*\int_C\frac{dy}{2\pi i}.
$$
In contrast to the original theory, no additional
factor of a finite renormalization appears here.

Similarly, for the cylinder amplitudes, we can show that the following
formulas hold:
\begin{eqnarray}
\hat{\oint}\frac{d\xi}{2\pi i}V^{\mbox{\scriptsize cyl}}_{1|1}(\xi|\xi') & = &
\hat{V}^{\mbox{\scriptsize cyl}~(1)}_Y(\xi')+O(a^{2/3}), \nonumber \\
\hat{\oint}\frac{d\xi}{2\pi i}
\hat{V}^{\mbox{\scriptsize cyl}~(3)}_Y(\xi_1,\xi,\xi_2) & = &
\hat{V}^{\mbox{\scriptsize cyl}}_{2|Y}(\xi_1;\xi_2)+O(a).
\end{eqnarray}

\end{document}